\documentclass[twocolumn]{aastex701} 

\usepackage[normalem]{ulem}
\usepackage{makecell}
\usepackage{graphics,xcolor}
\usepackage{amsmath}
\usepackage{soul}
\usepackage{listings}
\usepackage{comment}
\usepackage{CJK}

\shortauthors{Karim et al.}
\received{February 1, 2026}
\revised{May 14, 2026}
\accepted{May 22, 2026}
\submitjournal{The Astrophysical Journal}

\begin{document}

\title{Pushing the Limit of Asteroseismic Detection for Cool Dwarfs using TESS and Deep Learning
}

\author[0009-0005-8878-8756]{Waly M Z Karim}
\affiliation{Department of Physics, University of Rochester, Rochester, NY 14627, USA}
\affiliation{Department of Astronomy, California Institute of Technology, Pasadena, CA 91125, USA}
\email[show]{walym.zkarim64@gmail.com}  
\correspondingauthor{Waly M Z Karim}

\author[0000-0003-2102-3159]{Rocio Kiman}
\affil{Department of Astronomy, California Institute of Technology, Pasadena, CA 91125, USA}
\email{rociokiman@gmail.com}

\author[0000-0002-1988-143X]{Derek Buzasi}
\affil{Department of Astronomy and Astrophysics, University of Chicago, 5801 S. Ellis Ave., Chicago, IL 60637, USA}
\email{}

\author[0000-0002-8791-6286]{Cecilia Garraffo}
\affiliation{Center for Astrophysics | Harvard \& Smithsonian, 60 Garden Street, Cambridge, MA~02138, USA}
\affiliation{Institute for Applied Computational Science, Harvard University, 33 Oxford St., Cambridge, MA 02138, USA}
\email{}

\author[0009-0003-5349-6994]{Joshua D. Wing}
\affiliation{Center for Astrophysics | Harvard \& Smithsonian, 60 Garden Street, Cambridge, MA~02138, USA}
\email{}

\author[0000-0002-4544-0750]{Jim Fuller}
\affil{TAPIR, Mailcode 350-17, California Institute of Technology, Pasadena, CA 91125, USA}
\email{}

\author[0009-0007-6399-315X]{Benjamin J. Ricketts}
\affil{Anton Pannekoek Institute, University of Amsterdam, Science Park 904, Amsterdam, 1098 XH, Netherlands}
\affil{SRON, Niel Bohrweg 4, Leiden, 2033 CA, Netherlands}
\email{}

\author[0000-0001-8907-4089]{Viktor Khalack}
\affil{Université de Moncton, 18 Antonine-Maillet Ave, Moncton, NB E1A 3E9, Canada}
\email{viktor.khalack@umoncton.ca}

\author[0009-0001-9144-5907]{Sajia Shahrin Neha}
\affil{Department of Astrophysical Sciences, Princeton University, Princeton, NJ 08544}
\affil{Department of Astronomy, California Institute of Technology, Pasadena, CA 91125, USA}
\email{}
\begin{abstract}
Asteroseismology provides a powerful probe of stellar interiors by detecting stellar oscillations, including solar-like oscillations, which are stochastically excited by near-surface convection. While thousands of solar-like oscillators have been identified in evolved stars, only a limited number of main-sequence cool dwarfs have confirmed oscillations due to the low amplitudes of their signals. In this work, we train a convolutional autoencoder on TESS two-minute light curves to automatically identify solar-like oscillation features in cool dwarf main sequence and sub-giant stars. Using catalogs of confirmed oscillators for training and validation, our network achieves a classification accuracy of 99.8\% on the test set, along with Precision of 0.945, Recall of 0.998, and F1 Score of 0.971. From the Asteroseismic Target List, our model identifies 3463 potential solar-like oscillators (probability $>$\,50\%). After further analysis, we find a list of 24 candidate stars that have the potential to exhibit solar-like oscillations. Notably, several of these candidates occupy regions of the color–magnitude diagram that are accessible only through more resource-intensive radial velocity observations, thereby has the potential of extending the detection frontier of TESS-based asteroseismology. Our candidate catalog provides a valuable foundation for follow-up efforts aimed at expanding the sample of cool-dwarf solar-like oscillators. This will ultimately improve our understanding of stellar structure and evolution across the lower main sequence and strengthen the evidence for using deep learning techniques to study stellar light curves.

\end{abstract}

\keywords{Late-type stars, Asteroseismology, Convolutional neural networks, Variable stars, Time domain astronomy}

\section{Introduction} \label{sec:introduction}
Asteroseismology, the study of stellar oscillations, provides a powerful means to probe the internal structure and dynamics of stars \citep{ChristensenDalsgaard1983}. Recent progress in photometric time-series observation following space-based telescopes like the Wide-Field Infared Explorer \citep[WIRE,][]{Brunt2006}, the Microvariablity and Oscillations of Stars mission \citep[MOST,][]{Matthews2003}, the Convection, Rotation, and planetary Transits telescope \citep[CoRoT,][]{2006cosp...36.3749B}, Kepler \citep{Koch_2010}, and the Transiting Exoplanet Survey Satellite \citep[TESS,][]{2015JATIS...1a4003R} has revolutionized the field of asteroseismology and revealed fundamental properties of thousands of stars \citep[e.g.,][]{2014ApJS..214...27M, 2014A&A...569A..21L, 2015A&A...580A.141L, 2018ApJS..236...42Y, Y_ld_z_2019}. These missions have made several important asteroseismic discoveries, ranging from measuring global oscillation parameters of stars in different parts of the H-R diagram to obtaining fundamental stellar properties of metal-poor stars in our galaxy \citep{Hon_2021, marasco2025asteroseismologymetalpoorredgiants, 2010ApJ...723.1607H, Metcalfe_2012, Anders_2016, Valentini_2016}. Depending on the type of oscillation, these pulsating stars occupy different regions of the H-R diagram. Solar-like oscillators appear on the cool main sequence and red-giant branch, whereas classical pulsators such as $\delta$~Scuti, RR-Lyrae, and Cepheids lie along the instability strip. 
The low oscillation amplitudes of low-mass main sequence stars, particularly in the late K~dwarfs and M~dwarfs, have restricted the previous attempts to find oscillation signals in them \citep{2025AAS...24511104K, Rodr_guez_L_pez_2014, 2019FrASS...6...76R}.

G and K~dwarfs primarily exhibit solar-like oscillations, while M~dwarfs have been theorized to exhibit similar oscillation features \citep{Rodr_guez_L_pez_2012, Campante_2024, Hatt_2023}. Solar-like oscillations contain sequences of radial overtones that are stochastically excited by turbulent convection close to the stellar photosphere. They can be broadly characterized by two main global parameters: the frequency of maximum power ($\nu_{\text{max}}$) and the large frequency separation ($\Delta \nu$). $\nu_{\text{max}}$ describes the central frequency of the visible Gaussian envelope like power excess, and $\Delta \nu$ describes the approximately regular spacing between consecutive radial overtones for a given angular degree. A precise measurement of these oscillation parameters, combined with an independent measurement of the effective temperature, can be used to identify fundamental properties such as stellar mass, radius and age, with high accuracy \citep[e.g.,][]{2010ApJ...723.1583M, 2012ApJ...757...99S, 2012ApJ...760...32H, Guggenberger2016, 2016ApJ...832..121G, 2022A&A...657A..31M}. 

Any star having a near-surface convection zone can excite solar-like oscillations. Hence, this type of variability has been observed in the main-sequence \citep[e.g.,][]{2014ApJS..210....1C}, sub-giant \citep[e.g.,][]{2012A&A...543A..54A, 2022A&A...657A..31M} and red giant \citep[e.g.,][]{2010ApJ...713L.176B, 2018ApJS..236...42Y} branches of the H-R diagram. However, there have only been a handful of detections of solar-like oscillators in cool main sequence stars ($T_{\text{eff}} < 5500 \text{ K}$) due to their low luminosity and small oscillation amplitudes \citep{hekker2019scalingrelationssolarlikeoscillations, Campante_2024, Mathur_2019, Hatt_2023}. The probability of detecting solar-like oscillation in these part of the H-R diagram is further complicated by the fact that the least evolved stars have oscillation frequencies in the range from $10^3$\,$\mu \text{Hz}$ to $10^4$\,$\mu \text{Hz}$, which is close to the Nyquist limit of Kepler and CoRoT based photometry \citep{Benk__2016, 2013MNRAS.430.2986M}. 
All of these constraints combined have resulted in only of order \textbf{$ \sim100$} solar-like oscillators in the main sequence and sub-giant branch which are colder than the Sun~\citep{Hatt_2023, Campante_2024, 2008ApJ...687.1180A, 2017ApJS..229...30M, 2008ApJ...682.1370K}, compared to of order $\sim10^5$ red giant solar-like oscillators detected in a single study in~\cite{Hon_2021}.

TESS provides a unique opportunity to detect high-frequency, low-amplitude stellar oscillations owing to its short observing cadence, as low as 20 seconds as opposed to 1 min for Kepler, and its sensitivity at red optical wavelengths. 
TESS maximizes sky coverage by observing the sky in sectors, with individual targets typically monitored in multiple sectors. Each sector spans approximately 27.4 days, yielding light curves with baselines long enough to probe oscillatory signals in main sequence stars \citep{2025A&A...701A.285L}. During the nominal mission, TESS delivered light curves at 2 min cadence, corresponding to a Nyquist frequency of 4167\,$\mu\text{Hz}$, while the extended mission introduced a 20-second cadence mode, pushing the Nyquist limit to 25,000 $\mu\text{Hz}$. These capabilities make TESS exceptionally well-suited for detecting high-frequency, solar-like oscillations in cool dwarfs.
However, exploiting the full potential of such high-quality data requires searching oscillations across large samples of stars. 

Because solar-like oscillations are stochastically excited and intrinsically variable from star to star, constructing a general parametric model is challenging. For a list of existing parametric models \textbf{\citep[See ][]{2009CoAst.160...74H, 2010A&A...511A..46M, 2022JOSS....7.3331C, Mosser_2011, 2025ascl.soft04002N}}. 
Existing parametric approaches are primarily designed to extract seismic parameters once oscillations have been confirmed, rather than to detect their presence in the first place. This motivates the need for techniques that capture the characteristic features of solar-like oscillators while remaining robust to stochastic variability. One such example can be found in \citet{2009CoAst.160...74H} where the authors used an automated approach to calculate autocorrelation functions of power spectrum to look for signs of oscillations. However, it also required a final manual evaluation of each object, making it impractical for searching over a large catalog.

Deep learning offers an effective alternative for efficiently searching large datasets to identify solar-like oscillations. Recently, deep learning techniques have been widely adopted in astronomical classification problems both in time series and frequency domain \citep[e.g.,][]{Mahabal_2017, Naul_2017, Jamal_2020, 2021FrASS...8..168B, pérezgalarce2025selfregulatedconvolutionalneuralnetwork, Aguirre_2018, Becker_2020, 2021MNRAS.505..515Z, abdollahi_torabi_raeisi_rahvar_2022, 2021AJ....162..209A, Audenaert2025}. By training on known solar-like oscillators, a neural network (NN) can learn the underlying signal morphology and subsequently be applied to identify candidate oscillations in new targets as demonstrated by \cite{Hon_2018}, to find solar-like oscillation in red giants using Kepler. 

Despite these advances, contemporary architectures present practical limitations for detecting solar-like oscillations in TESS data. 
Many state-of-the-art models \citep[e.g.,][]{Naul_2017, Jamal_2020} assume fixed-length or phase-folded inputs (or segment long light curves into fixed-length slices), an approach that is problematic for TESS sector data: a single 27.4-day sector sampled at 2 min cadence contains around $2\times 10^4$ measurements (and about $1 \times 10^5$ measurements for a 20-second cadence sector). Thus, forcing fixed-length inputs typically requires aggressive rebinning, which reduces sensitivity to high-frequency signals, or segmentation, which reduces oscillation amplitude since we are using the Lomb-Scargle periodogram \citep{1976Ap&SS..39..447L} where amplitude can depend on the segment size \citep{2018ApJS..236...16V}. 
Also, phase folding does not reveal the oscillation features for solar-like oscillators (see Section~\ref{sec:lightcurve-vs-periodogram}). In contrast, periodograms cleanly separate the broadband granulation background from the oscillation power excess: the former is well described by Harvey-like/super-Lorentzian components and a low-frequency slope, while the latter appears as a Gaussian-like envelope centered near $\nu_{\text{max}}$ \citep[e.g.,][]{2014A&A...570A..41K,Mosser_2011}. 
Detecting new solar-like oscillators therefore reduces to learning these two spectral signatures, a task well suited to convolutional architectures that can recognize spatial patterns with high accuracy. 
This approach was pioneered for red giants by \citet{Hon_2018}, who trained Convolutional Neural Networks (CNN) on log-scale power-spectrum images to detect oscillation power excess with human-level performance. In this study, we modify this technique to find pulsation in main-sequence and sub-giant stars.

In this work, we present the first deep-learning study specifically targeting solar-like oscillations in cool main-sequence and sub-giant dwarfs using TESS 2 min cadence data. In our study, we have found very promising candidates in M, K, and G~dwarf regions of the color-magnitude diagram. Although we do not claim any detection, these candidates could be verified in follow-up observational efforts.
We trained a one-dimensional convolutional autoencoder that learns a compressed representation (also known as a latent space) of stellar periodograms. This latent space representation is coupled to a classifier head that distinguishes solar-like oscillators from a broad comparison set of other types of stellar variability, including other types of pulsators that do not exhibit solar-like oscillations such as $\delta$~Scuti, RR-Lyrae, and Cepheids. 
We then applied the trained network to a sample of around 91,000 stars cooler than G0, selected from the Asteroseismic Target List \citep[ATL,][]{hey2024precisetimedomainasteroseismologyrevised}.
Candidate detections were subsequently validated with \texttt{pySYD} \citep{2022JOSS....7.3331C}, which provided global asteroseismic parameters and diagnostic plots to confirm the variability source. To give a brief outline of this paper, we discuss our sample selection and data pre-processing in Section~\ref{sec:data}, and then introduce the architecture and training process of the neural network in Section~\ref{sec:network-and-training}. In Section~\ref{sec:new_candidates}, we discuss the target and candidate selection process and the various cuts we considered to make our candidate list, along with a comparison with previously detected solar-like oscillators. In Section~\ref{sec:discussion}, we present the most interesting objects in our list of candidates, followed by our conclusions in Section~\ref{sec:conclusion}. 

\section{Data} \label{sec:data}
To train our neural network, we used periodograms instead of a more conventional light curve approach, and Section~\ref{sec:lightcurve-vs-periodogram} is dedicated to justify this choice. We have also constructed a comprehensive training set composed of two main samples: 4,177 previously identified solar-like oscillators and 82,204 stars representing other types of pulsators, variable stars, and non-variable sources. Each of these samples is described in detail in Section~\ref{sec:solr-like-oscillators} and Section~\ref{sec:non-solr-like-oscillators}, respectively. In Section~\ref{sec:pre-processing}, we also describe the processing steps used to transform these samples into a suitable training dataset. 

\subsection{Light Curves versus Periodograms} \label{sec:lightcurve-vs-periodogram}

There are four standard representation of stellar light curves that can be used as inputs for neural network classification: full time-series, segmented time series, phase-folded light curves, and periodograms. We experimented with all four representations and chose to proceed with the periodogram-based approach for the remainder of our analysis. Below, we justify this choice by discussing the properties of each representation and the advantages of periodogram-based classification.

\cite{2021FrASS...8..168B} and \cite{Burhanudin_2021} used raw time series data to classify variable sources. The time series used in this study consisted of only 100 and 30 epochs respectively. However, TESS 2 min cadence data has a longer baseline, consisting of $\sim15,000$ points. Training a neural network with such a large input layer is computationally expensive and impractical. 

\begin{figure}
    \centering
    \includegraphics[width=0.98\linewidth]{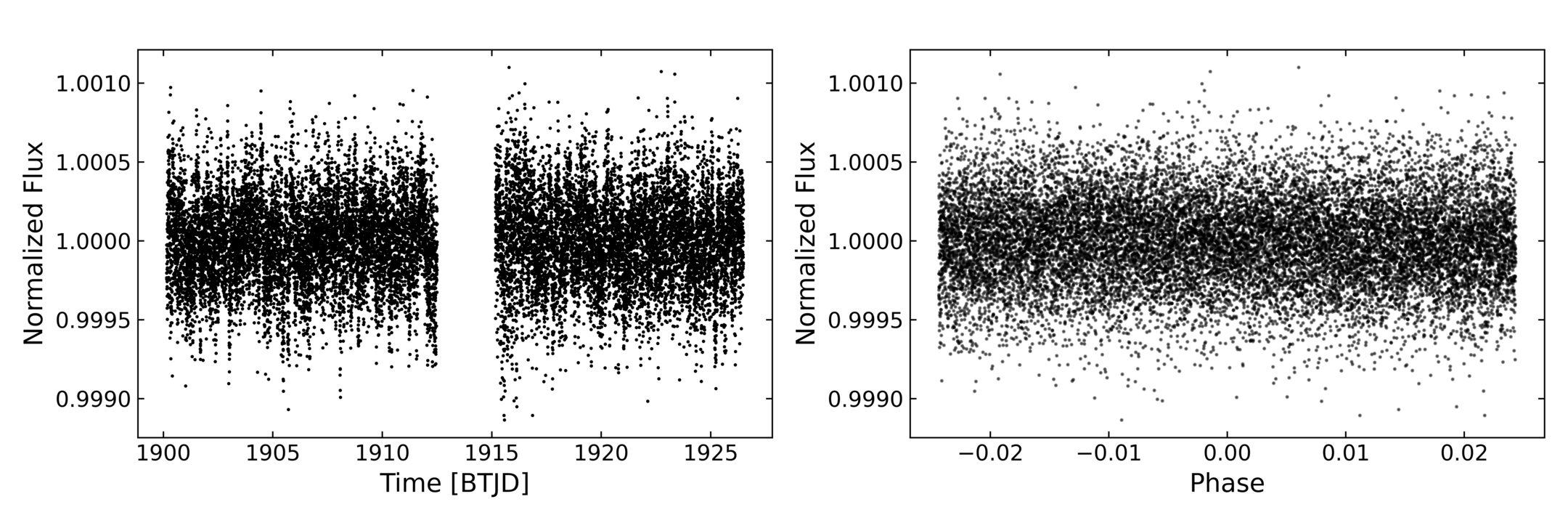}
    \caption{Light curves of solar-like oscillator before and after phase folding on its $\nu_{\text{max}}$. Left panel shows the raw light curves and right panel shows light curve after phase folding.}
    \label{fig:folding_before-after}
\end{figure}

The second method consists of phase folded light curves, as explored by \cite{Naul_2017} and \cite{Jamal_2020} where this technique has shown to reach up to 99\% and 97\% accuracies respectively. However, to phase fold the light curve, one needs to know the period of variability beforehand, which is often hard to accurately determine. Also, we have noticed that the phase folding on one frequency does not reveal a consistent oscillatory pattern because more than one oscillation mode is stochastically excited. Figure~\ref{fig:folding_before-after} shows an example where a light curve for a solar-like oscillator has been folded on the $\nu_{\text{max}}$ \citep{Hatt_2023}, and there is no noticeable variability pattern in the folded light curve. 

A third approach is to split the long baseline of the light curves into segments of smaller lengths, and have overlapping segments to create a running window effect that preserves the phase information. However, as discussed in \cite{Naul_2017}, if the light curves have high frequency variation in brightness, the neural network's ability to learn these variability patterns deteriorates significantly. We have also tried the approach of splitting the light curves into distinct segments without creating the running window effect. However, given the length of TESS 2 min cadence light curves, each light curve was split into 5 - 6 distinct signals, which made our training set 5 - 6 times larger. This resulted in slower training convergence, and the classification accuracies were low as well (about 80\%). 

The last approach is using the periodogram. The periodogram has the advantage that binning it to a smaller length does not remove the signal for any frequency range as long as it is not too aggressive. Also as our goal is to identify solar-like oscillators, and these have a particular shape in the periodogram (Figure~\ref{fig:solr-osc-example}), we found that it is the most efficient way to distinguish them. The NN can learn to find those features to distinguish our target signals from the others. Therefore, we decided to use a 1D periodogram, and re-bin it in logarithmic scale (which is equivalent to a log-log periodogram) to a fixed length of 4096 points to train our model.

\subsection{Solar-like Oscillators} \label{sec:solr-like-oscillators}
We trained our model on a set of previously identified solar-like oscillator to make it learn and recognize their periodogram shapes. The sample of solar-like oscillators was obtained from \cite{Hatt_2023} which consists of 4177 manually verified solar like oscillators (\texttt{SOLR}) from TESS 120-s and 20-s cadence data. 
The authors used a detection algorithm based on finding power-excess and regularly spaced radial overtones in the power-spectrum, followed by a manual verification of each candidate. This resulted in 3691 solar-like oscillators in 120-s cadence and 486 \texttt{SOLR} in 20-s cadence. 
The distribution of frequencies of maximum amplitude, $\nu_{max}$ of this sample ranges between $\sim 10$\,$\mu$Hz to $\sim 3000$\,$\mu$Hz.
Some of the \texttt{SOLR} from \cite{Hatt_2023} were found using 20 second cadence that do not show oscillations in 2 min light curves. We decided to remove these objects for uniformity across our training sample. 

\begin{figure}
    \centering
    \includegraphics[width=0.98\linewidth]{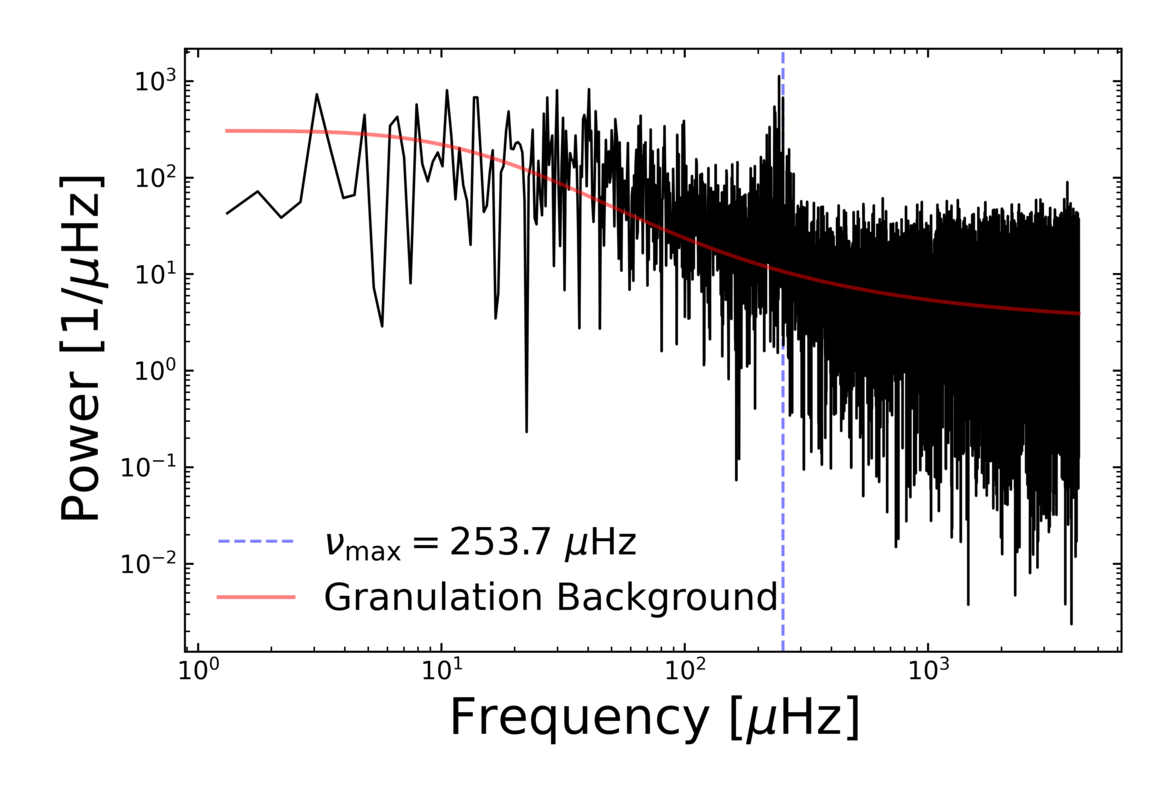}
    \caption{Power Spectrum of a solar-like oscillator of an evolved star with \texttt{TIC 141201954}. Granulation background fit is in red and oscillation power excess frequency is marked in a blue vertical dashed line. \textbf{The figure is just for schematic illustration, and does not serve any quantitative purpose.}}
    \label{fig:solr-osc-example}
\end{figure}

Most \texttt{SOLR} power spectra are characterized by a granulation-dominated slope at low frequencies, followed by a Gaussian-shaped envelope of excess power arising from stellar oscillations \citep{Hon_2018}. An example is shown in Figure~\ref{fig:solr-osc-example}, which just serves the purpose of schematic illustration, and does not serve any quantitative purpose. In the plot, the granulation component is highlighted in red, and the oscillation hump is clearly visible near the frequency of maximum power, indicated by a vertical dashed blue line.  These two distinct features are visually most apparent in log–log space, which makes it easily identifiable by a Convolutional Neural Network because the log–log scaling emphasizes the characteristic shapes of the granulation background and oscillation hump. Therefore, we bin our periodograms logarithmically before passing them to the neural network. This type of periodogram representation was first used by \citet{Hon_2018} to identify red giant oscillations, where the data were treated as 2D images in logarithmic space. In contrast, we adopt a 1D representation, which is significantly less computationally expensive while still preserving the relevant morphological information. Our deep learning model is trained to learn and recognize this characteristic shape (see Section~\ref{sec:network-and-training}) and is subsequently applied to search for oscillations in stars of previously unknown types (see Section~\ref{sec:new_candidates}).

\subsection{Non Solar-like Oscillators} \label{sec:non-solr-like-oscillators}

We obtained the sample of stars that do not show solar-like oscillations from \cite{balona2023identificationclassificationtessvariable}. 
This author provided visual classification of 120,000 variable stars observed by Kepler and TESS. For our sample, we selected  82,204 stars that were observed by TESS. These stars have both variable and non-variable light curves. We grouped the stars that do not show solar-like oscillations (\texttt{NonSOLR}) into four major classes: Pulsators (\texttt{PULS}), Rotators (\texttt{ROT}), Eclipsing Binaries (\texttt{EB}), and Non Variables (\texttt{NV}). The distribution of these four groups is shown in Table~\ref{tab:ns-stars}. These four categories were chosen as true negatives for training because they are the most abundant types of variable stars that do not show solar-like oscillations \citep{balona2023identificationclassificationtessvariable}.

\begin{deluxetable}{cccccc}
\tablecaption{Number of stars per class of light curve in our training and test sample combined.}
\tablenum{1}

\tablehead{
\colhead{SOLR} & \multicolumn{4}{c}{NonSOLR} & \colhead{} \\
\colhead{} & \colhead{PULS} & \colhead{ROT} & \colhead{EB} & \colhead{NV} & \colhead{}
}

\startdata
4177 & 15200 & 24688 & 5796 & 36520 & \\
\enddata
\label{tab:ns-stars}
\tablecomments{The SOLR column refers to the sample of known solar-like oscillators. The NonSOLR type consists of four variability classes: Pulsators (PULS) that are not solar-like, rotators (ROT), eclipsing binaries (EB), and non-variables (NV). }
\end{deluxetable}

\begin{figure}
    \centering
    \includegraphics[width=0.98\linewidth]{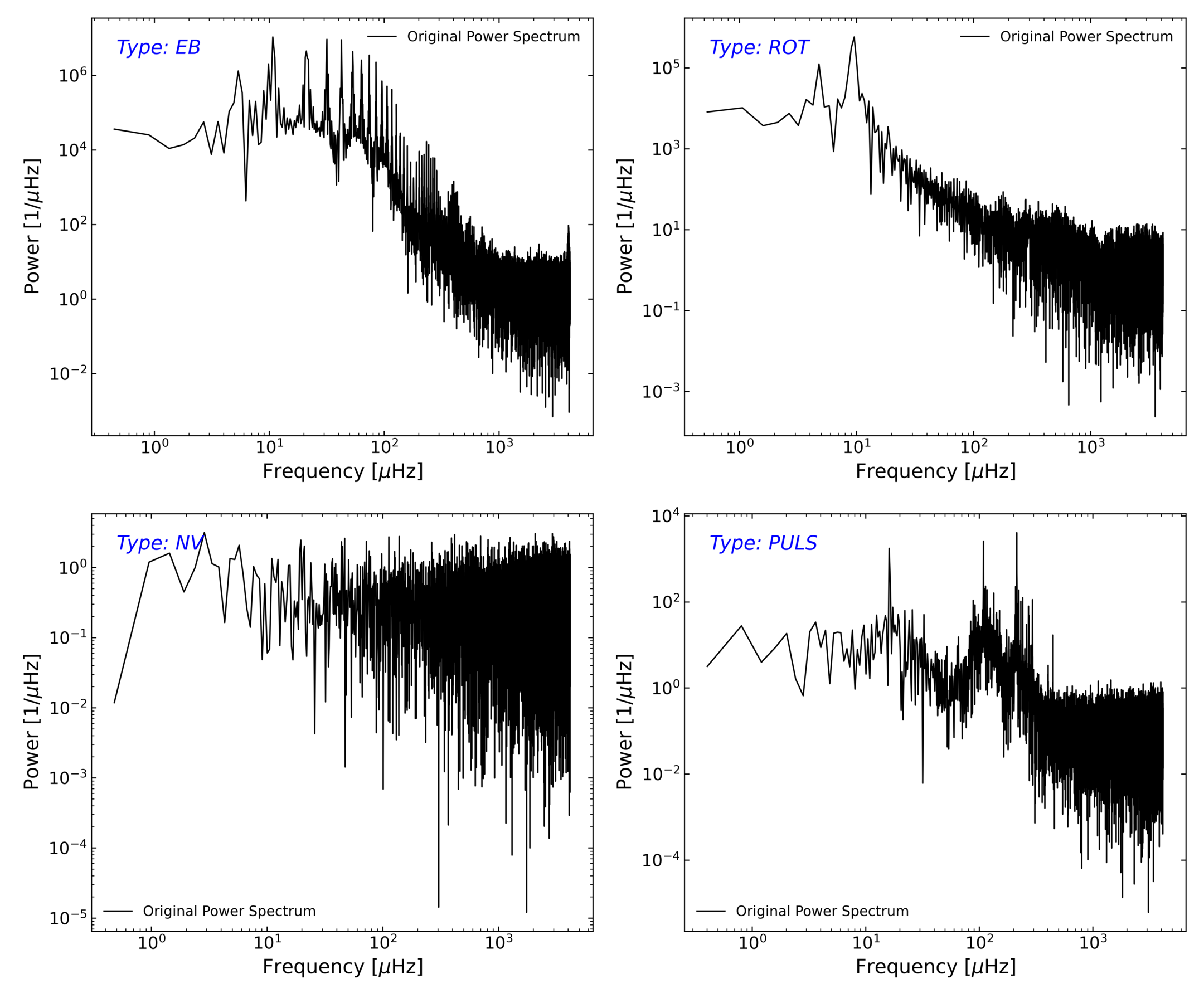}
    \caption{Example power spectrum of stars with no solar-like oscillation. Top left is eclipsing binary, and right is rotator. Bottom left is non variable and bottom right is a $\delta$~Scuti pulsator.}
    \label{fig:ns-stars}
\end{figure}

Figure~\ref{fig:ns-stars} shows examples of power spectrum from stars from different variability classes that do not show any solar-like oscillations. 
The top left panel of the figure shows an eclipsing binary, which has a very distinct shape of multiple harmonics. The top right panel shows a rotator, with a singular low frequency peak. The bottom left panel shows a non variable star with the characteristic flat white noise, followed by a $\delta$~Scuti pulsator on the bottom right panel, with multiple peaks above 20 $\mu$Hz. Visibly, these stars have different shapes from a solar-like oscillator (see Figure~\ref{fig:solr-osc-example}) since none of them have a combination of granulation slope and a gaussian-like power excess hump due to oscillations.

\subsection{Pre-processing} \label{sec:pre-processing}

In this section, we describe our data pre-processing techniques for both samples of \texttt{SOLR} and \texttt{NonSOLR}. First, we obtained the TIC ids of the \texttt{SOLR} and \texttt{NonSOLR} sample from the two catalogs described in Section~\ref{sec:solr-like-oscillators} and Section~\ref{sec:non-solr-like-oscillators}. 

For the \texttt{SOLR} dataset, we used light curves from all available TESS sectors to allow the model to learn both intrinsic signal characteristics and instrumental systematics. Each sector’s light curve was treated as an individual object, even if multiple sectors corresponded to the same star. This approach was mainly taken because for some stars, the oscillation signal becomes visible only after combining multiple sectors. However, stitching multiple sectors is known to produce systematics that might interfere with the training process \citep{Boyle2025}. It also helped us address the class imbalance between \texttt{SOLR}, which originally had only 4177 stars, and \texttt{NonSOLR}, which has 82,204 stars. After considering all sector light curves, the \texttt{SOLR} sample had 29,621 periodogram signals. By including all the sectors available for one star as separate stars, we are training the neural network on TESS systematics, as well as on recognizing the pulsation signal. This method is similar to the standard data augmentation technique of training a network on the same signal multiple times, but each time adding a small noise component to the signal. It has been shown to improve classification and generalization capabilities of neural network models\citep[e.g.,][]{kim2024noiseinjection}. 

For the \texttt{NonSOLR} sample, we downloaded only one sector per star. We did not use any criteria in choosing a sector, but rather just downloaded the first sector the star was observed in. 

For all light curves, we utilized the \texttt{lightkurve}\footnote{https://lightkurve.github.io/lightkurve/} package to download the time series. The signals were processed by TESS Science Processing Operation Center \citep[SPOC,][]{2016SPIE.9913E..3EJ}, which carries out simple aperture based photometry and removes any instrumental noise. Also, we only used TESS 120-s cadence light curves, since 20-s data is available for only a select number of stars, and has been shown to have different systematics than 120-s data \citep{2022AJ....163...79H}. Once we downloaded the light curves, we cleaned it using \texttt{remove\_outliers()} method 
which removed data points from the light curves residing beyond a 5$\sigma$ deviation from the mean flux.
As a next step in our pre-processing, we calculated the Lomb-Scargle periodogram using built-in \texttt{lightkurve.periodogram.Periodogram()} functionality with its default settings. These periodograms are used as the input to the NN (See Section~\ref{sec:lightcurve-vs-periodogram}).
To facilitate identifying the shape of solar-like oscillators by the neural network, we binned the periodogram on a logarithmic scale (See Section~\ref{sec:solr-like-oscillators}). The network also requires all input to be of specific length, so we binned the signal to 4096 points, where this number was chosen based on trial and error. Logarithmic binning reduces the frequency resolution of the periodogram. Given the Nyquist limit of 4167 $\mu \text{Hz}$, and the fact that each TESS 2 min periodgram has around 16,000 points, this binning results in a frequency resolution of about 4.7 $\mu \text{Hz}$ around 2000 $\mu \text{Hz}$ frequency and a resolution of around 7.1 $\mu \text{Hz}$ around 3000 $\mu \text{Hz}$ frequency. However, these resolutions are well above the mode separations of solar-like oscillation, which is in the order of $\sim 100\,\mu \text{Hz}$. Therefore, logarithmic binning does not pose a threat of losing the oscillation signal.

Once we obtained the signals binned to 4096 points, we used \texttt{MinMaxScaler()} function from \texttt{scipy}\footnote{https://scipy.org/} to normalize each signal between 0 to 1. This helps the neural network to have a smooth gradient flow and converge faster. Since we are only doing a binary classification, we assigned the stars to two different classes: \texttt{SOLR}, which shows solar-like oscillations and \texttt{NonSOLR}, which does not show solar-like oscillation, but might contain other types of variability.

\section{Network Architecture \& Training} \label{sec:network-and-training}

To enable variability classification, we developed a convolutional autoencoder designed to learn characteristic periodogram morphologies of stars across different variability types. The network compresses each input periodogram into a low-dimensional latent representation of length 128. A classification head is then attached to this latent space to distinguish between two primary categories: \texttt{SOLR} and \texttt{NonSOLR}. The network architecture was developed using \texttt{PyTorch}\footnote{https://pytorch.org/} framework in python. All network architectures and their full implementation is publicly available on GitHub\footnote{https://github.com/waleey/ae\_classifier.git}. 

The complete model was trained on the sample described in the previous Sections~\ref{sec:solr-like-oscillators}, \ref{sec:non-solr-like-oscillators} and \ref{sec:pre-processing}, and evaluated on an independent test set, achieving a classification accuracy of 99.8\%. In this section, we present the architectural design of the neural network and discuss its training, validation, and testing performance.

\subsection{Autoencoder} \label{sec:autoencoder}

\begin{figure}
    \centering
    \includegraphics[width=0.99\linewidth]{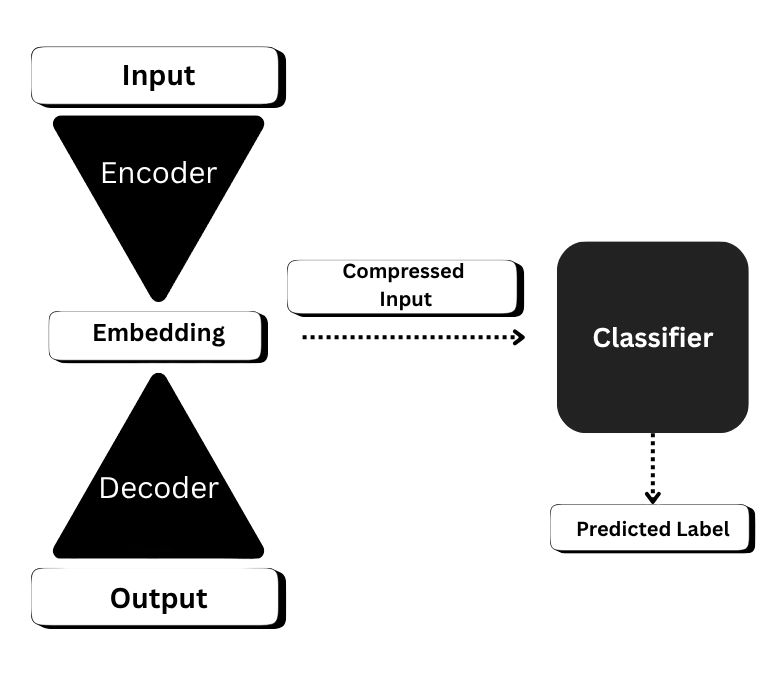}
    \caption{Schematic diagram of the autoencoder and classifier developed for this study.}
    \label{fig:autoencoder}
\end{figure}

An autoencoder is a neural network architecture widely used for unsupervised representation learning, dimensionality reduction, and data compression \citep{Hinton2006ReducingTD}. At its core, an autoencoder is trained to reconstruct its input data, thereby forcing the model to learn compact, informative representations of complex, high-dimensional inputs. This property makes autoencoders particularly suitable for analyzing large and structured astronomical datasets, where signals of interest (e.g., variability patterns) are often embedded in noisy, high-dimensional measurements.
As illustrated in Figure~\ref{fig:autoencoder}, a typical autoencoder is composed of three main components: an encoder, a latent embedding (or bottleneck layer), and a decoder. The encoder is often implemented using CNNs, which are well-suited for learning spatial and frequency-localized features \citep{LeCun1998GradientbasedLA}. It transforms the high-dimensional input data into a compact latent representation that preserves the most salient features. This intermediate representation is referred to as the embedding or latent space. The decoder, typically a mirror reflection of the encoder in structure, performs the inverse operation—reconstructing the original input from the embedding using a sequence of transposed convolutions and upsampling layers.
The training objective of an autoencoder is to minimize the reconstruction error between the original input and the output produced by the decoder, typically using mean squared error (MSE) or other suitable loss functions. By optimizing this reconstruction loss, the network learns to capture the essential structure of the data while discarding irrelevant variations such as noise.

In this work, we employed a convolutional autoencoder to learn a low-dimensional representation of periodogram signals associated with various stellar variability classes. These embeddings encapsulate key frequency-domain features characteristic of each variability type (e.g., solar-like oscillators, other pulsators, eclipsing binaries, rotational variables, and non-variables), providing a compressed but informative summary of the periodogram (See Section~\ref{sec:latent-space} for more discussion). Once trained, these embeddings serve as input features to a downstream classification network, which assigns labels to different variability types based on their learned frequency domain structures.
A simple representation of the network is in Figure~\ref{fig:autoencoder}, and a more detailed structure can be found in Section~\ref{sec:encoder}.


The autoencoder design used in this study was mainly inspired by the work of \cite{Ricketts_2023}, who developed an algorithm for mapping the X-ray variability of black hole binaries. This neural network was comprised of two main sections: encoder and decoder. Both sections had 9 convolutional layers in conjunction with pooling and dense layers to reduce the input dataset to lower dimension. 

We applied significant changes to the autoencoder to make it suitable for our case. We made the network deeper by adding 5 more convolutional layers in both encoder and decoder. This structure was chosen because of the large dimensionality of our input signal (4096) compared to smaller input dimension (256) of  \cite{Ricketts_2023}. Moreover, each convolutional block includes a \texttt{BatchNormalization} layer, which normalizes the output of the preceding layer to stabilize and accelerate training by improving gradient flow. For the convolutional and dense layers, we used \texttt{kaimin\_normal}~\citep{he2015delvingdeeprectifierssurpassing} weight initialization, and for \texttt{BatchNorm1d} layers, we used constant weight initialization technique. We also employed a \texttt{leaky\_relu} activation function to mitigate the vanishing gradient problem and ensure better feature propagation, particularly for large and diverse datasets. In addition, a \texttt{Dropout} layer with a probability of 0.4 is incorporated to reduce overfitting by preventing the network from relying too heavily on specific neurons and promoting more robust generalization.

To facilitate learning of both local and global frequency domain features in the periodogram signals, we adopted a varying set of kernel sizes in the convolution layers of encoder and decoder. A kernel size specifies how many neighboring data points a convolutional filter spans, determining the local region over which it extracts features. Smaller kernels (e.g., size 1 and 3) are well-suited to learn fine-grained, local features such as sharp peaks due to a rotation signals or oscillation power excess envelopes in solar-like oscillators~\citep{simonyan2015deepconvolutionalnetworkslargescale}. Large kernels (e.g., 7, 11, 15) allow the network to learn more extended patterns and global structures which was key in distinguishing the granulation background from other types of noise~\citep{yu2016multiscalecontextaggregationdilated}.

To facilitate dimensionality reduction, we have experimented with (1) strided convolution\footnote{https://www.deeplearningbook.org/contents/convnets.html} and (2) \texttt{MaxPooling}\footnote{https://pytorch.org/docs/torch.nn.MaxPool1d.html}. The strided convolution technique generated artifacts in the reconstructed signal (see e.g. \citealt{kinoshita2020fixedsmoothconvolutionallayer} for a similar discussion) and was unable to reconstruct the fast-varying features in the periodogram. \texttt{MaxPooling}, on the other hand, could accurately reconstruct those fast variations without producing any artifacts. Therefore, we decided to continue using \texttt{MaxPooling} as our dimensionality reduction technique.

The latent space dimension of 128 was chosen by means of trial and error. Initially, we started with a latent space dimension of 32, which yielded a classification accuracy of 87\% for solar-like oscillators. Then, we moved up to a 64-dimensional latent space, which resulted in a classification accuracy of 97\% for solar-like oscillators, but only 77\% classification accuracy for identifying stars that do not show solar-like oscillation, referred to as \texttt{NonSOLR} in this text. To investigate the low classification accuracy, we analyzed the reconstructed periodograms and found that although these low-dimensional latent spaces were able to capture the slowly varying granulation background, they were consistently missing the oscillation hump, which shows up as Gaussian envelopes. Finally, in a 128-dimensional latent space, we reached appreciable accuracy in the classification of both solar-like oscillators and other types of variables, and visual inspection of the reconstructed periodograms also revealed that the model improved at capturing those oscillation humps it was missing in a lower-dimensional latent space. We also tried making the latent space size 256-dimensional, and the accuracy of finding solar-like oscillators increased to 99.7\%, but the accuracy for finding other types of variables increased to 95\%. Although it reduced the number of false positives by 0.8\%, this larger latent space also increased the convergence time for training the model by about 30 minutes and required increased memory for storing the latent space representation of the input signal. Therefore, we decided to continue our analysis with a 128-dimensional representation.

For upsampling in the decoder, we used the cubic interpolation function provided by \texttt{PyTorch}\footnote{https://pytorch.org/}. We also explored other techniques such as strided convolution and \texttt{MaxUnpooling}, but both techniques generated artifacts in the reconstructed periodograms and gave higher reconstruction loss compared to the cubic interpolation approach. Therefore, we did not consider those techniques in our study. Also, to enhance the model's ability to focus on the region of the periodograms with the most information, we used attention layers which dynamically weights the importance of different input positions, allowing the model to focus on the most relevant parts of the sequence when making predictions. Attention layers are known for its ability to selectively amplifying important features and suppress noise \citep{Vaswani2017AttentionIA}.

To quantify our error in the reconstructed signal compared to the original input in autoencoder, we used the mean squared error (MSE) loss function. A minimal reconstruction loss signifies the decoder is able to successfully reconstruct the original input from the embedding layer, which in turn means the embedding contains the most important information about the variability patterns of the periodograms. We also tested the Mean Absolute Error (\texttt{MAE}) loss function, but did not see any noticeable improvement in the performance of the network and it was not considered further. 

\subsection{Classifier}

We classified the compressed representation of the input signal from the autoencoder in one of two categories: \texttt{SOLR} and \texttt{NonSOLR}. In order to handle classification tasks, the autoencoder has a classifier head attached to the embedding layer (See Figure~\ref{fig:autoencoder}). Therefore, the autoencoder works as an automated feature extraction network and the classifier is an estimation network that predicts the labels based on those extracted features. Our classifier is a standard \texttt{mlp} network \citep{1986Natur.323..533R} with two dense layers, one activation function and a dropout layer. 
This composite architecture of combining autoencoder with a classifier is inspired by the work of \cite{Jamal_2020}. The first dense layer has an input length of 128. This layer outputs a sequence of 16 points that the second dense layer encodes into one output number. During training and validation, the raw output value is passed on to the loss function to calculate the classification loss. Since we are doing a binary classification task, we have used \texttt{BCEWithLogits()} as our classifier loss function. \texttt{BCEWithLogits()} is given by the following equation:

\begin{equation}
\begin{split}
    \mathcal{L}_{\text{C}} = \frac{1}{N} \sum_{i=1}^{N} w_{y_i} \Big[ 
    & -y_i \cdot \log \left( \sigma(\hat{z}_i) \right) \\
    & - (1 - y_i) \cdot \log \left( 1 - \sigma(\hat{z}_i) \right) 
    \Big]
    \label{eq:class_loss}
\end{split}
\end{equation}

\noindent where, 

\begin{equation}
    \sigma(\hat{z}_i) = \frac{1}{1 + e^{-\hat{z}_i}}
\end{equation}

\noindent where $\hat{z_i}$ is the raw logit output from the model for i-th sample, $y_i \in \{0, 1 \}$ is the provided true label for the same sample, and N is the number of signals in the batch. Also, $w_{yi}$ is an extra weight factor multiplied to each element of a sample that depends on the sample's class. Thus for a given sample from a given class, each element has the same weight factor. This forces the loss function to put equal emphasis on the minority class (in our case \texttt{SOLR}) as the majority class (\texttt{NonSOLR}, see Table~\ref{tab:ns-stars}).
To see the detailed architecture of the classifier network, please refer to~\cite{Jamal_2020}.

\subsection{Training, Validation \& Testing}
We split the dataset using a 80--20 train-test split using a random number generator, which resulted in 92,896 objects in the training sample, and 23,224 signals in the test sample. This splitting was done based on unique TIC IDs because \texttt{SOLR} sample has signals from multiple sectors, and the same star being in both the training, validation, or test set will cause contamination between the sets, preventing correct generalization of the underlying signals. Finally, the training sample was further split into 70--30 training-validation set, resulting in 65,027 stars in the final training set and 27,868 stars in the validation set, using unique TIC IDs as discussed above. The training and validation set was used during the training phase of the NN, and the test set was kept hidden and only used to calculate prediction accuracy of the model.

\begin{figure}
    \centering
    \includegraphics[width=0.98\linewidth]{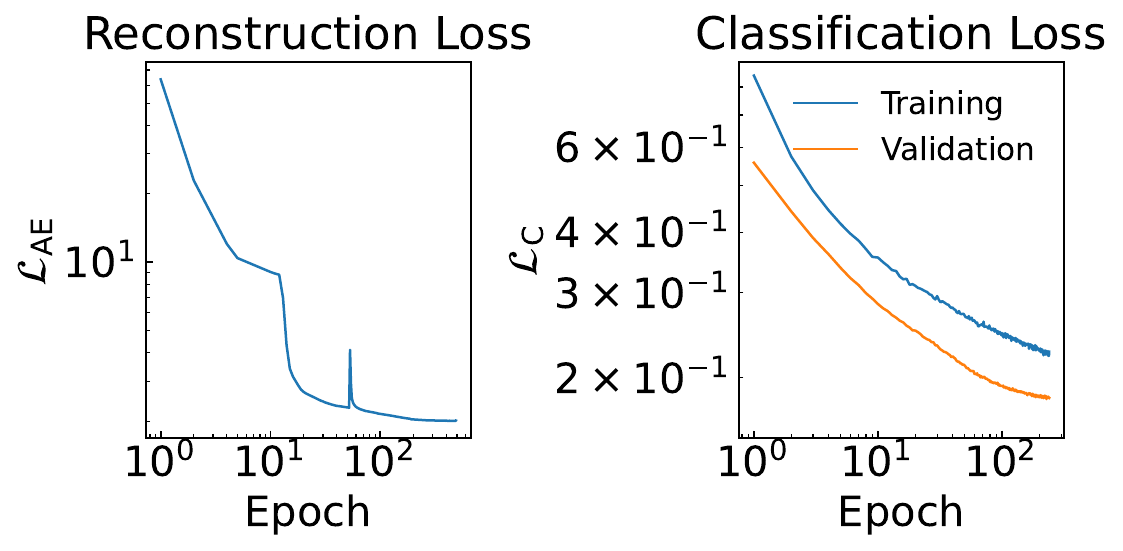}
    \caption{Loss as a function of epochs for autoencoder (left) and classifier (right). Only training loss is shown for autoencoder while both training and validation loss is shown for classifier. We computed validation loss for all epochs only for the classifier network.}
    \label{fig:loss-plot}
\end{figure}

We experimented with two different training schemes: sequential training and joint training. In sequential training, we trained only the autoencoder first by minimizing reconstruction loss given by MSE loss (see Section~\ref{sec:autoencoder}) and then trained the classifier to minimize classification loss given by Equation~\ref{eq:class_loss}. This gave the best performance for the neural network in predicting the labels in the test set, since the autoencoder worked as a stand alone feature extraction algorithm that contained the most important spatial features of the periodograms in the embedding space, and the classifier worked as an independent estimation network to predict the labels based on extracted features. 
We also tried training the network in the joint scheme where we trained both autoencoder and classifier simultaneously, and tried to minimize total loss given by the direct sum of MSE and classifier loss (Equation~\ref{eq:class_loss}). However, this produced poor reconstruction, resulting in almost 10 times higher reconstruction loss compared to sequential training (see Figure~\ref{fig:loss-plot}). One possible strategy to mitigate this issue during joint training is to apply a time-dependent weighting scheme to the two loss components, such that the reconstruction loss dominates the early stages of training. This allows the autoencoder to first learn the salient features of the input representation, after which the weights can be gradually balanced between the reconstruction and classification losses to improve classification performance. However, because the sequential training scheme already yielded robust and satisfactory results without the need for additional complexity, we adopt this approach for the present study.

The probability that a given star exhibits solar-like oscillations was obtained by applying a \texttt{softmax()} activation function to the classifier output. This operation maps the raw network logits to normalized probabilities in the range [0,1], where values closer to 1 indicate a higher likelihood of the star being a solar-like oscillator. For the computation of validation accuracy, we adopted a classification threshold of 0.5: stars with softmax probabilities greater than this threshold were labeled as \texttt{SOLR}, while those below were labeled as \texttt{NonSOLR}.

Our training procedure follows a standard neural-network optimization workflow, using gradient-based updates and empirically tuned hyperparameters. The neural network uses a gradient descent algorithm that updates the parameters to minimize loss function. For gradient descent, we have used Adam Optimizer~\citep{kingma2014adam} with learning rates $1 \times 10^{-3}$ for autoencoder training and $1 \times 10^{-4}$ for classifier training. Since using a fixed learning rate provided satisfactory results, we did not use any learning rate scheduler. All hyperparameters including learning rate, batch size, size of the embedding layer were selected by trial and error and can be found in Table~\ref{tab:hyperparams}. We also implemented an early stopping procedure based on the validation loss, meaning that training was halted when the validation loss failed to improve for a fixed number of epochs (patience), in order to prevent overfitting and ensure the model retained good generalization performance. The min\_delta and patience parameters for early stopping is also found based on trial and error and listed in Table~\ref{tab:hyperparams}. We also trained our network on Google Colab\footnote{https://colab.research.google.com/} T4 GPUs. The autoencoder training took 4 hour 24 minutes and the classifier training took around 2 hour 11 minutes.

\begin{deluxetable}{lc}
\tablenum{2}
\tablecaption{Hyperparameters used for training.\label{tab:hyperparams}}
\tablehead{
\colhead{Hyperparameter} & \colhead{Value}
}
\startdata
AE Learning Rate & $1\times10^{-3}$ \\
Classifier Learning Rate & $1\times10^{-4}$ \\
Batch Size & 64 \\
Embedding Size & 128 \\
min\_delta & $1\times10^{-4}$ \\
patience & 50 \\
\enddata
\end{deluxetable}

Figure~\ref{fig:loss-plot} displays the reconstruction and classification losses for the autoencoder and classifier, respectively. During the initial autoencoder training phase, we only record the training reconstruction loss, as validation is not performed until the autoencoder training concludes, given that the validation set is not used during autoencoder training. Instead, we monitor the training loss and apply early stopping once it plateaus. At that point, we freeze the autoencoder weights, and begin training the classifier on top of the learned representation. Since we are not training the autoencoder anymore, we only compute the validation loss for it, and the loss stayed stable at a value of 2.14, which is extremely close to its training loss before freezing. 

In the classification loss, shown in the right panel of Figure~\ref{fig:loss-plot}, we observed that the validation classification loss was slightly lower than the training loss. This is due to the use of dropout layers during training, which randomly deactivates neurons, effectively reducing model capacity. In contrast, dropout is disabled during validation, allowing the model to utilize its full capacity, which can result in slightly better performance. However, the difference between training and validation classification loss is minor and diminishes with more training epochs, suggesting it does not significantly affect model performance. We confirmed this by lowering the dropout probability, which reduced the gap between training and validation losses. Despite the small discrepancy, we retained the dropout layers to promote better generalization, as it improved our validation loss.

We trained the model till the early stopping procedure kicked in, as discussed above, and saved our trained model. Then, we used this model to calculate the classification metrics and confusion matrix on our test set that was hidden from the model during training phase. We evaluated the model performance using three standard classification metrics: precision, recall, and F1 score.
Precision is defined as the fraction of predicted positive samples that are truly positive, and thus quantifies the model’s ability to avoid false positives.
Recall is defined as the fraction of actual positive samples that are correctly identified, minimizing false negatives: an important consideration in our application where missing positive detections is particularly undesirable.
The F1 score represents the harmonic mean of precision and recall, providing a balanced measure of both metrics.
The formal definitions and corresponding values of these metrics for the validation dataset are summarized in Table~\ref{tab:val-metric}. 
Our precision is 0.945, meaning the model is able to avoid false positives with probability 0.945. A recall of 0.998 signifies the fraction of actually positive samples recovered during validation. Our F1 score is 0.971, meaning our model has high accuracy in recovering the true samples, as well as robust against negative samples. 

\begin{deluxetable}{lcc}
\tablenum{3}
\tablecaption{Validation metrics and their definitions.\label{tab:val-metric}}
\tablehead{
\colhead{Metric} & \colhead{Definition} & \colhead{Value}
}
\startdata
Precision & {\scriptsize $\displaystyle \frac{\mathrm{TP}}{\mathrm{TP} + \mathrm{FP}}$} & 0.945 \\
 & & \\
Recall & {\scriptsize $\displaystyle \frac{\mathrm{TP}}{\mathrm{TP} + \mathrm{FN}}$} & 0.998 \\
 & & \\
F1 Score & {\scriptsize $\displaystyle 2 \cdot \frac{\mathrm{Precision} \cdot \mathrm{Recall}}{\mathrm{Precision} + \mathrm{Recall}}$} & 0.971 \\
 & & \\
\enddata
\label{tab:val-metric}
\tablecomments{Accuracy metrics and their corresponding definitions. TP, FP, FN, and TN denote true positives, false positives, false negatives, and true negatives, respectively. All metrics are calculated on an independent test set.}
\end{deluxetable}

\begin{figure}
    \centering
    \includegraphics[width=0.98\linewidth]{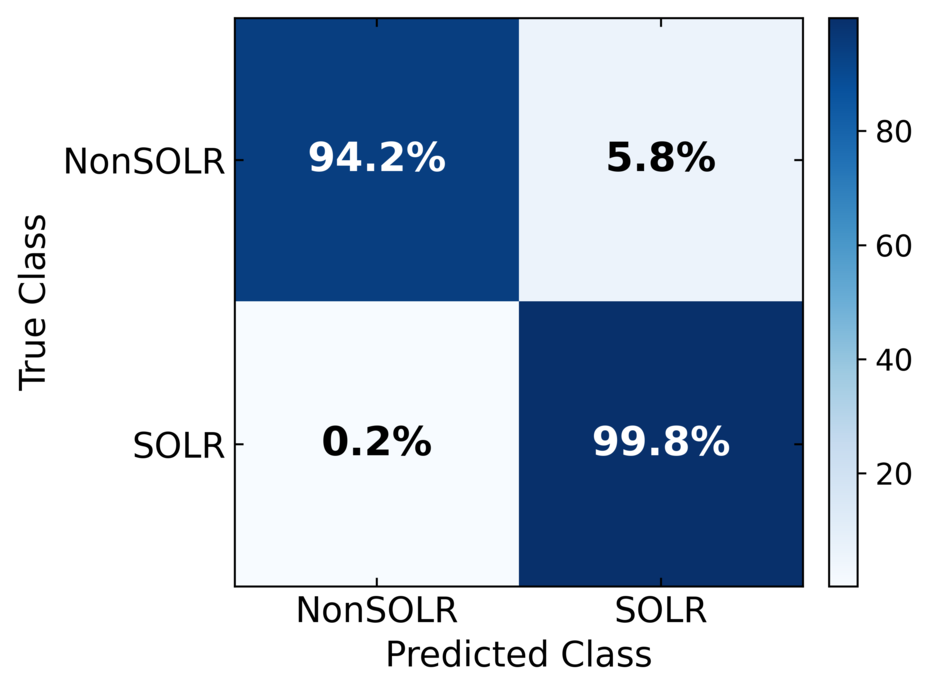}
    \caption{Confusion Matrix for test set. Here, \texttt{SOLR} refers to solar-like oscillator and \texttt{NonSOLR} refers to stars that do not exhibit solar-like oscillation. The color bar describes the fractional percentage in each panel between 0 and 1. True positive is 99.8\% and true negative is 94.2\%. False positive and false negative are 5.8\% and 0.2\% respectively.}
    \label{fig:confusion-matrix}
\end{figure}

As a final test of our model we estimated the values of the confusion matrix, which is shown in Figure~\ref{fig:confusion-matrix}. The model achieved 99.8\% accuracy in recovering the solar-like oscillators (\texttt{SOLR}) and 94.2\% accuracy for stars that do not have solar-like oscillation (\texttt{NonSOLR}). 5.8\% of the \texttt{NonSOLR} were misclassified as \texttt{SOLR}, and most of these false positives are non-variables. Upon investigating the periodogram of these non-variables, we found that they have a low-frequency granulation like slope in their periodogram, which likely made the network classify them as solar-like oscillator. On the other hand, 0.2\% of the \texttt{SOLR} type stars were misclassified as \texttt{NonSOLR}. These stars have a median $\nu_{\text{max}}$ of 300 $\mu\text{Hz}$. After investigating their periodograms, we found that for most stars, the low-frequency correlated noise and the granulation signal overlap with the oscillation signal, which makes it harder for the neural network to distinguish them. However, since we are interested in stars with comparatively higher frequency, this is within our acceptable margin of error.

\section{Finding New Pulsation Signals} \label{sec:new_candidates}

In this section, we outline the methodology employed to identify new solar-like oscillator candidates. We begin by describing the target sample used for the search, which was analyzed using the neural network architecture detailed in Section~\ref{sec:network-and-training}. We then present the newly identified candidates and characterize their properties. Most of our analysis was done on 2 min cadence data as discussed before, and we complemented this analysis with 20-sec data discussed in Section~\ref{20-sec-search}. Finally, we added a comparison with previously reported solar-like oscillators.

\subsection{Target Selection}
\label{sec:target-selection}
Our target sample was selected from the TESS asteroseismic target list \citep[ATL,][]{Campante_2017, Schofield_2019}, a list of asteroseismic targets compiled before the launch of TESS to streamline its target selection process for 2 min cadence. The authors of the ATL used fundamental stellar parameters and scaling relations to predict oscillation frequencies ($\nu_{\text{max}}$) and amplitudes of solar-like oscillations. Combining these values with the photometric noise characteristic of TESS, they calculated a detection probability following the method described in \cite{Chaplin_2011}. 
For our analysis, we used a revised version of the ATL \citep{hey2024precisetimedomainasteroseismologyrevised} that used an updated $\nu_{\text{max}}$ scaling relation considering surface gravity data available from \textit{Gaia} DR3 \citep{2016A&A...595A...1G, 2023A&A...674A...1G}, and an updated amplitude scaling relation derived from Kepler observations \citep{Huber_2011}. 

After cross-matching the entire ATL sample with \textit{Gaia} stellar parameters based on \textit{Gaia} ID, we selected stars with a $(G_{\rm BP}-G_{\rm RP}) \geq 0.78$, which corresponds approximately to stars colder than G0 dwarfs ($T_{\text{eff}} \le 5930\,\text{K}$) \footnote{https://www.pas.rochester.edu/ \~emamajek/} \citep{2013ApJS..208....9P}. We kept only stars in the main sequence and sub-giant branch, based on their position in the \textit{Gaia} color-magnitude diagram (CMD). We also applied a frequency cut based on the ATL predicted $\nu_{\text{max}}$ and selected stars with $\nu_{\text{max}} \leq 4167\,\mu\text{Hz}$, to remove stars with predicted frequency faster than the TESS 2 min cadence Nyquist limit of 4167 $\mu\text{Hz}$. We downloaded the SPOC light curves for the remaining 81,642 stars using their TIC IDs, and we pre-processed the signals in the same manner described in Section~\ref{sec:data}. We also removed the known solar-like oscillators that were used to train, validate, and test our model as described in Section~\ref{sec:network-and-training}.
Our final sample included periodograms of 81,642 stars, which were then passed on to the neural network, described in Section~\ref{sec:network-and-training}, to estimate their probability of having solar-like oscillations.

For each target, the neural network outputs a probability of having solar-like oscillations between 0 and 1. From this output, we found that the majority of high-probability targets are clustered near the sub-giant branch, and the probability gradually decreases towards the low-temperature region along the main-sequence. 


\begin{deluxetable*}{ccccccc}[t]
\tablecaption{Asteroseismic and stellar parameters of all candidates with Probability $\ge$ 0.5 of being a solar-like oscillator as determined by neural network algorithm.}
\tablehead{
\colhead{TIC} &
\colhead{$R_\star$} &
\colhead{$T_{\mathrm{eff}}$} &
\colhead{$\nu_{\max}$ (Predicted)} &
\colhead{SNR} &
\colhead{Probability}&
\colhead{Candidate}\\
\colhead{} &
\colhead{[$R_\odot$]} &
\colhead{[K]} &
\colhead{[$\mu$Hz]} &
\colhead{} &
\colhead{} &
\colhead{}
}
\startdata
34860555  & 1.264 & 6095 & 1323.83 & 0.621 & 0.76 & 0\\
8193616   & 1.249 & 5629 & 1896.17 & 1.094 & 0.59 & 0\\
6528048   & 1.418 & 5877 & 1818.08 & 1.322 & 0.67 & 0\\
4995808   & 1.830 & 5382 & 1065.62 & 3.664 & 0.55 & 1\\
9333731   & 1.892 & 5904 & 826.96  & 4.149 & 0.64 & 1\\
\enddata
\tablecomments{Columns are: TESS Input Catalog identifier (TIC), stellar radius $R_\star$, effective temperature $T_{\mathrm{eff}}$, frequency of maximum power $\nu_{\max}$, signal-to-noise ratio (SNR), and Probability of exhibiting solar-like oscillation as determined by the neural network (Probability). Finally, the Candidate columns denote whether the star is included in our candidate sample in Section~\ref{sec:new_candidates}. A full version of this table containing all 3463 candidates with probability $\ge $ 0.5 can be found online.}
\label{tab:0.5-mini}
\end{deluxetable*}

To remove targets that are unlikely to show solar-like oscillations, we selected stars with probability $\ge$ 0.5 (this cut was also used during training process), yielding a sample of 3463 candidates. This list of candidates can be found in Table~\ref{tab:0.5-mini}. 

\subsection{Vetting of Candidates}
\label{sec:new-candidates}

In this section, we describe the manual vetting process of the shortlisted target sample produced by the neural network. For the vetting process, we leveraged the known properties of solar-like oscillators and \texttt{pySYD} \citep{2022JOSS....7.3331C}, an open source Python package to automatically calculate global oscillation parameters, to further shortlist our candidate list. A step-by-step description of this process is given below.  

Solar-like oscillations are characterized by a Gaussian-like power excess envelope near the frequency of maximum power ($\nu_{\text{max}}$). Moreover, they have evenly spaced radial overtones for a fixed angular degree of oscillation mode. This results in evenly spaced peaks near $\nu_{\text{max}}$ in their periodogram. This feature can be revealed by calculating auto-correlation functions (acf) of their background-corrected periodogram and identifying periodic patterns. Moreover, this also means their \'echelle diagram will have vertical ridges for the right choice of large frequency separation ($\Delta \nu$). To identify these properties, we calculated the background-corrected power spectrum, $\nu_{\text{max}}$ and acf using \texttt{pySYD}. Furthermore, we calculated their \'{e}chelle diagram using the \texttt{echelle} package \citep{daniel_hey_2020_3629933}. 

In the subsequent stage of our analysis, we applied a series of selection criteria to refine the candidate list: (1) We required consistency between the $\nu_{\text{max}}$ values derived using \texttt{pySYD} and those predicted in the ATL catalog. The \texttt{pySYD} algorithm models the oscillation power excess as a Gaussian envelope characterized by a central frequency and width (1$\sigma$); candidates were retained if the ATL-predicted $\nu_{\text{max}}$ lay within this envelope; (2) We required the presence of periodicity in the acf; (3) We excluded targets affected by background contamination by conducting a 15–20 arcmin search around each star and comparing the light curves of nearby sources to identify potential blending or flux leakage; (4) Finally, we visually inspected the \'{e}chelle diagrams and retained only those stars exhibiting well-defined, vertically aligned ridges indicative of regular frequency spacing. In this process, we also removed the stars that were previously identified as solar-like oscillators and were not part of our training process (see Section~\ref{sec:prev-det}).

\begin{figure*}
    \centering
    \includegraphics[width=.99\textwidth]{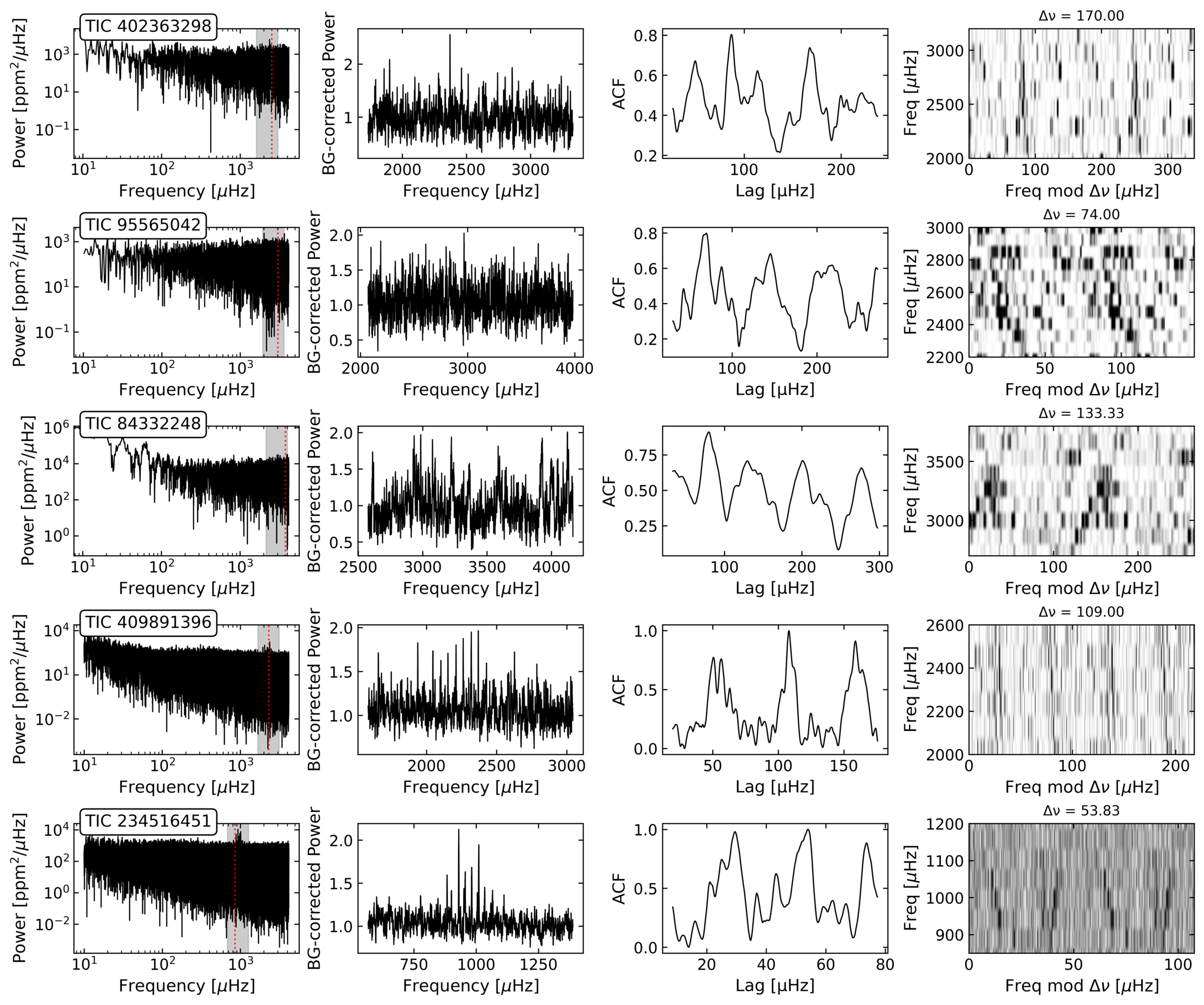}
    \caption{Diagnostic plots for five solar-like oscillator candidates from Figure~\ref{fig:teff-rad-plot} with TIC 402363298, 95565042, 84332248, 409891396, and 234516451 respectively. \textit{First Column}: power spectrum with region near observed $\nu_{\text{max}}$ is shaded in dark grey and predicted $\nu_{\text{max}}$ from scaling relation in dotted red line, \textit{Second Column}: background-corrected power spectrum zoomed-in on expected $\nu_{\text{max}}$, \textit{Third Column}: auto-correlation function, \textit{Fourth Column}: \'{e}chelle diagram. TIC 402363298 and 84332248 are in K~dwarf region, TIC 95565042, 409891396, and 234516451 in G~dwarf region.}
    \label{fig:pysyd-output}
\end{figure*}

\begin{figure}
    \centering
    \includegraphics[width=\linewidth]{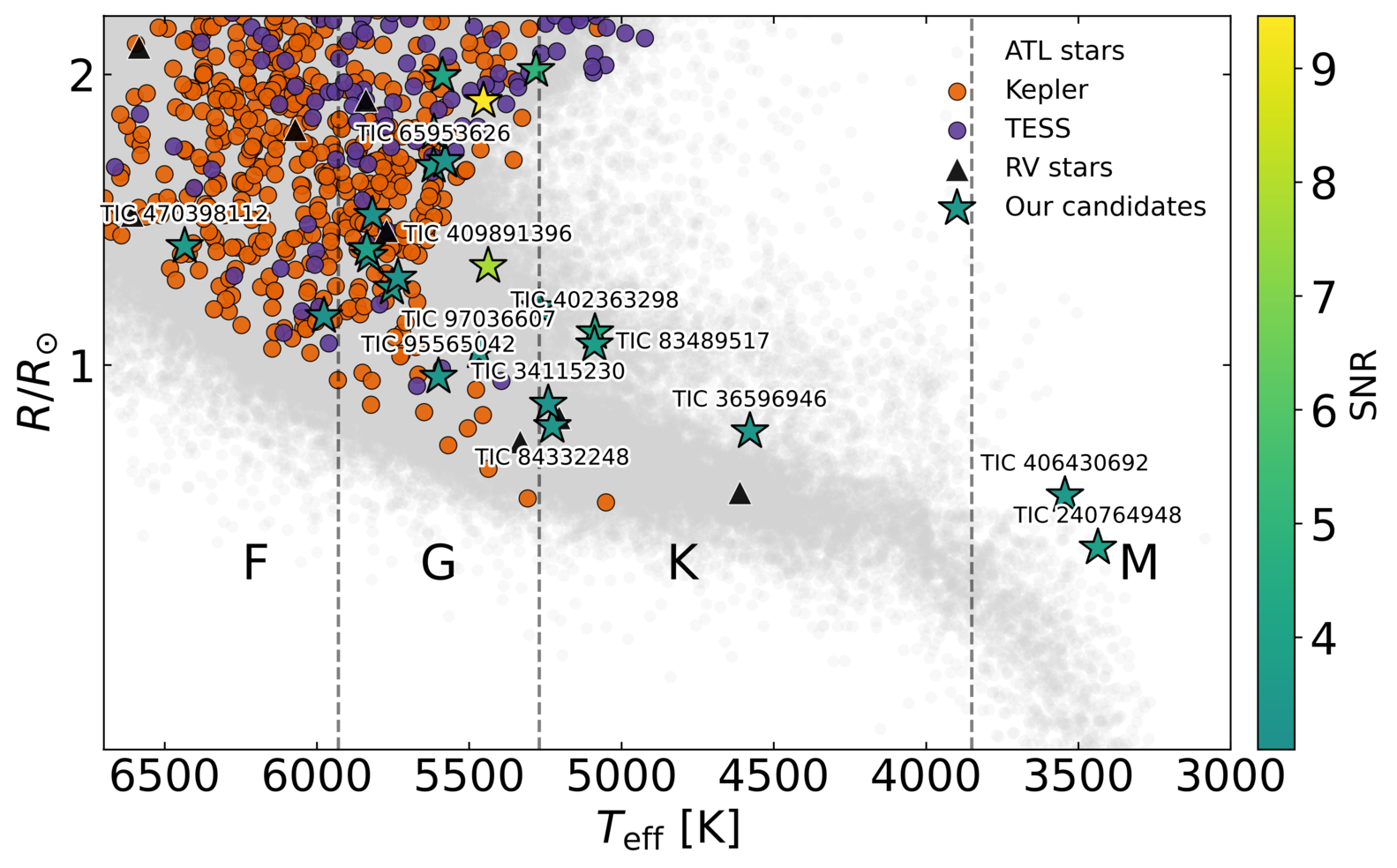}
    \caption{Stellar radius versus effective temperature plot for our candidates (star symbol) and previously discovered solar-like oscillators using TESS (violet circles), Kepler (orange circles), and radial velocity (black triangles). Our candidates are color-coded by their signal-to-noise ratio. TESS and Kepler stars were taken from \cite{Hatt_2023} and \cite{2017ApJS..229...30M}, respectively. Radial velocity candidates were obtained from \cite{2008ApJ...687.1180A}, \cite{2008ApJ...682.1370K} and references therein. Every star with TIC ID in the figure is discussed in Section~\ref{sec:discussion}. Stars without TIC IDs are more common targets, which are also discussed in Section~\ref{sec:discussion}. These candidates satisfy the following conditions: (1) existence of power excess near predicted $\nu_{\text{max}}$, (2) contains periodic auto-correlation function (3) no contamination from stars in the background, and (4) existence of vertical ridges in \'{e}chelle diagram.}
    \label{fig:teff-rad-plot}
\end{figure}

Figure~\ref{fig:pysyd-output} shows five examples of our manual vetting process and includes the full power spectra, background-corrected power spectra, autocorrelation functions, and \'{e}chelle diagrams for five main sequence G and K~dwarf candidates indicated on the radius-temperature plot (Figure~\ref{fig:teff-rad-plot}). The first column contains the power spectrum, where the expected $\nu_{\text{max}}$ from ATL catalog is represented by a dotted red line. The observed $\nu_{\text{max}}$ calculated using the \texttt{pySYD} package and the width of the gaussian envelope around it is shaded in gray. The second column contains the background-corrected power spectrum zoomed in near the observed $\nu_{\text{max}}$, followed by autocorrelation function in third column, and \'{e}chelle diagram in the fourth. 

All five candidates in Figure~\ref{fig:pysyd-output} have periodic auto-correlation function, and their \'{e}chelle diagrams show dark vertical ridges. The predicted $\nu_{\text{max}}$ for all five candidates, denoted by the dotted red line, agrees with the observed $\nu_{\text{max}}$ using \texttt{pySYD}. Diagnostic plots for all the rest of the interesting candidates are provided in the Appendix~\ref{sec:Appendix}.

This analysis greatly reduced our number of candidates. Out of the 3463 stars, 23 stars satisfied all four criteria, with a few exceptions that we discuss later in this section.
We did not exclude any candidates based on signal-to-noise ratio. This choice was made given that previous works have shown that for faint stars with low signal-to-noise ratio, oscillation features, such as $\Delta \nu$, can be revealed by a periodic acf, even when a reliable measurement of $\nu_{\text{max}}$ is not possible \citep[e.g.,][]{sayeed2025homogeneouscatalogoscillatingsolartype, 2014ApJS..210....1C}. Table~\ref{tab:candidate-table} contains the asteroseismic parameters $\nu_{\text{max}}$ and $\Delta \nu$ derived in this study along with fundamental stellar parameters of these 23 stars. This table also includes one candidate found using TESS 20 second search discussed in Section~\ref{20-sec-search}. To calculate the signal-to-noise ratio (SNR), we used running median of the periodogram as the noise level. In cases where the periodogram has a clear power excess, height of that peak was used as signal, otherwise height of the maximum peak near the region of expected $\nu_{\text{max}}$ was chosen to be the signal. The stellar parameters $\text{T}_{\text{eff}}$ and radius were taken from Gaia DR3 \citep{2023A&A...674A...1G} and Tess Input Catalog 
\citep{2018AJ....156..102S}.

A radius-temperature plot of our 23 candidates is shown in Figure~\ref{fig:teff-rad-plot} against previously discovered solar-like oscillators using photometry from TESS and Kepler, and radial velocity measurements. The solar-like oscillators from Kepler were taken from \cite{2017ApJS..229...30M}, which provided a catalog of 12,777 solar-like oscillator. The TESS discoveries of this figure are from \cite{Hatt_2023}, as described in Section~\ref{sec:data}, that added 4177 solar-like oscillator across main-sequence, sub-giant and giant branch in H-R diagram. The 12 radial velocity candidates are taken from \cite{2008ApJ...687.1180A}, \cite{2008ApJ...682.1370K} and references therein. The plot also includes our candidates that were found using 20 second cadence only, which will be discussed in Section~\ref{20-sec-search}.

Our stars are color-coded by their signal-to-noise ratio. As expected, the majority of candidates lie in the G-dwarf regime, with effective temperatures in the range 
$5270\,K \le T_{\text{eff}} \le 5930\,K$, consistent with most previous TESS detections. The impact of our work is most evident at cooler temperatures, particularly for  $T_{\text{eff}} \le 5000\,K$, where no solar-like oscillators had previously been identified using TESS. In this regime, we identify two main-sequence candidates (TIC 36596946 and TIC 418544237), as well as two cooler candidates located above the main sequence with $T_{\text{eff}} \le 3800\,K$ (TIC 406430692 and TIC 240764948). From left to right in the figure, the signal-to-noise ratio decreases systematically, as expected, reflecting the lower oscillation amplitudes of cooler stars.

\begin{deluxetable*}{cccccccc}[t]
\tablecaption{Asteroseismic and stellar parameters of candidate stars exhibiting periodic auto-correlation functions and dark vertical ridges in their \'{e}chelle diagrams.\label{tab:astero_params}}
\tablehead{
\colhead{TIC} &
\colhead{$R_\star$} &
\colhead{$T_{\mathrm{eff}}$} &
\colhead{$\nu_{\max}$} &
\colhead{$\Delta \nu$} &
\colhead{PE} &
\colhead{SNR} &
\colhead{Cadence} \\
\colhead{} &
\colhead{[$R_\odot$]} &
\colhead{[K]} &
\colhead{[$\mu$Hz]} &
\colhead{[$\mu$Hz]} &
\colhead{} &
\colhead{} &
\colhead{}
}
\startdata
84332248  & 0.864 & 5226.7 & 3070.21 & 133.33 & 1 & 3.443707238 & 2 min \\
314805964 & 1.147 & 5270.8 & 2274.71 & 115.41 & 1 & 3.439693346 & 2 min \\
65953626  & 1.608 & 5616.8 & 1222.75 & 66.33  & 1 & 3.626556000 & 2 min + 20 sec\\
402363298 & 1.081 & 5086.9 & 2312.52 & 170.00 & 0 & 4.252623103 & 2 min \\
409891396 & 1.266 & 5437.5 & 2400.00 & 109.00 & 1 & 7.771395930 & 2 min + 20 sec\\
9102735   & 1.627 & 5578.6 & 1142.96 & 67.80  & 1 & 3.267807716 & 2 min \\
9333731   & 1.992 & 5588.0 & 733.13  & 43.00  & 1 & 4.149027362 & 2 min \\
330014070 & 1.430 & 5817.6 & 1682.50 & 99.00  & 0 & 3.101131390 & 2 min \\
406430692 & 0.733 & 3543.9 & 1700.10 & 131.50 & 0 & 3.470821167 & 2 min \\
211227897 & 1.298 & 5828.1 & 1711.02 & 76.00  & 0 & 3.330173687 & 2 min \\
240764948 & 0.647 & 3435.7 & 2166.91 & 105.00 & 1 & 3.894182859 & 2 min \\
34115230  & 0.912 & 5240.3 & 3363.51 & 102.00 & 0 & 3.197691186 & 2 min \\
79279141  & 1.318 & 5835.9 & 1848.90 & 64.10  & 1 & 3.923868602 & 2 min \\
20887793  & 1.124 & 5976.8 & 2108.20 & 94.50  & 1 & 3.016192161 & 2 min \\
36596946  & 0.854 & 4578.2 & 3325.50 & 137.00 & 1 & 3.380654612 & 2 min \\
4995808   & 1.740 & 5615.8 & 952.65  & 36.00  & 1 & 3.664899277 & 2 min \\
83489517  & 1.050 & 5088.5 & 2234.99 & 87.50  & 0 & 3.710662560 & 2 min \\
92254955  & 1.202 & 5753.7 & 2250.00 & 103.00 & 0 & 3.660322086 & 2 min \\
95565042  & 0.973 & 5600.7 & 2774.66 & 74.00  & 1 & 3.379471087 & 2 min \\
97917479  & 1.232 & 5733.6 & 1288.84 & 80.00  & 0 & 3.202157115 & 2 min \\
333704881 & 2.021 & 5280.5 & 830.00  & 31.50  & 0 & 4.935405533 & 2 min \\
470398112 & 1.328 & 6433.8 & 1850.29 & 72.50  & 1 & 3.636444166 & 2 min \\
234516451 & 1.876 & 5453.0 & 977.63  & 53.83  & 1 & 9.460528    & 2 min + 20 sec \\
97036607 & 1.0327 & 5467.3 & 2241    & 75     & 0 & 3.488465    & 20 sec \\
\enddata
\tablecomments{Columns are: TESS Input Catalog identifier (TIC), stellar radius $R_\star$, effective temperature $T_{\mathrm{eff}}$, frequency of maximum power $\nu_{\max}$, large frequency separation $\Delta \nu$, presence of a visible power excess (PE; 1 = yes, 0 = no), and signal-to-noise ratio (SNR).}
\label{tab:candidate-table}
\end{deluxetable*}

\subsection{20 Second Search}
\label{20-sec-search}

Following our analysis of TESS 2 min cadence light curves using the neural network and \texttt{pySYD}, we additionally examined the available 20 second cadence light curves from TESS for the targets of the ATL catalog \citep{hey2024precisetimedomainasteroseismologyrevised}. From the ATL target sample described in Section~\ref{sec:target-selection}, 13,620 stars had 20 second cadence light curves in TESS. We first binned the light curves of these stars to 2 min cadence to make them compatible for our analysis using the neural network described in Section~\ref{sec:network-and-training}. After analyzing their signals using the neural network in the similar manner as their 2 min counterpart, 596 stars had a probability $>$ 0.5 of having solar-like oscillation. Then we manually vetted the candidates using \texttt{pySYD} and \texttt{echelle} package using the criteria outlined in Section~\ref{sec:new-candidates}. This process yielded four stars, three of which (TIC~65953626, TIC~409891396, and TIC~234516451) were already present in our candidate list based on the 2 min cadence analysis. These three candidates were also the only 2 min candidates that also had 20 second cadence, so this analysis strengthens their candidacy and does not discard any other 2 min candidates from Section~\ref{sec:new-candidates}.
The only new candidate from the 20 second cadence data is TIC~97036607. This new candidate is included in our radius-temperature plot in Figure~\ref{fig:teff-rad-plot} and in our candidate list in Table~\ref{tab:candidate-table}.

\begin{figure*}
    \centering
    \includegraphics[width=0.98\linewidth]{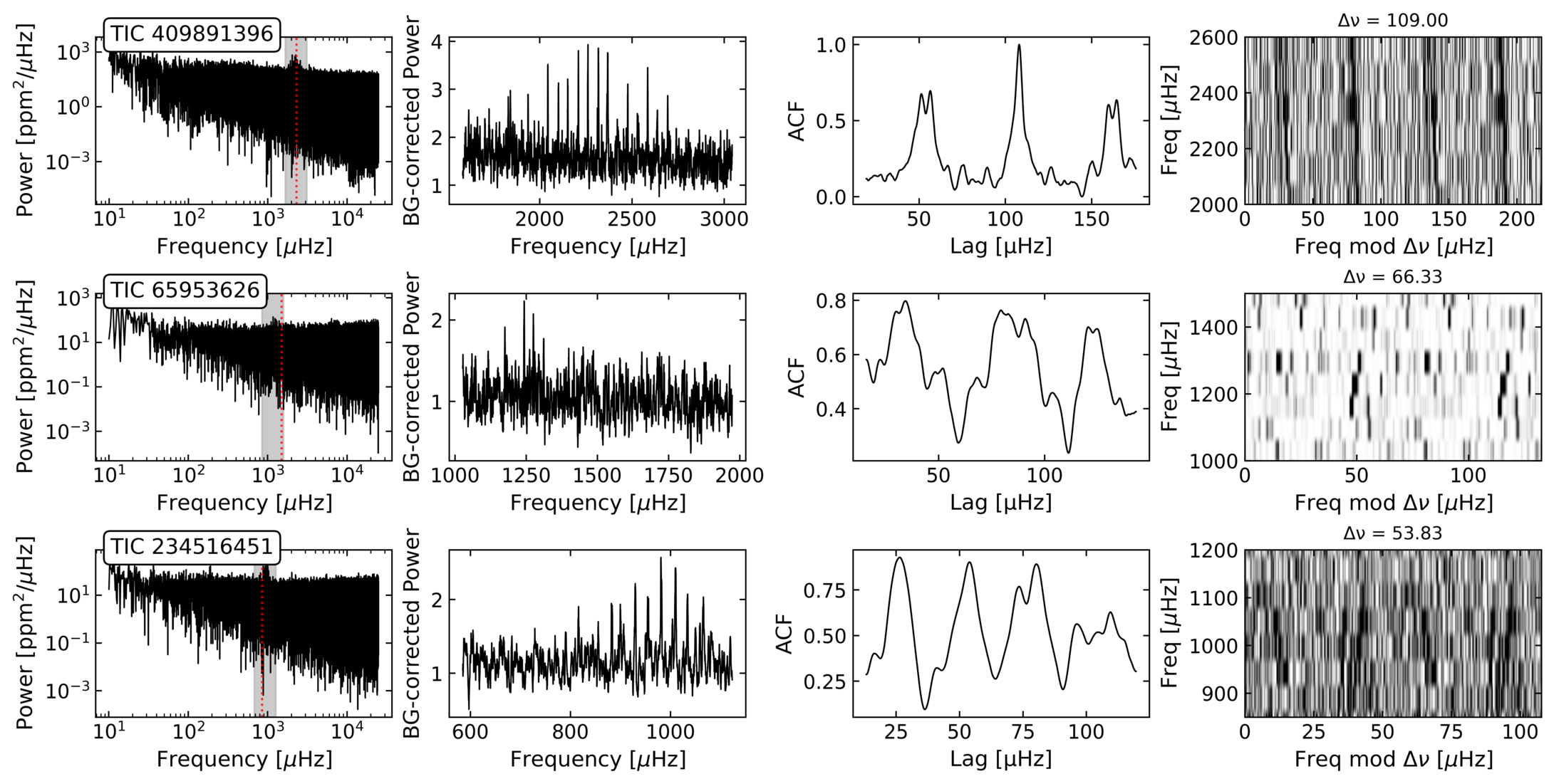}
    \caption{Diagnostic plots for two solar-like oscillator candidates from Figure~\ref{fig:teff-rad-plot} with TIC 65953626, and 409891396 using TESS 20 second cadence. \textit{First Column}: power spectrum with region near observed $\nu_{\text{max}}$ is shaded in dark grey and predicted $\nu_{\text{max}}$ from scaling relation in dotted red line, \textit{Second Column}: background-corrected power spectrum zoomed-in on expected $\nu_{\text{max}}$, \textit{Third Column}: auto-correlation function, \textit{Fourth Column}: \'{e}chelle diagram.}
    \label{fig:overlap-stars}
\end{figure*}

Diagnostic plots for TIC~65953626, TIC~409891396, and TIC~234516451, using the 20 second data, are shown in Figure~\ref{fig:overlap-stars}. As evident from these plots, the SNR of the first two candidates improved substantially to 4.029 and 9.089, respectively, compared to the values reported in Table~\ref{tab:candidate-table}, whereas the SNR of the last candidate decreased to 7.41303. In addition, the auto-correlation functions exhibit a more pronounced periodic structure, and the ridges in the \'{e}chelle diagrams are more clearly defined. For TIC~409891396 in particular, the background-corrected power spectrum reveals a greater number of radial overtones relative to the 2 min cadence results shown in Figure~\ref{fig:pysyd-output}. These improvements provide strong support for the robustness of the candidate list presented in Figure~\ref{fig:teff-rad-plot}. With the availability of 20 second cadence data for all candidates, the detection of solar-like oscillations in these stars can be further strengthened and, in some cases, confidently confirmed.

\begin{figure*}
    \centering
    \includegraphics[width=0.98\linewidth]{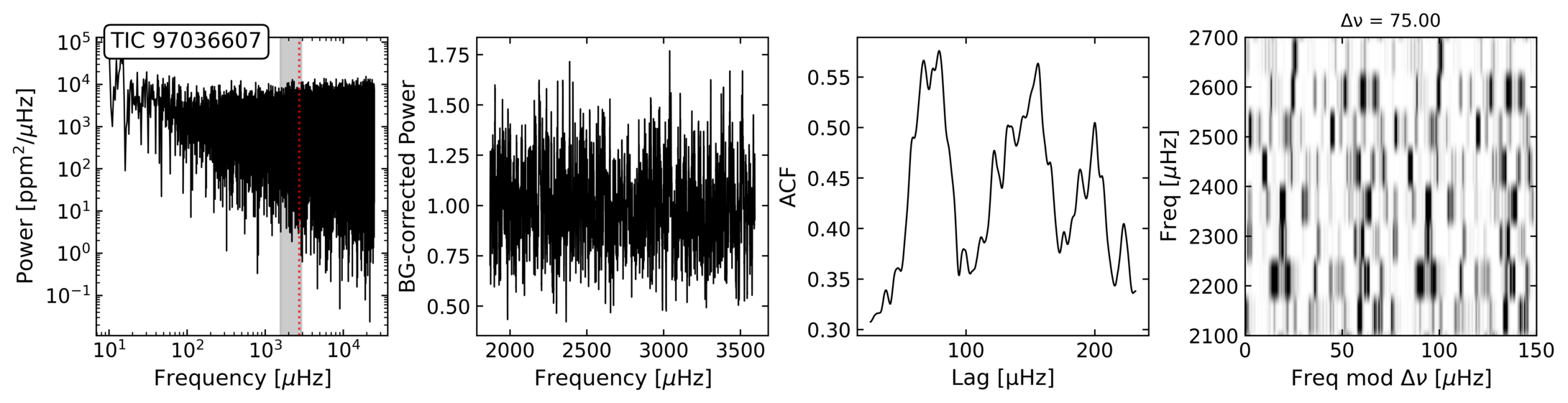}
    \caption{Same as Figure~\ref{fig:overlap-stars} for TIC~97036607. }
    \label{fig:20sec-new}
\end{figure*}

TIC~97036607 is a main-sequence late G~dwarf which has a low SNR (3.49), and its power spectrum does not have any visible power excess. However, the existence of periodic structures in the autocorrelation function and evidence of two dark ridges in \'{e}chelle diagram make it a compelling candidate. We also investigated the 2 min cadence data of this star, and found that although it was classified as solar-like oscillator by the neural network, its autocorrelation function did not show such periodic pattern and we did not find any ridges in its \'{e}chelle diagram. Only in 20 second cadence data was the oscillation signal strong enough to show these diagnostic features. A summary plot of these diagnostic features can be found in Figure~\ref{fig:20sec-new}.

\subsection{Comparison to Previous Detections}
\label{sec:prev-det}
In this section, we discuss the general overlap between the ATL catalog and previously detected solar-like oscillators using Kepler and radial velocity techniques. Then we point out how many of these candidates were selected in the sub-sample of ATL that we used to test the neural networ,k and how many of them were recovered as solar-like oscillators.

Our sample was taken from the ATL catalog, which initially had 1,985,986 stars. Then, we cross-matched these stars with \textit{Gaia}, and cleaned the sample based on $(G_{\rm BP}-G_{\rm RP})$ and position in the H-R diagram as described in Section~\ref{sec:target-selection}. Furthermore, we also removed the stars that were already in our training set described in Section~\ref{sec:data}. This resulted in 267,849 stars, out of which only 81,642 stars had TESS light curves that we passed on to the neural network to find new solar-like oscillators. In the test sample of this 81,642 stars, 10 stars were previously identified as solar-like oscillators in the literature using Kepler and radial velocity techniques \citep{2017ApJS..229...30M, 2008ApJ...687.1180A, 2008ApJ...682.1370K}. Our neural network was able to recover 7 of these stars as solar-like oscillators, with probability greater than 0.5. We then processed these 7 stars through the \texttt{pySYD} pipeline, and only one of them, which was originally detected using Kepler, satisfied our four criteria laid out in Section~\ref{sec:new_candidates} to classified as candidate for solar-like oscillator. A summary of these 10 stars, their diagnostic properties to exhibit solar-like oscillation, and fundamental properties can be found in Table~\ref{tab:prev-osc}.

\begin{deluxetable*}{cccccccc}[t]
\tablecaption{List of stars in target sample that were previously identified as solar-like oscillator.}
\tablehead{
\colhead{TIC} &
\colhead{$T_{\mathrm{eff}}$} &
\colhead{$\nu_{\max}$} &
\colhead{acf} &
\colhead{\'{E}chelle} &
\colhead{Recovered} &
\colhead{Detection} &
\colhead{Reference}\\
\colhead{} &
\colhead{[K]} &
\colhead{[$\mu$Hz]} &
\colhead{} &
\colhead{Diagram} &
\colhead{NN} &
\colhead{Method} &
\colhead{}
}
\startdata
419015728  & 5333.00 & 4244.57 & 0 & 0& 0 & RV & \cite{2008ApJ...687.1180A}\\
123317744  & 4486.75 & 760.68  & 0 & 0& 1 & Kepler & \cite{2017ApJS..229...30M}\\
399952145  & 5284.57 & 4328.43 & 0 & 0& 1 & Kepler & \cite{2017ApJS..229...30M}\\
27013540   & 5212.55 & 606.85  & 1 & 1& 1 & Kepler & \cite{2017ApJS..229...30M}\\
394172596  & 5058.00 & 6156.30 & 0 & 0& 1 & Kepler & \cite{2017ApJS..229...30M}\\
172424739  & 4289.22 & 3577.78 & 0 & 0& 1 & Kepler & \cite{2017ApJS..229...30M}\\
123234496  & 5270.85 & 6035.91 & 0 & 0& 1 & Kepler & \cite{2017ApJS..229...30M}\\
1674663309 & 5204.63 & 3046.97 & 0 & 0& 1 & RV & \cite{2008ApJ...682.1370K}\\
231698181  & 4560.30 & 4228.75 & 0 & 0& 0 & RV & \cite{Campante_2024}\\
406951876  & 5239.41 & 793     & 0 & 0& 0 & Kepler & \cite{2017ApJS..229...30M}\\
\enddata
\tablecomments{ The columns are: TESS Input Catalog(TIC), effective temperature $T_{\mathrm{eff}}$, frequency of maximum power $\nu_{\max}$, presence of periodic peaks in the autocorrelation function (acf), recovery of the solar-like oscillations using neural network (NN), and original detection method. The $\nu_{\max}$ of the stars were obtained from the respective studies.}
\label{tab:prev-osc}
\end{deluxetable*}

TIC 419015728 and TIC 231698181 were identified as solar-like oscillator in \cite{2008ApJ...687.1180A} and \cite{Campante_2024} using radial velocity technique, but our neural network could not recover them. Their $\nu_{\text{max}}$ is 4244.57\,$\mu\text{Hz}$ and 4228.75\,$\mu\text{Hz}$ respectively, which is beyond the Nyquist limit of 4167 $\mu\text{Hz}$ in TESS 2 min cadence. Therefore, it is not surprising that we did not see any oscillation signal in these two stars. Both of these stars have a SNR of 2.5 \citep{2009A&A...494..237T, Campante_2024}, but the Nyquist limit of TESS 2 min cadence makes it unlikely to observe this signal. Moreover, they do not currently have 20 sec cadence in TESS, so we could not verify if the oscillation actually becomes visible in TESS.

TIC 406951876 is a K0 Giant star with temperature 5239 K. It has a $\nu_{\text{max}}$ of 793 $\mu\text{Hz}$ which is well within the reach of TESS 2 min cadence. However, it was not identified by the neural network as exhibiting solar-like oscillations. From \cite{hey2024precisetimedomainasteroseismologyrevised}, the reported oscillation amplitude using Kepler data for this star is 8.95 in arbitrary unit (a.u.), but in TESS 2 min data, the noise amplitude calculated from the median noise floor of the power spectrum has a value of 18.01 (a.u.), making it unlikely to find the oscillation signal.

Among the stars we have recovered using the neural network, only TIC 27013540 passed the four criteria we used to confirm the candidacy of solar-like oscillation mentioned in Section~\ref{sec:new-candidates}. This star has a $\nu_{\text{max}}$ of 606\,$\mu \text{Hz}$ (compared to $\nu_{\text{max}}$ 576$\mu \text{Hz}$ which we recovered), which is well within the range of TESS 2 min light curves. It also shows a periodic autocorrelation function and dark vertical ridges in its \'{e}chelle diagram, which satisfies our candidacy requirements. 

TIC 399952145, TIC 394172596, and TIC 231698181 have $\nu_{\text{max}}$ above the Nyquist limit of 4167 $\mu\text{Hz}$ for TESS 2 min cadence, which makes it very hard to observe the oscillation signal in the light curves. However, these super-Nyquist oscillation signals can reveal themselves as alias peaks in the power spectrum at a frequency lower than the Nyquist limit. This is consistent with our other observation of super-Nyquist oscillation using the neural network discussed in Section~\ref{sec:discussion}. Also, the neural network is trained to not only find the oscillation signal, but also the granulation background, which usually resides in the low-frequency regime of the power spectrum, enabling the neural network to identify the granulation shape. 

For the remaining stars in Table~\ref{tab:prev-osc}, the values of $\nu_{\text{max}}$ lie well within the Nyquist limit of the TESS 2 min cadence, enabling the neural network to detect their variability pattern in the power spectrum created by oscillation modes. However, these stars did not satisfy the additional selection criteria outlined in Section~\ref{sec:new-candidates}; specifically, they exhibited neither significant periodicity in their autocorrelation functions nor the characteristic dark vertical ridges in their \'{e}chelle diagrams. Nonetheless, the ability of the neural network to recover these oscillation signatures from TESS 2 min light curves, serves as a valuable validation of the model’s capacity to reliably identify subtle oscillatory features.

\section{Discussion of Interesting Objects} \label{sec:discussion}

In the previous section we presented our solar-like oscillator candidates in both main sequence and sub-giant branch. This section includes the discussion of all the objects that have their TIC IDs in Figure~\ref{fig:teff-rad-plot}, which includes the most interesting objects given their low temperature. For each interesting object, we first discuss the strength of its candidacy, followed by a discussion of its fundamental stellar properties such as temperature and radius, and finally compare it with oscillators previously found in the same region of the radius-temperature plot.

\begin{figure*}
    \centering
    \includegraphics[width=0.9\linewidth]{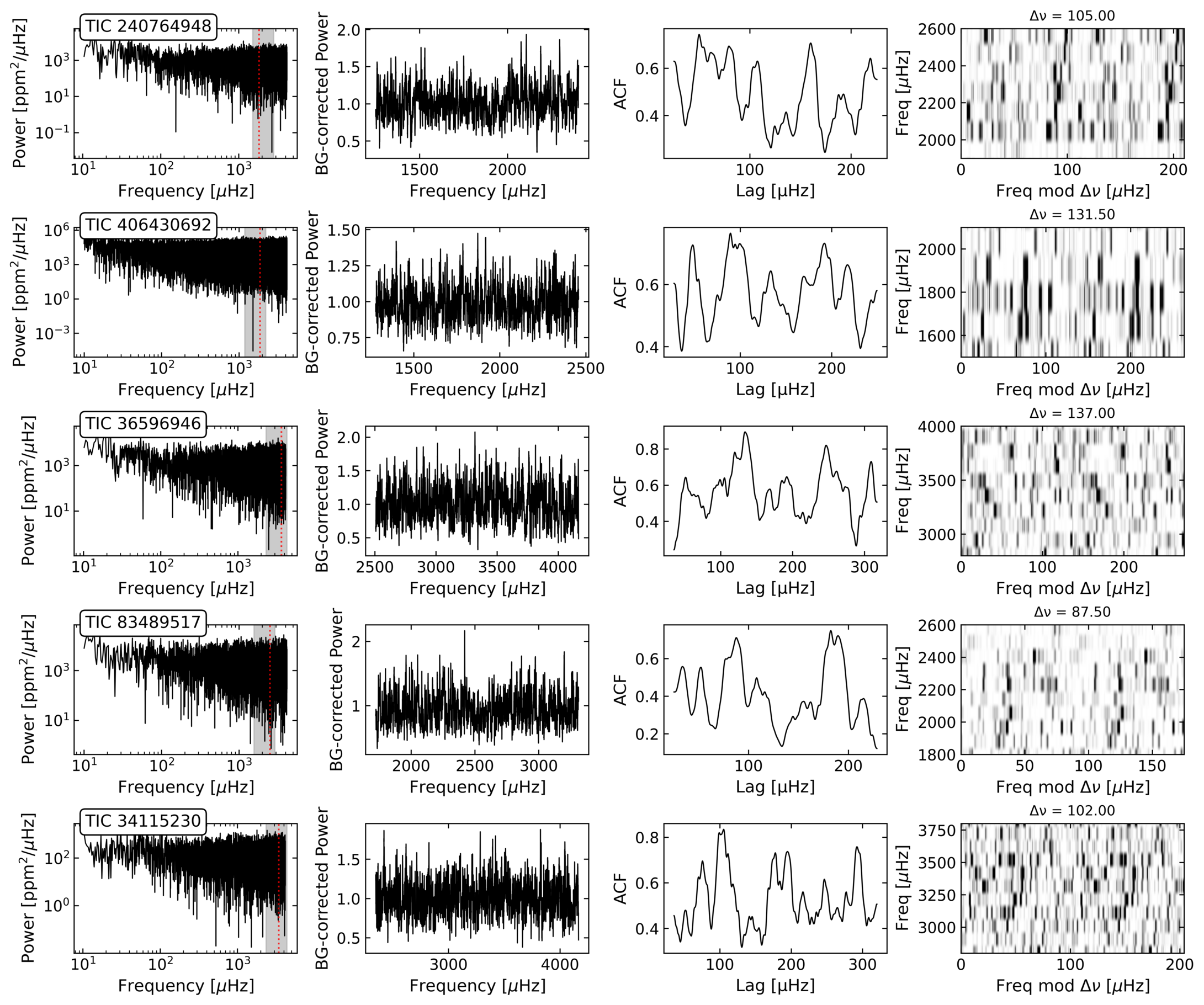}
    \caption{Diagnostic plots of TIC 240764948, TIC 406430692, TIC 36596946, TIC 83489517, and TIC 34115230. \textit{First Column}: power spectrum with region near observed $\nu_{\text{max}}$ is shaded in dark grey and predicted $\nu_{\text{max}}$ from scaling relation in dotted red line, \textit{Second Column}: background-corrected power spectrum zoomed-in on expected $\nu_{\text{max}}$, \textit{Third Column}: auto-correlation function, \textit{Fourth Column}: \'{e}chelle diagram. The plots are in ascending order of temperature. The objects are presented in ascending order of temperature.}
    \label{fig:187}
\end{figure*}

TIC 240764948 is an M~dwarf residing right above the main sequence in the temperature-vs-radius plot shown in Figure~\ref{fig:teff-rad-plot}. Its power spectrum (Figure~\ref{fig:187}, row 1) does not have a visible power excess, but the predicted $\nu_{\text{max}}$ falls within the shaded area. Its autocorrelation function has repeating peaks at a frequency interval of around 50 $\mu\text{Hz}$, although the shape of the peaks are not uniform across the frequency range. The strongest evidence supporting the existence of solar-like oscillations is the presence of dark vertical ridges in its \'{e}chelle diagram, with one angular oscillation mode visible around $\Delta\nu \approx 95\,\mu\text{Hz}$. This star has a nearby star at a distance of 4.4 arcseconds found through visual inspection. The neighbor, Gaia DR3 2596359407478746112, has a similar effective temperature (3880 K), parallax (23.80 and 23.84 mas) and proper motion ((62.16, -4.13) and (64.97, -8.62) mas/yr), making it a potential M~dwarf binary system. As part of our effort to check for contamination in light curves from background stars, we used \texttt{tess\_photometry} (M. Pedersen in prep) to extract the light curves of these two stars directly from their target pixel files. However, we did not find any differences between the two light curves as the stars were practically blended together. Therefore, the oscillation signal could be coming from either of the stars. This is an interesting candidate for follow up with 20-second cadence and/or radial velocity to find oscillation, as it would be the coldest star to date with pulsation signal (see e.g. \cite{Campante_2024} for the coldest pulsating star to date).

This candidate exhibits a larger radius and a lower ATL-predicted $\nu_{\text{max}}$ (1827.26 $\mu\text{Hz}$) than is typical for main-sequence stars of this spectral type. Given its effective temperature, it is unlikely that the star has evolved into a sub-giant. A more plausible explanation is flux contamination from a binary companion, which could bias the inferred stellar radius upward and consequently lower the predicted $\nu_{\text{max}}$ in ATL. An alternative scenario is that the star is interacting with its companion and has become inflated, leading to an increased radius and reduced $\nu_{\text{max}}$. However, given their projected separation of 4.4 arcseconds, it is unlikely that the system is within the dynamical range required for such interaction. One interesting possibility in this case could be TIC 240764948 being part of an unresolved binary system, which contributed to its increase in radius calculation and lower predicted $\nu_{\text{max}}$.
A third possibility is that the candidate is a pre-main-sequence star, which could be plausible if it were located near a star-forming region. However, using \texttt{BANYAN} $\Sigma$ \citep{2015ApJ...798...73G}, we have verified that this star does not lie within a young moving group.

TIC 406430692 is another M~dwarf candidate that also lies slightly above the main sequence in Figure~\ref{fig:teff-rad-plot}. It shares the same diagnostic properties as TIC 240764948. Its \'{e}chelle diagram (Figure~\ref{fig:187}, row 2) has one clearly visible angular mode around 80 $\mu\text{Hz}$, and there are some dark vertical patches near 20 $\mu\text{Hz}$, which could be indicative of another angular mode, but we do not have enough signal strength to resolve it clearly. 
This star has a nearby red giant (TIC 406430687) about 65 arcseconds away. Therefore, we extracted the light curve for TIC 406430687 directly from the target pixel file by carefully selecting a mask that contains light only from this star using \texttt{tess\_photometry} (M. Pedersen in prep), and re-ran our analysis using \texttt{pySYD} to see if a signal similar to TIC 406430692 appears with a larger amplitude. However, we did not find the same signal re-appearing, and concluded that TIC 406430692 does not have any contamination from the red giant.

Just like TIC 240764948, this candidate also exhibits a larger radius and a lower ATL-predicted $\nu_{\text{max}}$. Although it does not have a resolved equal-mass binary companion, its high RUWE value (2.9) suggests the presence of an unresolved binary system. In such a scenario, excess flux from the companion would bias the inferred stellar radius upward and consequently lower the predicted $\nu_{\text{max}}$. Similar effects may also arise if the star is inflated due to tidal interaction with a close companion, as observed for TIC 406430692. Because the system is unresolved, it is difficult to conclusively rule out either of these scenarios, and further follow-up observations are required for verification. Finally, as with the previous candidate, we have verified that this star does not lie within a star-forming region.

Several of our candidates are main-sequence K dwarfs located near previously confirmed solar-like oscillators in the radius–temperature plot (Figure~\ref{fig:teff-rad-plot}). These stars exhibit consistent asteroseismic diagnostics, including periodic peaks in their autocorrelation functions and vertical ridges in their \'{e}chelle diagrams, despite differences in their predicted and observed $\nu_{\text{max}}$ values.

TIC~84332248 and TIC~34115230 form a closely related pair of main-sequence K dwarfs. Both exhibit well-defined autocorrelation periodicity and vertical structures in their \'{e}chelle diagrams, consistent with p-mode oscillations (See Figure~\ref{fig:pysyd-output}, row 3 and Figure~\ref{fig:187}, row 5). TIC~84332248 shows particularly strong ridges and a measured $\Delta\nu = 133\,\mu$Hz, placing it near the known oscillator 70~Oph ($\Delta\nu = 161,\mu$Hz; \citealt{2006A&A...450..695C}). TIC~34115230, while also located near 70~Oph in Figure~\ref{fig:teff-rad-plot}, has a smaller spacing ($\Delta\nu = 102,\mu$Hz), consistent with its larger inferred radius. Together, these two targets probe a region of parameter space with only one Kepler detection and no prior TESS discoveries, making them especially important for follow-up.

TIC~95565042 is a G-dwarf main-sequence candidate that displays a clean, periodic autocorrelation function and a single dominant ridge in its \'{e}chelle diagram, likely corresponding to a radial or dipole mode (Figure~\ref{fig:pysyd-output}, row 2). Although a small number of TESS detections exist in this region of Figure~\ref{fig:teff-rad-plot}, their density remains low, reinforcing the significance of this candidate.

A second group consists of stars that have evolved off the main sequence and lie on or near the subgiant branch. These objects generally show lower $\nu_{\text{max}}$ values and broader mode structures but still exhibit multiple independent diagnostics consistent with solar-like oscillations.

TIC~36596946 lies close to $\epsilon$~Ind in the radius–temperature plane but has evolved slightly, resulting in a lower predicted $\nu_{\text{max}}$ of $3663,\mu$Hz. Its observed power excess near $3300,\mu$Hz, combined with periodic autocorrelation peaks and weak but discernible \'{e}chelle ridges with $\Delta\nu$ in the $20$–$30\,\mu$Hz range (Figure~\ref{fig:187}, row 3), makes it a strong oscillation candidate. Confirming oscillations in this star would extend TESS detections into a region of the color–magnitude diagram with no previous asteroseismic identifications.

TIC~83489517 and TIC~402363298 occupy a similar location slightly above the main sequence and share broadly comparable diagnostic behavior. Both show agreement between ATL-predicted and \texttt{pySYD}-derived $\nu_{\text{max}}$ values, visible power excess near $2400\,\mu$Hz, periodic autocorrelation functions, and vertical ridges in their \'{e}chelle diagrams indicative of low-degree modes (Figure~\ref{fig:187}, row 4, and Figure~\ref{fig:pysyd-output}, row 1). TIC~83489517 exhibits a particularly clear ridge near $\Delta\nu \approx 40\,\mu$Hz, with tentative evidence for an additional mode at higher spacing, while TIC~402363298 shows ridges near $100\,\mu$Hz. The difference in inferred mode spacing, despite their proximity in the color–magnitude diagram, is likely driven by low signal-to-noise limitations that affect overtone identification \citep[e.g.,][]{2014ApJS..210....1C}.

\begin{figure*}
    \centering
    \includegraphics[width=0.9\linewidth]{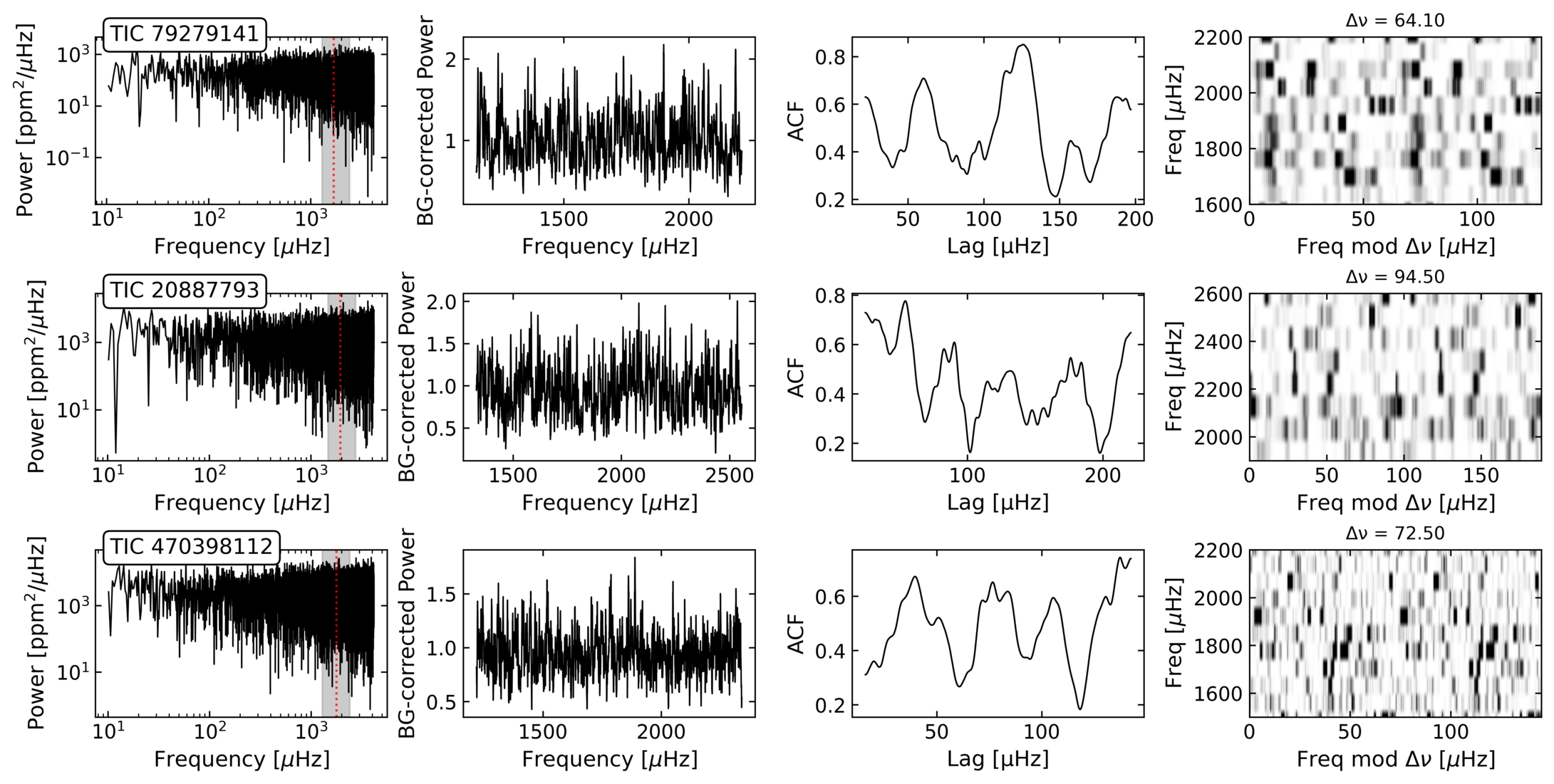}
    \caption{The same at at Figure~10, but for TIC 79279141, TIC 20887793, and TIC 470398112.}
    \label{fig:141}
\end{figure*}

TIC~470398112 is the hottest star in our sample, an F-type dwarf with $T_{\mathrm{eff}} = 6433$~K. Its diagnostic plots show a strong power excess near the predicted $\nu_{\text{max}}$, a periodic autocorrelation function, and prominent \'{e}chelle ridges (Appendix~\ref{sec:Appendix}, Figure~\ref{fig:141}, row 3). The unusually pronounced oscillation power, relative to cooler targets, highlights the robustness of the neural network in identifying solar-like oscillations across a wide range of stellar temperatures and evolutionary states.

The rest of the targets in Figure~\ref{fig:teff-rad-plot} are in the G~dwarf regime, where the density of previous TESS and Kepler discovery of solar-like oscillator is the highest. Their diagnostic plots are included in Appendix~\ref{sec:Appendix}.  These stars satisfy the criteria to exhibit solar-like oscillation as discussed in Section~\ref{sec:new-candidates}. They have periodic peaks in their autocorrelation functions with varying degrees of periodicity. TIC 409891396 and 333704881, for example, have clear periodic structure, while TIC 249763443 has periodic peaks, but the shape of the peaks are not strictly periodic, and varies from peak to peak. However, all these candidates have dark vertical ridges in their \'{e}chelle diagram. TIC 249763443 and TIC 85435539 show signs of existence of multiple angular degree in their \'{e}chelle diagram with the presence of double vertical ridges. The \'{e}chelle diagram of TIC 187316023 shows existence of ridges that are stretched slightly towards the right, which could be result from stellar activity \citep{Hatt_2024}.

\begin{figure}
    \centering
    \includegraphics[width=0.99\linewidth]{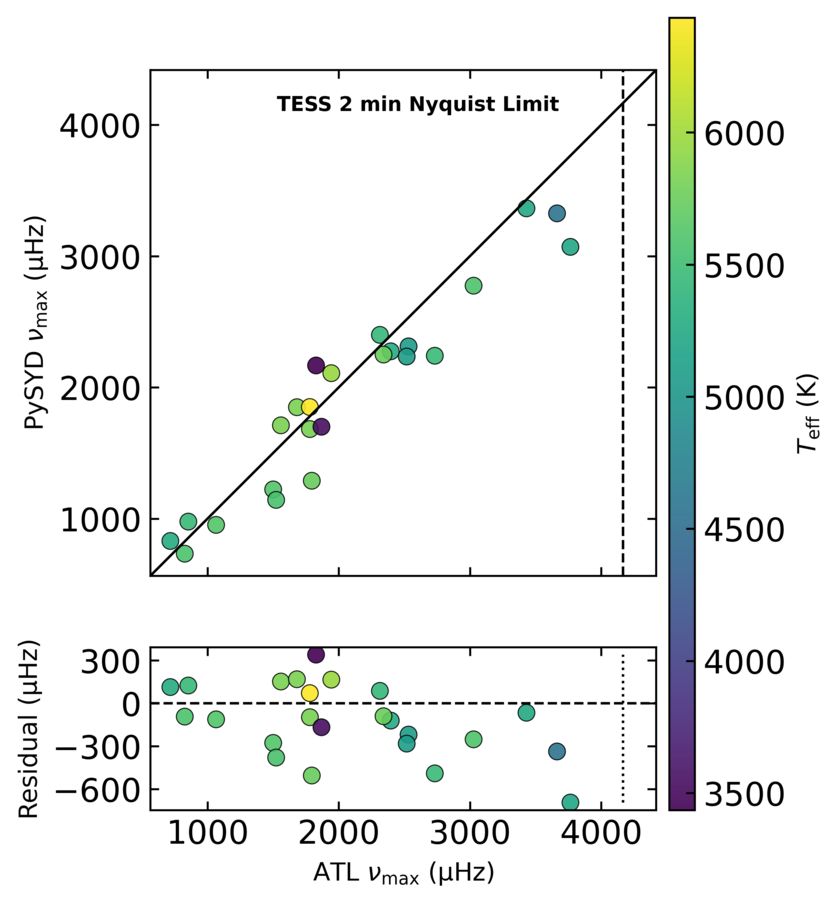}
    \caption{Comparison of predicted $\nu_{\text{max}}$ from ATL catalog \citep{hey2024precisetimedomainasteroseismologyrevised} and obseved $\nu_{\text{max}}$ using \texttt{pySYD}. The stars are color-coded based on their effective temperature. All the candidates with ther TIC IDs are discussed in this section.}
    \label{fig:numax-comparison}
\end{figure}

A comparison between the ATL predicted $\nu_{\text{max}}$ and the observed $\nu_{\text{max}}$ values measured using \texttt{pySYD} for all our candidates is shown in Figure~\ref{fig:numax-comparison}. The stars are color-coded according to their effective temperatures. In general, we find good agreement between the two frequencies with a mean difference of 124.12 $\mu\text{Hz}$. However, for stars with higher frequencies, specially for $\nu_{\text{max}} > 3600\,\mu\text{Hz}$, the difference grows larger. This frequency range lies near the Nyquist limit for TESS 2 min cadence data (4167~$\mu\text{Hz}$, shown in a vertical black dashed line), suggesting that the temporal resolution of these light curves may be insufficient for reliable $\nu_{\text{max}}$ measurements in this regime. Moreover, if the true $\nu_{\text{max}}$ of the star is over the Nyquist limit for TESS, we might only be able to observe an alias of that signal at a lower frequency range \citep{2014MNRAS.445..946C}, which could be the case for the two candidates close to the Nyquist limit of TESS where we underpredict their $\nu_{\text{max}}$ using \texttt{pySYD}.

\begin{figure}
    \centering
    \includegraphics[width=0.99\linewidth]{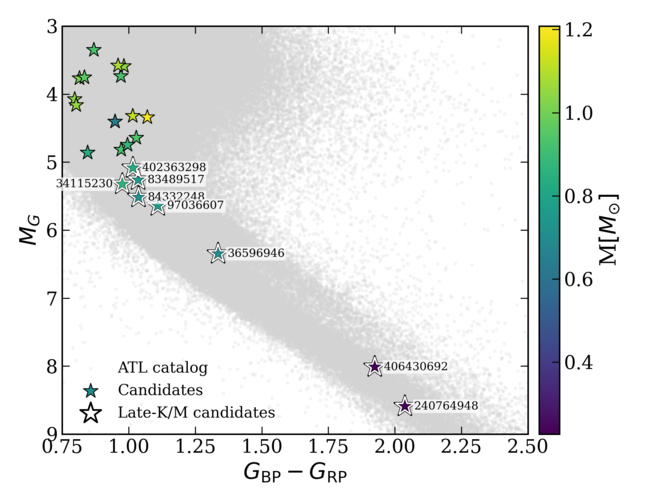}
    \caption{Color-Magnitude Diagram of the candidate stars from Figure~\ref{fig:teff-rad-plot}. Stars are color-coded based on their derived masses using observed $\nu_{\text{max}}$ from \texttt{pySYD} and scaling relations from~\cite{hekker2019scalingrelationssolarlikeoscillations}. }
    \label{fig:cmd-mass}
\end{figure}

Figure~\ref{fig:cmd-mass} contains the color-magnitude diagram (CMD) of the candidate stars color coded by their stellar mass derived using the our observed $\nu_{\text{max}}$. We have obtained the radius and temperature of each candidate from the ATL catalog~\citep{hey2024precisetimedomainasteroseismologyrevised}, and scaling relations from~\cite{hekker2019scalingrelationssolarlikeoscillations}. We have refrained from using our derived value of $\Delta\nu$ because low signal-to-noise ratio especially for late K~dwarfs and early M~dwarfs have prevented a reliable measurements of large frequency separation, which is consistent with previous studies~\citep{sayeed2025homogeneouscatalogoscillatingsolartype}. Without knowing $\Delta\nu$, we can only derive one of the three fundamental stellar parameters: mass, radius, and temperature, and have to use previously determined values for the other two. We choose to derive mass and use previously determined values of radius and temperature. The error in each mass measurement is calculated using standard error propagation and 1$\sigma$ deviations of $\nu_{\text{max}}$ derived using \texttt{pySYD}. 
In Figure~\ref{fig:cmd-mass}, the derived stellar masses of the two early M~dwarfs, TIC 240764948 and TIC 406430692, are $0.23\pm 0.03\, \text{M}_{\odot}$ and $0.23\pm 0.04\, \text{M}_{\odot}$. These two masses are the same up to two significant digits because their $\nu_{\text{max}}$ are very close due to their proximity in the CMD. The two main sequence K~dwarfs, TIC 84332248 and TIC 34115230, were found to have masses of $\nu_{\text{max}}$ to be $0.70\pm 0.10\,\text{M}_{\odot}$ and $0.86\pm 0.13\,\text{M}_{\odot}$. TIC 36596946, which lies very close to $\epsilon$ Ind in Figure~\ref{fig:teff-rad-plot}, was found to have a mass of  $0.69^{\pm 0.10\,\text{M}_{\odot}}$, which is consistent with the typical stellar mass for this location in the CMD. Finally, the two slightly evolved K~dwarfs, TIC~83489517 and TIC~402363298 were found to have stellar masses of $0.75\pm 0.11\,\text{M}_{\odot}$ and $0.82\pm 0.12\,\text{M}_{\odot}$, respectively. For an estimate of the derived stellar masses for the rest of the candidates, please refer to the color bar of Figure~\ref{fig:cmd-mass}. The derived stellar masses using observed frequencies of maximum oscillation ($\nu_{\text{max}}$) and position in the color-magnitude diagram are consistent with our hypothesis for the reason of oscillations. Therefore, this provides a consistency check for our candidate stars and make them more interesting for follow-up observations that might be able to confirm their candidacy.

\section{Conclusion} 
\label{sec:conclusion}

In this study, we developed a convolutional autoencoder to identify solar-like oscillations primarily in main-sequence cool dwarf stars using TESS 120~s cadence data. The network consists of two principal components: an autoencoder, which learns and compresses the morphological features of the input signal into a lower-dimensional representation, and a classifier, which assigns a label based on the compressed features. This architecture was originally adapted from \citet{Ricketts_2023}, but was substantially modified for the purposes of this work. In particular, we deepened the network, incorporated attention mechanisms, implemented pooling layers for dimensionality reduction within the encoder, and employed bicubic interpolation for upsampling in the decoder. Furthermore, we used periodograms binned in logarithmic frequency space as input to the network, as this representation highlights the characteristic morphology of solar-like oscillations—combining the low-frequency granulation slope with the Gaussian-like power excess associated with the oscillation envelope.

We trained the network on a catalog of known solar like oscillator from \cite{Hatt_2023}. Other types of variable and non variable stars were taken from \cite{balona2023identificationclassificationtessvariable} to make the neural network robust against a diverse class of variable signals. We found that our neural network achieves a validation accuracy of 99.8\%. 

We used the trained network to find solar-like oscillation signals in new targets from the Asteroseismic Target List \citep{hey2024precisetimedomainasteroseismologyrevised}. We found 3463 stars that were flagged as solar-like oscillator by the neural network with a probability higher than 0.5. We next used the \texttt{pySYD} package to calculate $\nu_{\text{max}}$, background corrected power spectrum, autocorrelation function and \'{e}chelle diagram. We shortlisted our candidate list based on (1) consistency of observed $\nu_{\text{max}}$ with prediction from scaling relation, (2) periodic autocorrelation function, (3) no contamination from light of background stars, and (4) vertical ridges in \'{e}chelle diagram. 
After applying these four cuts, we were left with 24 stars that we propose as candidates to exhibit solar like oscillation. A complete list of these candidates are available in Table~\ref{tab:candidate-table} and shown in Figure~\ref{fig:teff-rad-plot}. Given the low SNR of these candidates, we need follow-up observations to confirm the existence of solar-like oscillations.

Among the stars analyzed in this study, TIC~240764948 and TIC~406430692 are M-dwarf candidates that satisfy all detection criteria for solar-like oscillations. Both objects lie slightly above the main sequence in the radius–effective temperature plane, which could result from the existence of unresolved equal-mass binary companions in these systems. TIC~240764948 has a nearby companion with nearly identical effective temperature, proper motion, and parallax, indicating that it is likely part of an M-dwarf binary system. Two additional candidates TIC~84332248, and TIC~34115230—are K-dwarf main-sequence stars positioned near previously identified solar-like oscillators in the radius-temperature plot. TIC~95565042, a late main-sequence G-dwarf, is of particular interest because only two TESS detections of solar-like oscillations have been reported at lower temperatures. Finally, TIC~83489517 and TIC~402363298 appear to be slightly evolved K-dwarfs that have begun ascending the subgiant branch. These two objects are especially intriguing, as no prior TESS asteroseismic detections have been reported in this region of the color–magnitude diagram.

In summary, this work shows the power of deep-learning models to identify signals in light-curves that might be missed by visual inspection only. It has the potential of extending the detection limits of TESS 2 min cadence observations, given that we confirm the candidates in follow-up studies, by identifying solar-like oscillator candidates that were previously accessible only through substantially more resource-intensive techniques such as radial velocity measurements. Several of the detected candidates occupy regions of the color–magnitude diagram where TESS asteroseismic detections have so far been rare, underscoring the significance of this sample. The candidate list presented here provides valuable targets for future radial velocity and higher cadence photometry follow-up observations, which are better suited to detect high-frequency, low-amplitude oscillations that may remain unresolved in TESS 2 min data.

\begin{acknowledgments}
This research was partially funded through the Caltech-JPL President’s and Director’s Research \& Development Fund Program. This paper includes data collected by the TESS mission. Funding for the TESS mission is provided by the NASA's Science Mission Directorate. This paper includes data collected with the TESS mission, obtained from the MAST data archive at the Space Telescope Science Institute (STScI)~\citep{shaw2025impactmastdataarchive}. Funding for the TESS mission is provided by the NASA Explorer Program. STScI is operated by the Association of Universities for Research in Astronomy, Inc., under NASA contract NAS 5–26555. I also sincerely acknowledge May G. Pedersen's comment and feedback on our analysis and candidate selection. We also sincerely acknowledge Greg Hallinan's support towards this study during the summer of 2024, and William Chaplin and Sarbani Basu's support in devising this project.

\end{acknowledgments}

\begin{contribution}

W. K. primarily led the writing, analysis, and interpretation of the paper, and developed the new software tools used in this study. R. K. led the supervision of this project. D. B. provided regular feedback and helped the interpretation of the results. C. G. and J. W. helped with the development of the software tools. J. F. provided guidance and feedback during SURF program. S. N. helped with the comparison between different light curve analysis techniques. 

\end{contribution}


\software{
Astropy \citep{2013A&A...558A..33A,2018AJ....156..123A,2022ApJ...935..167A},
Lightkurve \citep{2018ascl.soft12013L},
NumPy \citep{2020Natur.585..357H},
Pandas \citep{reback2020pandas},
Matplotlib \citep{2007CSE.....9...90H},
SciPy \citep{2020NatMe..17..261V},
pySYD \citep{michelucci2022introductionautoencoders},
\texttt{echelle}~\citep{daniel_hey_2020_3629933},
\texttt{tess\_photometry} (M. G. Pedersen in prep),
PyTorch \citep{paszke2019pytorchimperativestylehighperformance}
}

\bibliographystyle{aasjournalv7}
\bibliography{bibliography}{}

@ARTICLE{2022ApJ...935..167A,
       author = {{Astropy Collaboration} and {Price-Whelan}, Adrian M. and {Lim}, Pey Lian and {Earl}, Nicholas and {Starkman}, Nathaniel and {Bradley}, Larry and {Shupe}, David L. and {Patil}, Aarya A. and {Corrales}, Lia and {Brasseur}, C.~E. and {N{\"o}the}, Maximilian and {Donath}, Axel and {Tollerud}, Erik and {Morris}, Brett M. and {Ginsburg}, Adam and {Vaher}, Eero and {Weaver}, Benjamin A. and {Tocknell}, James and {Jamieson}, William and {van Kerkwijk}, Marten H. and {Robitaille}, Thomas P. and {Merry}, Bruce and {Bachetti}, Matteo and {G{\"u}nther}, H. Moritz and {Aldcroft}, Thomas L. and {Alvarado-Montes}, Jaime A. and {Archibald}, Anne M. and {B{\'o}di}, Attila and {Bapat}, Shreyas and {Barentsen}, Geert and {Baz{\'a}n}, Juanjo and {Biswas}, Manish and {Boquien}, M{\'e}d{\'e}ric and {Burke}, D.~J. and {Cara}, Daria and {Cara}, Mihai and {Conroy}, Kyle E. and {Conseil}, Simon and {Craig}, Matthew W. and {Cross}, Robert M. and {Cruz}, Kelle L. and {D'Eugenio}, Francesco and {Dencheva}, Nadia and {Devillepoix}, Hadrien A.~R. and {Dietrich}, J{\"o}rg P. and {Eigenbrot}, Arthur Davis and {Erben}, Thomas and {Ferreira}, Leonardo and {Foreman-Mackey}, Daniel and {Fox}, Ryan and {Freij}, Nabil and {Garg}, Suyog and {Geda}, Robel and {Glattly}, Lauren and {Gondhalekar}, Yash and {Gordon}, Karl D. and {Grant}, David and {Greenfield}, Perry and {Groener}, Austen M. and {Guest}, Steve and {Gurovich}, Sebastian and {Handberg}, Rasmus and {Hart}, Akeem and {Hatfield-Dodds}, Zac and {Homeier}, Derek and {Hosseinzadeh}, Griffin and {Jenness}, Tim and {Jones}, Craig K. and {Joseph}, Prajwel and {Kalmbach}, J. Bryce and {Karamehmetoglu}, Emir and {Ka{\l}uszy{\'n}ski}, Miko{\l}aj and {Kelley}, Michael S.~P. and {Kern}, Nicholas and {Kerzendorf}, Wolfgang E. and {Koch}, Eric W. and {Kulumani}, Shankar and {Lee}, Antony and {Ly}, Chun and {Ma}, Zhiyuan and {MacBride}, Conor and {Maljaars}, Jakob M. and {Muna}, Demitri and {Murphy}, N.~A. and {Norman}, Henrik and {O'Steen}, Richard and {Oman}, Kyle A. and {Pacifici}, Camilla and {Pascual}, Sergio and {Pascual-Granado}, J. and {Patil}, Rohit R. and {Perren}, Gabriel I. and {Pickering}, Timothy E. and {Rastogi}, Tanuj and {Roulston}, Benjamin R. and {Ryan}, Daniel F. and {Rykoff}, Eli S. and {Sabater}, Jose and {Sakurikar}, Parikshit and {Salgado}, Jes{\'u}s and {Sanghi}, Aniket and {Saunders}, Nicholas and {Savchenko}, Volodymyr and {Schwardt}, Ludwig and {Seifert-Eckert}, Michael and {Shih}, Albert Y. and {Jain}, Anany Shrey and {Shukla}, Gyanendra and {Sick}, Jonathan and {Simpson}, Chris and {Singanamalla}, Sudheesh and {Singer}, Leo P. and {Singhal}, Jaladh and {Sinha}, Manodeep and {Sip{\H{o}}cz}, Brigitta M. and {Spitler}, Lee R. and {Stansby}, David and {Streicher}, Ole and {{\v{S}}umak}, Jani and {Swinbank}, John D. and {Taranu}, Dan S. and {Tewary}, Nikita and {Tremblay}, Grant R. and {de Val-Borro}, Miguel and {Van Kooten}, Samuel J. and {Vasovi{\'c}}, Zlatan and {Verma}, Shresth and {de Miranda Cardoso}, Jos{\'e} Vin{\'\i}cius and {Williams}, Peter K.~G. and {Wilson}, Tom J. and {Winkel}, Benjamin and {Wood-Vasey}, W.~M. and {Xue}, Rui and {Yoachim}, Peter and {Zhang}, Chen and {Zonca}, Andrea and {Astropy Project Contributors}},
        title = "{The Astropy Project: Sustaining and Growing a Community-oriented Open-source Project and the Latest Major Release (v5.0) of the Core Package}",
      journal = {\apj},
         year = 2022,
        month = aug,
       volume = {935},
       number = {2},
          eid = {167},
        pages = {167},
          doi = {10.3847/1538-4357/ac7c74},
archivePrefix = {arXiv},
       eprint = {2206.14220},
 primaryClass = {astro-ph.IM},
       adsurl = {https://ui.adsabs.harvard.edu/abs/2022ApJ...935..167A}
}

@ARTICLE{2018AJ....156..123A,
       author = {{Astropy Collaboration} and {Price-Whelan}, A.~M. and {Sip{\H{o}}cz}, B.~M. and {G{\"u}nther}, H.~M. and {Lim}, P.~L. and {Crawford}, S.~M. and {Conseil}, S. and {Shupe}, D.~L. and {Craig}, M.~W. and {Dencheva}, N. and {Ginsburg}, A. and {VanderPlas}, J.~T. and {Bradley}, L.~D. and {P{\'e}rez-Su{\'a}rez}, D. and {de Val-Borro}, M. and {Aldcroft}, T.~L. and {Cruz}, K.~L. and {Robitaille}, T.~P. and {Tollerud}, E.~J. and {Ardelean}, C. and {Babej}, T. and {Bach}, Y.~P. and {Bachetti}, M. and {Bakanov}, A.~V. and {Bamford}, S.~P. and {Barentsen}, G. and {Barmby}, P. and {Baumbach}, A. and {Berry}, K.~L. and {Biscani}, F. and {Boquien}, M. and {Bostroem}, K.~A. and {Bouma}, L.~G. and {Brammer}, G.~B. and {Bray}, E.~M. and {Breytenbach}, H. and {Buddelmeijer}, H. and {Burke}, D.~J. and {Calderone}, G. and {Cano Rodr{\'\i}guez}, J.~L. and {Cara}, M. and {Cardoso}, J.~V.~M. and {Cheedella}, S. and {Copin}, Y. and {Corrales}, L. and {Crichton}, D. and {D'Avella}, D. and {Deil}, C. and {Depagne}, {\'E}. and {Dietrich}, J.~P. and {Donath}, A. and {Droettboom}, M. and {Earl}, N. and {Erben}, T. and {Fabbro}, S. and {Ferreira}, L.~A. and {Finethy}, T. and {Fox}, R.~T. and {Garrison}, L.~H. and {Gibbons}, S.~L.~J. and {Goldstein}, D.~A. and {Gommers}, R. and {Greco}, J.~P. and {Greenfield}, P. and {Groener}, A.~M. and {Grollier}, F. and {Hagen}, A. and {Hirst}, P. and {Homeier}, D. and {Horton}, A.~J. and {Hosseinzadeh}, G. and {Hu}, L. and {Hunkeler}, J.~S. and {Ivezi{\'c}}, {\v{Z}}. and {Jain}, A. and {Jenness}, T. and {Kanarek}, G. and {Kendrew}, S. and {Kern}, N.~S. and {Kerzendorf}, W.~E. and {Khvalko}, A. and {King}, J. and {Kirkby}, D. and {Kulkarni}, A.~M. and {Kumar}, A. and {Lee}, A. and {Lenz}, D. and {Littlefair}, S.~P. and {Ma}, Z. and {Macleod}, D.~M. and {Mastropietro}, M. and {McCully}, C. and {Montagnac}, S. and {Morris}, B.~M. and {Mueller}, M. and {Mumford}, S.~J. and {Muna}, D. and {Murphy}, N.~A. and {Nelson}, S. and {Nguyen}, G.~H. and {Ninan}, J.~P. and {N{\"o}the}, M. and {Ogaz}, S. and {Oh}, S. and {Parejko}, J.~K. and {Parley}, N. and {Pascual}, S. and {Patil}, R. and {Patil}, A.~A. and {Plunkett}, A.~L. and {Prochaska}, J.~X. and {Rastogi}, T. and {Reddy Janga}, V. and {Sabater}, J. and {Sakurikar}, P. and {Seifert}, M. and {Sherbert}, L.~E. and {Sherwood-Taylor}, H. and {Shih}, A.~Y. and {Sick}, J. and {Silbiger}, M.~T. and {Singanamalla}, S. and {Singer}, L.~P. and {Sladen}, P.~H. and {Sooley}, K.~A. and {Sornarajah}, S. and {Streicher}, O. and {Teuben}, P. and {Thomas}, S.~W. and {Tremblay}, G.~R. and {Turner}, J.~E.~H. and {Terr{\'o}n}, V. and {van Kerkwijk}, M.~H. and {de la Vega}, A. and {Watkins}, L.~L. and {Weaver}, B.~A. and {Whitmore}, J.~B. and {Woillez}, J. and {Zabalza}, V. and {Astropy Contributors}},
        title = "{The Astropy Project: Building an Open-science Project and Status of the v2.0 Core Package}",
      journal = {\aj},
         year = 2018,
        month = sep,
       volume = {156},
       number = {3},
          eid = {123},
        pages = {123},
          doi = {10.3847/1538-3881/aabc4f},
archivePrefix = {arXiv},
       eprint = {1801.02634},
 primaryClass = {astro-ph.IM},
       adsurl = {https://ui.adsabs.harvard.edu/abs/2018AJ....156..123A}
}

@ARTICLE{2013A&A...558A..33A,
       author = {{Astropy Collaboration} and {Robitaille}, Thomas P. and
         {Tollerud}, Erik J. and {Greenfield}, Perry and {Droettboom}, Michael and
         {Bray}, Erik and {Aldcroft}, Tom and {Davis}, Matt and
         {Ginsburg}, Adam and {Price-Whelan}, Adrian M. and
         {Kerzendorf}, Wolfgang E. and {Conley}, Alexander and {Crighton}, Neil and
         {Barbary}, Kyle and {Muna}, Demitri and {Ferguson}, Henry and
         {Grollier}, Fr{\'e}d{\'e}ric and {Parikh}, Madhura M. and
         {Nair}, Prasanth H. and {Unther}, Hans M. and {Deil}, Christoph and
         {Woillez}, Julien and {Conseil}, Simon and {Kramer}, Roban and
         {Turner}, James E.~H. and {Singer}, Leo and {Fox}, Ryan and
         {Weaver}, Benjamin A. and {Zabalza}, Victor and {Edwards}, Zachary I. and
         {Azalee Bostroem}, K. and {Burke}, D.~J. and {Casey}, Andrew R. and
         {Crawford}, Steven M. and {Dencheva}, Nadia and {Ely}, Justin and
         {Jenness}, Tim and {Labrie}, Kathleen and {Lim}, Pey Lian and
         {Pierfederici}, Francesco and {Pontzen}, Andrew and {Ptak}, Andy and
         {Refsdal}, Brian and {Servillat}, Mathieu and {Streicher}, Ole},
        title = "{Astropy: A community Python package for astronomy}",
      journal = {\aap},
         year = "2013",
        month = "Oct",
       volume = {558},
          eid = {A33},
        pages = {A33},
          doi = {10.1051/0004-6361/201322068},
archivePrefix = {arXiv},
       eprint = {1307.6212},
 primaryClass = {astro-ph.IM},
       adsurl = {https://ui.adsabs.harvard.edu/abs/2013A&A...558A..33A}
}

@ARTICLE{2015JATIS...1a4003R,
       author = {{Ricker}, George R. and {Winn}, Joshua N. and {Vanderspek}, Roland and {Latham}, David W. and {Bakos}, G{\'a}sp{\'a}r {\'A}. and {Bean}, Jacob L. and {Berta-Thompson}, Zachory K. and {Brown}, Timothy M. and {Buchhave}, Lars and {Butler}, Nathaniel R. and {Butler}, R. Paul and {Chaplin}, William J. and {Charbonneau}, David and {Christensen-Dalsgaard}, J{\o}rgen and {Clampin}, Mark and {Deming}, Drake and {Doty}, John and {De Lee}, Nathan and {Dressing}, Courtney and {Dunham}, Edward W. and {Endl}, Michael and {Fressin}, Francois and {Ge}, Jian and {Henning}, Thomas and {Holman}, Matthew J. and {Howard}, Andrew W. and {Ida}, Shigeru and {Jenkins}, Jon M. and {Jernigan}, Garrett and {Johnson}, John Asher and {Kaltenegger}, Lisa and {Kawai}, Nobuyuki and {Kjeldsen}, Hans and {Laughlin}, Gregory and {Levine}, Alan M. and {Lin}, Douglas and {Lissauer}, Jack J. and {MacQueen}, Phillip and {Marcy}, Geoffrey and {McCullough}, Peter R. and {Morton}, Timothy D. and {Narita}, Norio and {Paegert}, Martin and {Palle}, Enric and {Pepe}, Francesco and {Pepper}, Joshua and {Quirrenbach}, Andreas and {Rinehart}, Stephen A. and {Sasselov}, Dimitar and {Sato}, Bun'ei and {Seager}, Sara and {Sozzetti}, Alessandro and {Stassun}, Keivan G. and {Sullivan}, Peter and {Szentgyorgyi}, Andrew and {Torres}, Guillermo and {Udry}, Stephane and {Villasenor}, Joel},
        title = "{Transiting Exoplanet Survey Satellite (TESS)}",
      journal = {Journal of Astronomical Telescopes, Instruments, and Systems},
         year = 2015,
        month = jan,
       volume = {1},
          eid = {014003},
        pages = {014003},
          doi = {10.1117/1.JATIS.1.1.014003},
       adsurl = {https://ui.adsabs.harvard.edu/abs/2015JATIS...1a4003R}
}

@misc{balona2023identificationclassificationtessvariable,
      title={Identification and classification of TESS variable stars}, 
      author={Luis A. Balona},
      year={2023},
      eprint={2212.10776},
      archivePrefix={arXiv},
      primaryClass={astro-ph.SR},
      url={https://arxiv.org/abs/2212.10776}, 
}

@ARTICLE{1976Ap&SS..39..447L,
       author = {{Lomb}, N.~R.},
        title = "{Least-Squares Frequency Analysis of Unequally Spaced Data}",
      journal = {\apss},
         year = 1976,
        month = feb,
       volume = {39},
       number = {2},
        pages = {447-462},
          doi = {10.1007/BF00648343},
       adsurl = {https://ui.adsabs.harvard.edu/abs/1976Ap&SS..39..447L}
}

@article{Ricketts_2023,
   title={Mapping the X-ray variability of GRS 1915 + 105 with machine learning},
   volume={523},
   ISSN={1365-2966},
   url={http://dx.doi.org/10.1093/mnras/stad1332},
   DOI={10.1093/mnras/stad1332},
   number={2},
   journal={Monthly Notices of the Royal Astronomical Society},
   publisher={Oxford University Press (OUP)},
   author={Ricketts, Benjamin J and Steiner, James F and Garraffo, Cecilia and Remillard, Ronald A and Huppenkothen, Daniela},
   year={2023},
   month=may, pages={1946–1966} }

@misc{michelucci2022introductionautoencoders,
      title={An Introduction to Autoencoders}, 
      author={Umberto Michelucci},
      year={2022},
      eprint={2201.03898},
      archivePrefix={arXiv},
      primaryClass={cs.LG},
      url={https://arxiv.org/abs/2201.03898}, 
}

@article{Rodr_guez_L_pez_2012,
   title={Pulsations in M dwarf stars},
   volume={419},
   ISSN={1745-3933},
   url={http://dx.doi.org/10.1111/j.1745-3933.2011.01174.x},
   DOI={10.1111/j.1745-3933.2011.01174.x},
   number={1},
   journal={Monthly Notices of the Royal Astronomical Society: Letters},
   publisher={Oxford University Press (OUP)},
   author={Rodríguez-López, C. and MacDonald, J. and Moya, A.},
   year={2012},
   month=jan, pages={L44–L48} }

@ARTICLE{2016A&A...595A...1G,
       author = {{Gaia Collaboration} and {Prusti}, T. and {de Bruijne}, J.~H.~J. and {Brown}, A.~G.~A. and {Vallenari}, A. and {Babusiaux}, C. and {Bailer-Jones}, C.~A.~L. and {Bastian}, U. and {Biermann}, M. and {Evans}, D.~W. and {Eyer}, L. and {Jansen}, F. and {Jordi}, C. and {Klioner}, S.~A. and {Lammers}, U. and {Lindegren}, L. and {Luri}, X. and {Mignard}, F. and {Milligan}, D.~J. and {Panem}, C. and {Poinsignon}, V. and {Pourbaix}, D. and {Randich}, S. and {Sarri}, G. and {Sartoretti}, P. and {Siddiqui}, H.~I. and {Soubiran}, C. and {Valette}, V. and {van Leeuwen}, F. and {Walton}, N.~A. and {Aerts}, C. and {Arenou}, F. and {Cropper}, M. and {Drimmel}, R. and {H{\o}g}, E. and {Katz}, D. and {Lattanzi}, M.~G. and {O'Mullane}, W. and {Grebel}, E.~K. and {Holland}, A.~D. and {Huc}, C. and {Passot}, X. and {Bramante}, L. and {Cacciari}, C. and {Casta{\~n}eda}, J. and {Chaoul}, L. and {Cheek}, N. and {De Angeli}, F. and {Fabricius}, C. and {Guerra}, R. and {Hern{\'a}ndez}, J. and {Jean-Antoine-Piccolo}, A. and {Masana}, E. and {Messineo}, R. and {Mowlavi}, N. and {Nienartowicz}, K. and {Ord{\'o}{\~n}ez-Blanco}, D. and {Panuzzo}, P. and {Portell}, J. and {Richards}, P.~J. and {Riello}, M. and {Seabroke}, G.~M. and {Tanga}, P. and {Th{\'e}venin}, F. and {Torra}, J. and {Els}, S.~G. and {Gracia-Abril}, G. and {Comoretto}, G. and {Garcia-Reinaldos}, M. and {Lock}, T. and {Mercier}, E. and {Altmann}, M. and {Andrae}, R. and {Astraatmadja}, T.~L. and {Bellas-Velidis}, I. and {Benson}, K. and {Berthier}, J. and {Blomme}, R. and {Busso}, G. and {Carry}, B. and {Cellino}, A. and {Clementini}, G. and {Cowell}, S. and {Creevey}, O. and {Cuypers}, J. and {Davidson}, M. and {De Ridder}, J. and {de Torres}, A. and {Delchambre}, L. and {Dell'Oro}, A. and {Ducourant}, C. and {Fr{\'e}mat}, Y. and {Garc{\'\i}a-Torres}, M. and {Gosset}, E. and {Halbwachs}, J. -L. and {Hambly}, N.~C. and {Harrison}, D.~L. and {Hauser}, M. and {Hestroffer}, D. and {Hodgkin}, S.~T. and {Huckle}, H.~E. and {Hutton}, A. and {Jasniewicz}, G. and {Jordan}, S. and {Kontizas}, M. and {Korn}, A.~J. and {Lanzafame}, A.~C. and {Manteiga}, M. and {Moitinho}, A. and {Muinonen}, K. and {Osinde}, J. and {Pancino}, E. and {Pauwels}, T. and {Petit}, J. -M. and {Recio-Blanco}, A. and {Robin}, A.~C. and {Sarro}, L.~M. and {Siopis}, C. and {Smith}, M. and {Smith}, K.~W. and {Sozzetti}, A. and {Thuillot}, W. and {van Reeven}, W. and {Viala}, Y. and {Abbas}, U. and {Abreu Aramburu}, A. and {Accart}, S. and {Aguado}, J.~J. and {Allan}, P.~M. and {Allasia}, W. and {Altavilla}, G. and {{\'A}lvarez}, M.~A. and {Alves}, J. and {Anderson}, R.~I. and {Andrei}, A.~H. and {Anglada Varela}, E. and {Antiche}, E. and {Antoja}, T. and {Ant{\'o}n}, S. and {Arcay}, B. and {Atzei}, A. and {Ayache}, L. and {Bach}, N. and {Baker}, S.~G. and {Balaguer-N{\'u}{\~n}ez}, L. and {Barache}, C. and {Barata}, C. and {Barbier}, A. and {Barblan}, F. and {Baroni}, M. and {Barrado y Navascu{\'e}s}, D. and {Barros}, M. and {Barstow}, M.~A. and {Becciani}, U. and {Bellazzini}, M. and {Bellei}, G. and {Bello Garc{\'\i}a}, A. and {Belokurov}, V. and {Bendjoya}, P. and {Berihuete}, A. and {Bianchi}, L. and {Bienaym{\'e}}, O. and {Billebaud}, F. and {Blagorodnova}, N. and {Blanco-Cuaresma}, S. and {Boch}, T. and {Bombrun}, A. and {Borrachero}, R. and {Bouquillon}, S. and {Bourda}, G. and {Bouy}, H. and {Bragaglia}, A. and {Breddels}, M.~A. and {Brouillet}, N. and {Br{\"u}semeister}, T. and {Bucciarelli}, B. and {Budnik}, F. and {Burgess}, P. and {Burgon}, R. and {Burlacu}, A. and {Busonero}, D. and {Buzzi}, R. and {Caffau}, E. and {Cambras}, J. and {Campbell}, H. and {Cancelliere}, R. and {Cantat-Gaudin}, T. and {Carlucci}, T. and {Carrasco}, J.~M. and {Castellani}, M. and {Charlot}, P. and {Charnas}, J. and {Charvet}, P. and {Chassat}, F. and {Chiavassa}, A. and {Clotet}, M. and {Cocozza}, G. and {Collins}, R.~S. and {Collins}, P. and {Costigan}, G. and {Crifo}, F. and {Cross}, N.~J.~G. and {Crosta}, M. and {Crowley}, C. and {Dafonte}, C. and {Damerdji}, Y. and {Dapergolas}, A. and {David}, P. and {David}, M. and {De Cat}, P. and {de Felice}, F. and {de Laverny}, P. and {De Luise}, F. and {De March}, R. and {de Martino}, D. and {de Souza}, R. and {Debosscher}, J. and {del Pozo}, E. and {Delbo}, M. and {Delgado}, A. and {Delgado}, H.~E. and {di Marco}, F. and {Di Matteo}, P. and {Diakite}, S. and {Distefano}, E. and {Dolding}, C. and {Dos Anjos}, S. and {Drazinos}, P. and {Dur{\'a}n}, J. and {Dzigan}, Y. and {Ecale}, E. and {Edvardsson}, B. and {Enke}, H. and {Erdmann}, M. and {Escolar}, D. and {Espina}, M. and {Evans}, N.~W. and {Eynard Bontemps}, G. and {Fabre}, C. and {Fabrizio}, M. and {Faigler}, S. and {Falc{\~a}o}, A.~J. and {Farr{\`a}s Casas}, M. and {Faye}, F. and {Federici}, L. and {Fedorets}, G. and {Fern{\'a}ndez-Hern{\'a}ndez}, J. and {Fernique}, P. and {Fienga}, A. and {Figueras}, F. and {Filippi}, F. and {Findeisen}, K. and {Fonti}, A. and {Fouesneau}, M. and {Fraile}, E. and {Fraser}, M. and {Fuchs}, J. and {Furnell}, R. and {Gai}, M. and {Galleti}, S. and {Galluccio}, L. and {Garabato}, D. and {Garc{\'\i}a-Sedano}, F. and {Gar{\'e}}, P. and {Garofalo}, A. and {Garralda}, N. and {Gavras}, P. and {Gerssen}, J. and {Geyer}, R. and {Gilmore}, G. and {Girona}, S. and {Giuffrida}, G. and {Gomes}, M. and {Gonz{\'a}lez-Marcos}, A. and {Gonz{\'a}lez-N{\'u}{\~n}ez}, J. and {Gonz{\'a}lez-Vidal}, J.~J. and {Granvik}, M. and {Guerrier}, A. and {Guillout}, P. and {Guiraud}, J. and {G{\'u}rpide}, A. and {Guti{\'e}rrez-S{\'a}nchez}, R. and {Guy}, L.~P. and {Haigron}, R. and {Hatzidimitriou}, D. and {Haywood}, M. and {Heiter}, U. and {Helmi}, A. and {Hobbs}, D. and {Hofmann}, W. and {Holl}, B. and {Holland}, G. and {Hunt}, J.~A.~S. and {Hypki}, A. and {Icardi}, V. and {Irwin}, M. and {Jevardat de Fombelle}, G. and {Jofr{\'e}}, P. and {Jonker}, P.~G. and {Jorissen}, A. and {Julbe}, F. and {Karampelas}, A. and {Kochoska}, A. and {Kohley}, R. and {Kolenberg}, K. and {Kontizas}, E. and {Koposov}, S.~E. and {Kordopatis}, G. and {Koubsky}, P. and {Kowalczyk}, A. and {Krone-Martins}, A. and {Kudryashova}, M. and {Kull}, I. and {Bachchan}, R.~K. and {Lacoste-Seris}, F. and {Lanza}, A.~F. and {Lavigne}, J. -B. and {Le Poncin-Lafitte}, C. and {Lebreton}, Y. and {Lebzelter}, T. and {Leccia}, S. and {Leclerc}, N. and {Lecoeur-Taibi}, I. and {Lemaitre}, V. and {Lenhardt}, H. and {Leroux}, F. and {Liao}, S. and {Licata}, E. and {Lindstr{\o}m}, H.~E.~P. and {Lister}, T.~A. and {Livanou}, E. and {Lobel}, A. and {L{\"o}ffler}, W. and {L{\'o}pez}, M. and {Lopez-Lozano}, A. and {Lorenz}, D. and {Loureiro}, T. and {MacDonald}, I. and {Magalh{\~a}es Fernandes}, T. and {Managau}, S. and {Mann}, R.~G. and {Mantelet}, G. and {Marchal}, O. and {Marchant}, J.~M. and {Marconi}, M. and {Marie}, J. and {Marinoni}, S. and {Marrese}, P.~M. and {Marschalk{\'o}}, G. and {Marshall}, D.~J. and {Mart{\'\i}n-Fleitas}, J.~M. and {Martino}, M. and {Mary}, N. and {Matijevi{\v{c}}}, G. and {Mazeh}, T. and {McMillan}, P.~J. and {Messina}, S. and {Mestre}, A. and {Michalik}, D. and {Millar}, N.~R. and {Miranda}, B.~M.~H. and {Molina}, D. and {Molinaro}, R. and {Molinaro}, M. and {Moln{\'a}r}, L. and {Moniez}, M. and {Montegriffo}, P. and {Monteiro}, D. and {Mor}, R. and {Mora}, A. and {Morbidelli}, R. and {Morel}, T. and {Morgenthaler}, S. and {Morley}, T. and {Morris}, D. and {Mulone}, A.~F. and {Muraveva}, T. and {Musella}, I. and {Narbonne}, J. and {Nelemans}, G. and {Nicastro}, L. and {Noval}, L. and {Ord{\'e}novic}, C. and {Ordieres-Mer{\'e}}, J. and {Osborne}, P. and {Pagani}, C. and {Pagano}, I. and {Pailler}, F. and {Palacin}, H. and {Palaversa}, L. and {Parsons}, P. and {Paulsen}, T. and {Pecoraro}, M. and {Pedrosa}, R. and {Pentik{\"a}inen}, H. and {Pereira}, J. and {Pichon}, B. and {Piersimoni}, A.~M. and {Pineau}, F. -X. and {Plachy}, E. and {Plum}, G. and {Poujoulet}, E. and {Pr{\v{s}}a}, A. and {Pulone}, L. and {Ragaini}, S. and {Rago}, S. and {Rambaux}, N. and {Ramos-Lerate}, M. and {Ranalli}, P. and {Rauw}, G. and {Read}, A. and {Regibo}, S. and {Renk}, F. and {Reyl{\'e}}, C. and {Ribeiro}, R.~A. and {Rimoldini}, L. and {Ripepi}, V. and {Riva}, A. and {Rixon}, G. and {Roelens}, M. and {Romero-G{\'o}mez}, M. and {Rowell}, N. and {Royer}, F. and {Rudolph}, A. and {Ruiz-Dern}, L. and {Sadowski}, G. and {Sagrist{\`a} Sell{\'e}s}, T. and {Sahlmann}, J. and {Salgado}, J. and {Salguero}, E. and {Sarasso}, M. and {Savietto}, H. and {Schnorhk}, A. and {Schultheis}, M. and {Sciacca}, E. and {Segol}, M. and {Segovia}, J.~C. and {Segransan}, D. and {Serpell}, E. and {Shih}, I. -C. and {Smareglia}, R. and {Smart}, R.~L. and {Smith}, C. and {Solano}, E. and {Solitro}, F. and {Sordo}, R. and {Soria Nieto}, S. and {Souchay}, J. and {Spagna}, A. and {Spoto}, F. and {Stampa}, U. and {Steele}, I.~A. and {Steidelm{\"u}ller}, H. and {Stephenson}, C.~A. and {Stoev}, H. and {Suess}, F.~F. and {S{\"u}veges}, M. and {Surdej}, J. and {Szabados}, L. and {Szegedi-Elek}, E. and {Tapiador}, D. and {Taris}, F. and {Tauran}, G. and {Taylor}, M.~B. and {Teixeira}, R. and {Terrett}, D. and {Tingley}, B. and {Trager}, S.~C. and {Turon}, C. and {Ulla}, A. and {Utrilla}, E. and {Valentini}, G. and {van Elteren}, A. and {Van Hemelryck}, E. and {van Leeuwen}, M. and {Varadi}, M. and {Vecchiato}, A. and {Veljanoski}, J. and {Via}, T. and {Vicente}, D. and {Vogt}, S. and {Voss}, H. and {Votruba}, V. and {Voutsinas}, S. and {Walmsley}, G. and {Weiler}, M. and {Weingrill}, K. and {Werner}, D. and {Wevers}, T. and {Whitehead}, G. and {Wyrzykowski}, {\L}. and {Yoldas}, A. and {{\v{Z}}erjal}, M. and {Zucker}, S. and {Zurbach}, C. and {Zwitter}, T. and {Alecu}, A. and {Allen}, M. and {Allende Prieto}, C. and {Amorim}, A. and {Anglada-Escud{\'e}}, G. and {Arsenijevic}, V. and {Azaz}, S. and {Balm}, P. and {Beck}, M. and {Bernstein}, H. -H. and {Bigot}, L. and {Bijaoui}, A. and {Blasco}, C. and {Bonfigli}, M. and {Bono}, G. and {Boudreault}, S. and {Bressan}, A. and {Brown}, S. and {Brunet}, P. -M. and {Bunclark}, P. and {Buonanno}, R. and {Butkevich}, A.~G. and {Carret}, C. and {Carrion}, C. and {Chemin}, L. and {Ch{\'e}reau}, F. and {Corcione}, L. and {Darmigny}, E. and {de Boer}, K.~S. and {de Teodoro}, P. and {de Zeeuw}, P.~T. and {Delle Luche}, C. and {Domingues}, C.~D. and {Dubath}, P. and {Fodor}, F. and {Fr{\'e}zouls}, B. and {Fries}, A. and {Fustes}, D. and {Fyfe}, D. and {Gallardo}, E. and {Gallegos}, J. and {Gardiol}, D. and {Gebran}, M. and {Gomboc}, A. and {G{\'o}mez}, A. and {Grux}, E. and {Gueguen}, A. and {Heyrovsky}, A. and {Hoar}, J. and {Iannicola}, G. and {Isasi Parache}, Y. and {Janotto}, A. -M. and {Joliet}, E. and {Jonckheere}, A. and {Keil}, R. and {Kim}, D. -W. and {Klagyivik}, P. and {Klar}, J. and {Knude}, J. and {Kochukhov}, O. and {Kolka}, I. and {Kos}, J. and {Kutka}, A. and {Lainey}, V. and {LeBouquin}, D. and {Liu}, C. and {Loreggia}, D. and {Makarov}, V.~V. and {Marseille}, M.~G. and {Martayan}, C. and {Martinez-Rubi}, O. and {Massart}, B. and {Meynadier}, F. and {Mignot}, S. and {Munari}, U. and {Nguyen}, A. -T. and {Nordlander}, T. and {Ocvirk}, P. and {O'Flaherty}, K.~S. and {Olias Sanz}, A. and {Ortiz}, P. and {Osorio}, J. and {Oszkiewicz}, D. and {Ouzounis}, A. and {Palmer}, M. and {Park}, P. and {Pasquato}, E. and {Peltzer}, C. and {Peralta}, J. and {P{\'e}turaud}, F. and {Pieniluoma}, T. and {Pigozzi}, E. and {Poels}, J. and {Prat}, G. and {Prod'homme}, T. and {Raison}, F. and {Rebordao}, J.~M. and {Risquez}, D. and {Rocca-Volmerange}, B. and {Rosen}, S. and {Ruiz-Fuertes}, M.~I. and {Russo}, F. and {Sembay}, S. and {Serraller Vizcaino}, I. and {Short}, A. and {Siebert}, A. and {Silva}, H. and {Sinachopoulos}, D. and {Slezak}, E. and {Soffel}, M. and {Sosnowska}, D. and {Strai{\v{z}}ys}, V. and {ter Linden}, M. and {Terrell}, D. and {Theil}, S. and {Tiede}, C. and {Troisi}, L. and {Tsalmantza}, P. and {Tur}, D. and {Vaccari}, M. and {Vachier}, F. and {Valles}, P. and {Van Hamme}, W. and {Veltz}, L. and {Virtanen}, J. and {Wallut}, J. -M. and {Wichmann}, R. and {Wilkinson}, M.~I. and {Ziaeepour}, H. and {Zschocke}, S.},
        title = "{The Gaia mission}",
      journal = {\aap},
         year = 2016,
        month = nov,
       volume = {595},
          eid = {A1},
        pages = {A1},
          doi = {10.1051/0004-6361/201629272},
archivePrefix = {arXiv},
       eprint = {1609.04153},
 primaryClass = {astro-ph.IM},
       adsurl = {https://ui.adsabs.harvard.edu/abs/2016A&A...595A...1G}
}

@ARTICLE{2023A&A...674A...1G,
       author = {{Gaia Collaboration} and {Vallenari}, A. and {Brown}, A.~G.~A. and {Prusti}, T. and {de Bruijne}, J.~H.~J. and {Arenou}, F. and {Babusiaux}, C. and {Biermann}, M. and {Creevey}, O.~L. and {Ducourant}, C. and {Evans}, D.~W. and {Eyer}, L. and {Guerra}, R. and {Hutton}, A. and {Jordi}, C. and {Klioner}, S.~A. and {Lammers}, U.~L. and {Lindegren}, L. and {Luri}, X. and {Mignard}, F. and {Panem}, C. and {Pourbaix}, D. and {Randich}, S. and {Sartoretti}, P. and {Soubiran}, C. and {Tanga}, P. and {Walton}, N.~A. and {Bailer-Jones}, C.~A.~L. and {Bastian}, U. and {Drimmel}, R. and {Jansen}, F. and {Katz}, D. and {Lattanzi}, M.~G. and {van Leeuwen}, F. and {Bakker}, J. and {Cacciari}, C. and {Casta{\~n}eda}, J. and {De Angeli}, F. and {Fabricius}, C. and {Fouesneau}, M. and {Fr{\'e}mat}, Y. and {Galluccio}, L. and {Guerrier}, A. and {Heiter}, U. and {Masana}, E. and {Messineo}, R. and {Mowlavi}, N. and {Nicolas}, C. and {Nienartowicz}, K. and {Pailler}, F. and {Panuzzo}, P. and {Riclet}, F. and {Roux}, W. and {Seabroke}, G.~M. and {Sordo}, R. and {Th{\'e}venin}, F. and {Gracia-Abril}, G. and {Portell}, J. and {Teyssier}, D. and {Altmann}, M. and {Andrae}, R. and {Audard}, M. and {Bellas-Velidis}, I. and {Benson}, K. and {Berthier}, J. and {Blomme}, R. and {Burgess}, P.~W. and {Busonero}, D. and {Busso}, G. and {C{\'a}novas}, H. and {Carry}, B. and {Cellino}, A. and {Cheek}, N. and {Clementini}, G. and {Damerdji}, Y. and {Davidson}, M. and {de Teodoro}, P. and {Nu{\~n}ez Campos}, M. and {Delchambre}, L. and {Dell'Oro}, A. and {Esquej}, P. and {Fern{\'a}ndez-Hern{\'a}ndez}, J. and {Fraile}, E. and {Garabato}, D. and {Garc{\'\i}a-Lario}, P. and {Gosset}, E. and {Haigron}, R. and {Halbwachs}, J. -L. and {Hambly}, N.~C. and {Harrison}, D.~L. and {Hern{\'a}ndez}, J. and {Hestroffer}, D. and {Hodgkin}, S.~T. and {Holl}, B. and {Jan{\ss}en}, K. and {Jevardat de Fombelle}, G. and {Jordan}, S. and {Krone-Martins}, A. and {Lanzafame}, A.~C. and {L{\"o}ffler}, W. and {Marchal}, O. and {Marrese}, P.~M. and {Moitinho}, A. and {Muinonen}, K. and {Osborne}, P. and {Pancino}, E. and {Pauwels}, T. and {Recio-Blanco}, A. and {Reyl{\'e}}, C. and {Riello}, M. and {Rimoldini}, L. and {Roegiers}, T. and {Rybizki}, J. and {Sarro}, L.~M. and {Siopis}, C. and {Smith}, M. and {Sozzetti}, A. and {Utrilla}, E. and {van Leeuwen}, M. and {Abbas}, U. and {{\'A}brah{\'a}m}, P. and {Abreu Aramburu}, A. and {Aerts}, C. and {Aguado}, J.~J. and {Ajaj}, M. and {Aldea-Montero}, F. and {Altavilla}, G. and {{\'A}lvarez}, M.~A. and {Alves}, J. and {Anders}, F. and {Anderson}, R.~I. and {Anglada Varela}, E. and {Antoja}, T. and {Baines}, D. and {Baker}, S.~G. and {Balaguer-N{\'u}{\~n}ez}, L. and {Balbinot}, E. and {Balog}, Z. and {Barache}, C. and {Barbato}, D. and {Barros}, M. and {Barstow}, M.~A. and {Bartolom{\'e}}, S. and {Bassilana}, J. -L. and {Bauchet}, N. and {Becciani}, U. and {Bellazzini}, M. and {Berihuete}, A. and {Bernet}, M. and {Bertone}, S. and {Bianchi}, L. and {Binnenfeld}, A. and {Blanco-Cuaresma}, S. and {Blazere}, A. and {Boch}, T. and {Bombrun}, A. and {Bossini}, D. and {Bouquillon}, S. and {Bragaglia}, A. and {Bramante}, L. and {Breedt}, E. and {Bressan}, A. and {Brouillet}, N. and {Brugaletta}, E. and {Bucciarelli}, B. and {Burlacu}, A. and {Butkevich}, A.~G. and {Buzzi}, R. and {Caffau}, E. and {Cancelliere}, R. and {Cantat-Gaudin}, T. and {Carballo}, R. and {Carlucci}, T. and {Carnerero}, M.~I. and {Carrasco}, J.~M. and {Casamiquela}, L. and {Castellani}, M. and {Castro-Ginard}, A. and {Chaoul}, L. and {Charlot}, P. and {Chemin}, L. and {Chiaramida}, V. and {Chiavassa}, A. and {Chornay}, N. and {Comoretto}, G. and {Contursi}, G. and {Cooper}, W.~J. and {Cornez}, T. and {Cowell}, S. and {Crifo}, F. and {Cropper}, M. and {Crosta}, M. and {Crowley}, C. and {Dafonte}, C. and {Dapergolas}, A. and {David}, M. and {David}, P. and {de Laverny}, P. and {De Luise}, F. and {De March}, R. and {De Ridder}, J. and {de Souza}, R. and {de Torres}, A. and {del Peloso}, E.~F. and {del Pozo}, E. and {Delbo}, M. and {Delgado}, A. and {Delisle}, J. -B. and {Demouchy}, C. and {Dharmawardena}, T.~E. and {Di Matteo}, P. and {Diakite}, S. and {Diener}, C. and {Distefano}, E. and {Dolding}, C. and {Edvardsson}, B. and {Enke}, H. and {Fabre}, C. and {Fabrizio}, M. and {Faigler}, S. and {Fedorets}, G. and {Fernique}, P. and {Fienga}, A. and {Figueras}, F. and {Fournier}, Y. and {Fouron}, C. and {Fragkoudi}, F. and {Gai}, M. and {Garcia-Gutierrez}, A. and {Garcia-Reinaldos}, M. and {Garc{\'\i}a-Torres}, M. and {Garofalo}, A. and {Gavel}, A. and {Gavras}, P. and {Gerlach}, E. and {Geyer}, R. and {Giacobbe}, P. and {Gilmore}, G. and {Girona}, S. and {Giuffrida}, G. and {Gomel}, R. and {Gomez}, A. and {Gonz{\'a}lez-N{\'u}{\~n}ez}, J. and {Gonz{\'a}lez-Santamar{\'\i}a}, I. and {Gonz{\'a}lez-Vidal}, J.~J. and {Granvik}, M. and {Guillout}, P. and {Guiraud}, J. and {Guti{\'e}rrez-S{\'a}nchez}, R. and {Guy}, L.~P. and {Hatzidimitriou}, D. and {Hauser}, M. and {Haywood}, M. and {Helmer}, A. and {Helmi}, A. and {Sarmiento}, M.~H. and {Hidalgo}, S.~L. and {Hilger}, T. and {H{\l}adczuk}, N. and {Hobbs}, D. and {Holland}, G. and {Huckle}, H.~E. and {Jardine}, K. and {Jasniewicz}, G. and {Jean-Antoine Piccolo}, A. and {Jim{\'e}nez-Arranz}, {\'O}. and {Jorissen}, A. and {Juaristi Campillo}, J. and {Julbe}, F. and {Karbevska}, L. and {Kervella}, P. and {Khanna}, S. and {Kontizas}, M. and {Kordopatis}, G. and {Korn}, A.~J. and {K{\'o}sp{\'a}l}, {\'A}. and {Kostrzewa-Rutkowska}, Z. and {Kruszy{\'n}ska}, K. and {Kun}, M. and {Laizeau}, P. and {Lambert}, S. and {Lanza}, A.~F. and {Lasne}, Y. and {Le Campion}, J. -F. and {Lebreton}, Y. and {Lebzelter}, T. and {Leccia}, S. and {Leclerc}, N. and {Lecoeur-Taibi}, I. and {Liao}, S. and {Licata}, E.~L. and {Lindstr{\o}m}, H.~E.~P. and {Lister}, T.~A. and {Livanou}, E. and {Lobel}, A. and {Lorca}, A. and {Loup}, C. and {Madrero Pardo}, P. and {Magdaleno Romeo}, A. and {Managau}, S. and {Mann}, R.~G. and {Manteiga}, M. and {Marchant}, J.~M. and {Marconi}, M. and {Marcos}, J. and {Marcos Santos}, M.~M.~S. and {Mar{\'\i}n Pina}, D. and {Marinoni}, S. and {Marocco}, F. and {Marshall}, D.~J. and {Martin Polo}, L. and {Mart{\'\i}n-Fleitas}, J.~M. and {Marton}, G. and {Mary}, N. and {Masip}, A. and {Massari}, D. and {Mastrobuono-Battisti}, A. and {Mazeh}, T. and {McMillan}, P.~J. and {Messina}, S. and {Michalik}, D. and {Millar}, N.~R. and {Mints}, A. and {Molina}, D. and {Molinaro}, R. and {Moln{\'a}r}, L. and {Monari}, G. and {Mongui{\'o}}, M. and {Montegriffo}, P. and {Montero}, A. and {Mor}, R. and {Mora}, A. and {Morbidelli}, R. and {Morel}, T. and {Morris}, D. and {Muraveva}, T. and {Murphy}, C.~P. and {Musella}, I. and {Nagy}, Z. and {Noval}, L. and {Oca{\~n}a}, F. and {Ogden}, A. and {Ordenovic}, C. and {Osinde}, J.~O. and {Pagani}, C. and {Pagano}, I. and {Palaversa}, L. and {Palicio}, P.~A. and {Pallas-Quintela}, L. and {Panahi}, A. and {Payne-Wardenaar}, S. and {Pe{\~n}alosa Esteller}, X. and {Penttil{\"a}}, A. and {Pichon}, B. and {Piersimoni}, A.~M. and {Pineau}, F. -X. and {Plachy}, E. and {Plum}, G. and {Poggio}, E. and {Pr{\v{s}}a}, A. and {Pulone}, L. and {Racero}, E. and {Ragaini}, S. and {Rainer}, M. and {Raiteri}, C.~M. and {Rambaux}, N. and {Ramos}, P. and {Ramos-Lerate}, M. and {Re Fiorentin}, P. and {Regibo}, S. and {Richards}, P.~J. and {Rios Diaz}, C. and {Ripepi}, V. and {Riva}, A. and {Rix}, H. -W. and {Rixon}, G. and {Robichon}, N. and {Robin}, A.~C. and {Robin}, C. and {Roelens}, M. and {Rogues}, H.~R.~O. and {Rohrbasser}, L. and {Romero-G{\'o}mez}, M. and {Rowell}, N. and {Royer}, F. and {Ruz Mieres}, D. and {Rybicki}, K.~A. and {Sadowski}, G. and {S{\'a}ez N{\'u}{\~n}ez}, A. and {Sagrist{\`a} Sell{\'e}s}, A. and {Sahlmann}, J. and {Salguero}, E. and {Samaras}, N. and {Sanchez Gimenez}, V. and {Sanna}, N. and {Santove{\~n}a}, R. and {Sarasso}, M. and {Schultheis}, M. and {Sciacca}, E. and {Segol}, M. and {Segovia}, J.~C. and {S{\'e}gransan}, D. and {Semeux}, D. and {Shahaf}, S. and {Siddiqui}, H.~I. and {Siebert}, A. and {Siltala}, L. and {Silvelo}, A. and {Slezak}, E. and {Slezak}, I. and {Smart}, R.~L. and {Snaith}, O.~N. and {Solano}, E. and {Solitro}, F. and {Souami}, D. and {Souchay}, J. and {Spagna}, A. and {Spina}, L. and {Spoto}, F. and {Steele}, I.~A. and {Steidelm{\"u}ller}, H. and {Stephenson}, C.~A. and {S{\"u}veges}, M. and {Surdej}, J. and {Szabados}, L. and {Szegedi-Elek}, E. and {Taris}, F. and {Taylor}, M.~B. and {Teixeira}, R. and {Tolomei}, L. and {Tonello}, N. and {Torra}, F. and {Torra}, J. and {Torralba Elipe}, G. and {Trabucchi}, M. and {Tsounis}, A.~T. and {Turon}, C. and {Ulla}, A. and {Unger}, N. and {Vaillant}, M.~V. and {van Dillen}, E. and {van Reeven}, W. and {Vanel}, O. and {Vecchiato}, A. and {Viala}, Y. and {Vicente}, D. and {Voutsinas}, S. and {Weiler}, M. and {Wevers}, T. and {Wyrzykowski}, {\L}. and {Yoldas}, A. and {Yvard}, P. and {Zhao}, H. and {Zorec}, J. and {Zucker}, S. and {Zwitter}, T.},
        title = "{Gaia Data Release 3. Summary of the content and survey properties}",
      journal = {\aap},
         year = 2023,
        month = jun,
       volume = {674},
          eid = {A1},
        pages = {A1},
          doi = {10.1051/0004-6361/202243940},
archivePrefix = {arXiv},
       eprint = {2208.00211},
 primaryClass = {astro-ph.GA},
       adsurl = {https://ui.adsabs.harvard.edu/abs/2023A&A...674A...1G}
}

@MISC{2018ascl.soft12013L,
   author = {{Lightkurve Collaboration} and {Cardoso}, J.~V.~d.~M. and
             {Hedges}, C. and {Gully-Santiago}, M. and {Saunders}, N. and
             {Cody}, A.~M. and {Barclay}, T. and {Hall}, O. and
             {Sagear}, S. and {Turtelboom}, E. and {Zhang}, J. and
             {Tzanidakis}, A. and {Mighell}, K. and {Coughlin}, J. and
             {Bell}, K. and {Berta-Thompson}, Z. and {Williams}, P. and
             {Dotson}, J. and {Barentsen}, G.},
    title = "{Lightkurve: Kepler and TESS time series analysis in Python}",
howpublished = {Astrophysics Source Code Library},
     year = 2018,
    month = dec,
archivePrefix = "ascl",
   eprint = {1812.013},
   adsurl = {http://adsabs.harvard.edu/abs/2018ascl.soft12013L},
}

@ARTICLE{Hatt_2023,
       author = {{Hatt}, Emily and {Nielsen}, Martin B. and {Chaplin}, William J. and {Ball}, Warrick H. and {Davies}, Guy R. and {Bedding}, Timothy R. and {Buzasi}, Derek L. and {Chontos}, Ashley and {Huber}, Daniel and {Kayhan}, Cenk and {Li}, Yaguang and {White}, Timothy R. and {Cheng}, Chen and {Metcalfe}, Travis S. and {Stello}, Dennis},
        title = "{Catalogue of solar-like oscillators observed by TESS in 120-s and 20-s cadence}",
      journal = {\aap},
         year = 2023,
        month = jan,
       volume = {669},
          eid = {A67},
        pages = {A67},
          doi = {10.1051/0004-6361/202244579},
archivePrefix = {arXiv},
       eprint = {2210.09109},
 primaryClass = {astro-ph.SR},
       adsurl = {https://ui.adsabs.harvard.edu/abs/2023A&A...669A..67H}
}

@article{Schofield_2019,
   title={The Asteroseismic Target List for Solar-like Oscillators Observed in 2 minute Cadence with the Transiting Exoplanet Survey Satellite},
   volume={241},
   ISSN={1538-4365},
   url={http://dx.doi.org/10.3847/1538-4365/ab04f5},
   DOI={10.3847/1538-4365/ab04f5},
   number={1},
   journal={The Astrophysical Journal Supplement Series},
   publisher={American Astronomical Society},
   author={Schofield, Mathew and Chaplin, William J. and Huber, Daniel and Campante, Tiago L. and Davies, Guy R. and Miglio, Andrea and Ball, Warrick H. and Appourchaux, Thierry and Basu, Sarbani and Bedding, Timothy R. and Christensen-Dalsgaard, Jørgen and Creevey, Orlagh and García, Rafael A. and Handberg, Rasmus and Kawaler, Steven D. and Kjeldsen, Hans and Latham, David W. and Lund, Mikkel N. and Metcalfe, Travis S. and Ricker, George R. and Serenelli, Aldo and Aguirre, Victor Silva and Stello, Dennis and Vanderspek, Roland},
   year={2019},
   month=mar, pages={12} }

@article{Hon_2018,
   title={Detecting Solar-like Oscillations in Red Giants with Deep Learning},
   volume={859},
   ISSN={1538-4357},
   url={http://dx.doi.org/10.3847/1538-4357/aabfdb},
   DOI={10.3847/1538-4357/aabfdb},
   number={1},
   journal={The Astrophysical Journal},
   publisher={American Astronomical Society},
   author={Hon, Marc and Stello, Dennis and Zinn, Joel C.},
   year={2018},
   month=may, pages={64} }

@article{Koch_2010,
   title={KEPLER MISSION
                    DESIGN, REALIZED PHOTOMETRIC PERFORMANCE, AND EARLY SCIENCE},
   volume={713},
   ISSN={2041-8213},
   url={http://dx.doi.org/10.1088/2041-8205/713/2/L79},
   DOI={10.1088/2041-8205/713/2/l79},
   number={2},
   journal={The Astrophysical Journal},
   publisher={American Astronomical Society},
   author={Koch, David G. and Borucki, William J. and Basri, Gibor and Batalha, Natalie M. and Brown, Timothy M. and Caldwell, Douglas and Christensen-Dalsgaard, Jørgen and Cochran, William D. and DeVore, Edna and Dunham, Edward W. and Gautier, Thomas N. and Geary, John C. and Gilliland, Ronald L. and Gould, Alan and Jenkins, Jon and Kondo, Yoji and Latham, David W. and Lissauer, Jack J. and Marcy, Geoffrey and Monet, David and Sasselov, Dimitar and Boss, Alan and Brownlee, Donald and Caldwell, John and Dupree, Andrea K. and Howell, Steve B. and Kjeldsen, Hans and Meibom, Søren and Morrison, David and Owen, Tobias and Reitsema, Harold and Tarter, Jill and Bryson, Stephen T. and Dotson, Jessie L. and Gazis, Paul and Haas, Michael R. and Kolodziejczak, Jeffrey and Rowe, Jason F. and Van Cleve, Jeffrey E. and Allen, Christopher and Chandrasekaran, Hema and Clarke, Bruce D. and Li, Jie and Quintana, Elisa V. and Tenenbaum, Peter and Twicken, Joseph D. and Wu, Hayley},
   year={2010},
   month=mar, pages={L79–L86} }

@article{kim2024noiseinjection,
  author    = {Kim, Gi Il and Chung, Kyung},
  title     = {Extraction of Features for Time Series Classification Using Noise Injection},
  journal   = {Sensors (Basel)},
  volume    = {24},
  number    = {19},
  pages     = {6402},
  year      = {2024},
  month     = {Oct},
  doi       = {10.3390/s24196402},
  pmid      = {39409442},
  pmcid     = {PMC11478850}
}

@INPROCEEDINGS{2016SPIE.9913E..3EJ,
       author = {{Jenkins}, Jon M. and {Twicken}, Joseph D. and {McCauliff}, Sean and {Campbell}, Jennifer and {Sanderfer}, Dwight and {Lung}, David and {Mansouri-Samani}, Masoud and {Girouard}, Forrest and {Tenenbaum}, Peter and {Klaus}, Todd and {Smith}, Jeffrey C. and {Caldwell}, Douglas A. and {Chacon}, A.~D. and {Henze}, Christopher and {Heiges}, Cory and {Latham}, David W. and {Morgan}, Edward and {Swade}, Daryl and {Rinehart}, Stephen and {Vanderspek}, Roland},
        title = "{The TESS science processing operations center}",
    booktitle = {Software and Cyberinfrastructure for Astronomy IV},
         year = 2016,
       editor = {{Chiozzi}, Gianluca and {Guzman}, Juan C.},
       series = {Society of Photo-Optical Instrumentation Engineers (SPIE) Conference Series},
       volume = {9913},
        month = aug,
          eid = {99133E},
        pages = {99133E},
          doi = {10.1117/12.2233418},
       adsurl = {https://ui.adsabs.harvard.edu/abs/2016SPIE.9913E..3EJ}
}

@ARTICLE{2021FrASS...8..168B,
       author = {{Bassi}, Saksham and {Sharma}, Kaushal and {Gomekar}, Atharva},
        title = "{Classification of Variable Stars Light Curves using Long Short Term Memory Network}",
      journal = {Frontiers in Astronomy and Space Sciences},
         year = 2021,
        month = nov,
       volume = {8},
          eid = {168},
        pages = {168},
          doi = {10.3389/fspas.2021.718139},
       adsurl = {https://ui.adsabs.harvard.edu/abs/2021FrASS...8..168B}
}

@article{Burhanudin_2021,
   title={Light-curve classification with recurrent neural networks for GOTO: dealing with imbalanced data},
   volume={505},
   ISSN={1365-2966},
   url={http://dx.doi.org/10.1093/mnras/stab1545},
   DOI={10.1093/mnras/stab1545},
   number={3},
   journal={Monthly Notices of the Royal Astronomical Society},
   publisher={Oxford University Press (OUP)},
   author={Burhanudin, U F and Maund, J R and Killestein, T and Ackley, K and Dyer, M J and Lyman, J and Ulaczyk, K and Cutter, R and Mong, Y-L and Steeghs, D and Galloway, D K and Dhillon, V and O’Brien, P and Ramsay, G and Noysena, K and Kotak, R and Breton, R P and Nuttall, L and Pallé, E and Pollacco, D and Thrane, E and Awiphan, S and Chote, P and Chrimes, A and Daw, E and Duffy, C and Eyles-Ferris, R and Gompertz, B and Heikkilä, T and Irawati, P and Kennedy, M R and Levan, A and Littlefair, S and Makrygianni, L and Mata-Sánchez, D and Mattila, S and McCormac, J and Mkrtichian, D and Mullaney, J and Sawangwit, U and Stanway, E and Starling, R and Strøm, P and Tooke, S and Wiersema, K},
   year={2021},
   month=may, pages={4345–4361} }

@article{Naul_2017,
   title={A recurrent neural network for classification of unevenly sampled variable stars},
   volume={2},
   ISSN={2397-3366},
   url={http://dx.doi.org/10.1038/s41550-017-0321-z},
   DOI={10.1038/s41550-017-0321-z},
   number={2},
   journal={Nature Astronomy},
   publisher={Springer Science and Business Media LLC},
   author={Naul, Brett and Bloom, Joshua S. and Pérez, Fernando and van der Walt, Stéfan},
   year={2017},
   month=nov, pages={151–155} }

@article{Jamal_2020,
   title={On Neural Architectures for Astronomical Time-series Classification with Application to Variable Stars},
   volume={250},
   ISSN={1538-4365},
   url={http://dx.doi.org/10.3847/1538-4365/aba8ff},
   DOI={10.3847/1538-4365/aba8ff},
   number={2},
   journal={The Astrophysical Journal Supplement Series},
   publisher={American Astronomical Society},
   author={Jamal, Sara and Bloom, Joshua S.},
   year={2020},
   month=oct, pages={30} }

@article{Hinton2006ReducingTD,
  title={Reducing the dimensionality of data with neural networks},
  author={Geoffrey E. Hinton and Ruslan R. Salakhutdinov},
  journal={Science},
  year={2006},
  volume={313},
  pages={504–507},
  doi={10.1126/science.1127647}
}

@article{LeCun1998GradientbasedLA,
  title={Gradient-based learning applied to document recognition},
  author={Yann LeCun and Léon Bottou and Yoshua Bengio and Patrick Haffner},
  journal={Proceedings of the IEEE},
  year={1998},
  volume={86},
  number={11},
  pages={2278--2324},
  doi={10.1109/5.726791}
}

@inproceedings{Vaswani2017AttentionIA,
  title={Attention is All You Need},
  author={Ashish Vaswani and Noam Shazeer and Niki Parmar and Jakob Uszkoreit and Llion Jones and Aidan N. Gomez and Lukasz Kaiser and Illia Polosukhin},
  booktitle={Advances in Neural Information Processing Systems (NeurIPS)},
  year={2017},
  pages={5998--6008}
}

@misc{he2015delvingdeeprectifierssurpassing,
      title={Delving Deep into Rectifiers: Surpassing Human-Level Performance on ImageNet Classification}, 
      author={Kaiming He and Xiangyu Zhang and Shaoqing Ren and Jian Sun},
      year={2015},
      eprint={1502.01852},
      archivePrefix={arXiv},
      primaryClass={cs.CV},
      url={https://arxiv.org/abs/1502.01852}, 
}

@article{kingma2014adam,
  title={Adam: A Method for Stochastic Optimization},
  author={Kingma, Diederik P and Ba, Jimmy},
  journal={arXiv preprint arXiv:1412.6980},
  year={2014}
}

@article{ChristensenDalsgaard1983,
  author    = {J{\o}rgen Christensen-Dalsgaard and S{\o}ren Frandsen},
  title     = {Solar models and oscillations},
  journal   = {Solar Physics},
  year      = {1983},
  volume    = {82},
  pages     = {469--486},
  doi       = {10.1007/BF00872687}
}

@INPROCEEDINGS{2006cosp...36.3749B,
       author = {{Baglin}, A. and {Auvergne}, M. and {Boisnard}, L. and {Lam-Trong}, T. and {Barge}, P. and {Catala}, C. and {Deleuil}, M. and {Michel}, E. and {Weiss}, W.},
        title = "{CoRoT: a high precision photometer for stellar ecolution and exoplanet finding}",
    booktitle = {36th COSPAR Scientific Assembly},
         year = 2006,
       volume = {36},
        month = jan,
        pages = {3749},
       adsurl = {https://ui.adsabs.harvard.edu/abs/2006cosp...36.3749B}
}

@ARTICLE{2014ApJS..214...27M,
       author = {{Metcalfe}, T.~S. and {Creevey}, O.~L. and {Do{\u{g}}an}, G. and {Mathur}, S. and {Xu}, H. and {Bedding}, T.~R. and {Chaplin}, W.~J. and {Christensen-Dalsgaard}, J. and {Karoff}, C. and {Trampedach}, R. and {Benomar}, O. and {Brown}, B.~P. and {Buzasi}, D.~L. and {Campante}, T.~L. and {{\c{C}}elik}, Z. and {Cunha}, M.~S. and {Davies}, G.~R. and {Deheuvels}, S. and {Derekas}, A. and {Di Mauro}, M.~P. and {Garc{\'\i}a}, R.~A. and {Guzik}, J.~A. and {Howe}, R. and {MacGregor}, K.~B. and {Mazumdar}, A. and {Montalb{\'a}n}, J. and {Monteiro}, M.~J.~P.~F.~G. and {Salabert}, D. and {Serenelli}, A. and {Stello}, D. and {Ste\&{\c{s}}acute} and {licki}, M. and {Suran}, M.~D. and {Y{\i}ld{\i}z}, M. and {Aksoy}, C. and {Elsworth}, Y. and {Gruberbauer}, M. and {Guenther}, D.~B. and {Lebreton}, Y. and {Molaverdikhani}, K. and {Pricopi}, D. and {Simoniello}, R. and {White}, T.~R.},
        title = "{Properties of 42 Solar-type Kepler Targets from the Asteroseismic Modeling Portal}",
      journal = {\apjs},
         year = 2014,
        month = oct,
       volume = {214},
       number = {2},
          eid = {27},
        pages = {27},
          doi = {10.1088/0067-0049/214/2/27},
archivePrefix = {arXiv},
       eprint = {1402.3614},
 primaryClass = {astro-ph.SR},
       adsurl = {https://ui.adsabs.harvard.edu/abs/2014ApJS..214...27M}
}

@ARTICLE{2014A&A...569A..21L,
       author = {{Lebreton}, Y. and {Goupil}, M.~J.},
        title = "{Asteroseismology for ``{\`a} la carte'' stellar age-dating and weighing. Age and mass of the CoRoT exoplanet host HD 52265}",
      journal = {\aap},
         year = 2014,
        month = sep,
       volume = {569},
          eid = {A21},
        pages = {A21},
          doi = {10.1051/0004-6361/201423797},
archivePrefix = {arXiv},
       eprint = {1406.0652},
 primaryClass = {astro-ph.SR},
       adsurl = {https://ui.adsabs.harvard.edu/abs/2014A&A...569A..21L}
}

@ARTICLE{2015A&A...580A.141L,
       author = {{Lagarde}, N. and {Miglio}, A. and {Eggenberger}, P. and {Morel}, T. and {Montalb{\'a}n}, J. and {Mosser}, B. and {Rodrigues}, T.~S. and {Girardi}, L. and {Rainer}, M. and {Poretti}, E. and {Barban}, C. and {Hekker}, S. and {Kallinger}, T. and {Valentini}, M. and {Carrier}, F. and {Hareter}, M. and {Mantegazza}, L. and {Elsworth}, Y. and {Michel}, E. and {Baglin}, A.},
        title = "{Models of red giants in the CoRoT asteroseismology fields combining asteroseismic and spectroscopic constraints}",
      journal = {\aap},
         year = 2015,
        month = aug,
       volume = {580},
          eid = {A141},
        pages = {A141},
          doi = {10.1051/0004-6361/201525856},
archivePrefix = {arXiv},
       eprint = {1505.01529},
 primaryClass = {astro-ph.SR},
       adsurl = {https://ui.adsabs.harvard.edu/abs/2015A&A...580A.141L}
}

@article{Y_ld_z_2019,
   title={Fundamental properties of Kepler and CoRoT targets – IV. Masses and radii from frequencies of minimum Δν and their implications},
   volume={489},
   ISSN={1365-2966},
   url={http://dx.doi.org/10.1093/mnras/stz2223},
   DOI={10.1093/mnras/stz2223},
   number={2},
   journal={Monthly Notices of the Royal Astronomical Society},
   publisher={Oxford University Press (OUP)},
   author={Yıldız, M and Çelik Orhan, Z and Kayhan, C},
   year={2019},
   month=aug, pages={1753–1769} }

@ARTICLE{2012ApJ...757...99S,
       author = {{Silva Aguirre}, V. and {Casagrande}, L. and {Basu}, S. and {Campante}, T.~L. and {Chaplin}, W.~J. and {Huber}, D. and {Miglio}, A. and {Serenelli}, A.~M. and {Ballot}, J. and {Bedding}, T.~R. and {Christensen-Dalsgaard}, J. and {Creevey}, O.~L. and {Elsworth}, Y. and {Garc{\'\i}a}, R.~A. and {Gilliland}, R.~L. and {Hekker}, S. and {Kjeldsen}, H. and {Mathur}, S. and {Metcalfe}, T.~S. and {Monteiro}, M.~J.~P.~F.~G. and {Mosser}, B. and {Pinsonneault}, M.~H. and {Stello}, D. and {Weiss}, A. and {Tenenbaum}, P. and {Twicken}, J.~D. and {Uddin}, K.},
        title = "{Verifying Asteroseismically Determined Parameters of Kepler Stars Using Hipparcos Parallaxes: Self-consistent Stellar Properties and Distances}",
      journal = {\apj},
         year = 2012,
        month = sep,
       volume = {757},
       number = {1},
          eid = {99},
        pages = {99},
          doi = {10.1088/0004-637X/757/1/99},
archivePrefix = {arXiv},
       eprint = {1208.6294},
 primaryClass = {astro-ph.SR},
       adsurl = {https://ui.adsabs.harvard.edu/abs/2012ApJ...757...99S}
}

@ARTICLE{2012ApJ...760...32H,
       author = {{Huber}, D. and {Ireland}, M.~J. and {Bedding}, T.~R. and {Brand{\~a}o}, I.~M. and {Piau}, L. and {Maestro}, V. and {White}, T.~R. and {Bruntt}, H. and {Casagrande}, L. and {Molenda-{\.Z}akowicz}, J. and {Silva Aguirre}, V. and {Sousa}, S.~G. and {Barclay}, T. and {Burke}, C.~J. and {Chaplin}, W.~J. and {Christensen-Dalsgaard}, J. and {Cunha}, M.~S. and {De Ridder}, J. and {Farrington}, C.~D. and {Frasca}, A. and {Garc{\'\i}a}, R.~A. and {Gilliland}, R.~L. and {Goldfinger}, P.~J. and {Hekker}, S. and {Kawaler}, S.~D. and {Kjeldsen}, H. and {McAlister}, H.~A. and {Metcalfe}, T.~S. and {Miglio}, A. and {Monteiro}, M.~J.~P.~F.~G. and {Pinsonneault}, M.~H. and {Schaefer}, G.~H. and {Stello}, D. and {Stumpe}, M.~C. and {Sturmann}, J. and {Sturmann}, L. and {ten Brummelaar}, T.~A. and {Thompson}, M.~J. and {Turner}, N. and {Uytterhoeven}, K.},
        title = "{Fundamental Properties of Stars Using Asteroseismology from Kepler and CoRoT and Interferometry from the CHARA Array}",
      journal = {\apj},
         year = 2012,
        month = nov,
       volume = {760},
       number = {1},
          eid = {32},
        pages = {32},
          doi = {10.1088/0004-637X/760/1/32},
archivePrefix = {arXiv},
       eprint = {1210.0012},
 primaryClass = {astro-ph.SR},
       adsurl = {https://ui.adsabs.harvard.edu/abs/2012ApJ...760...32H}
}

@article{Guggenberger2016,
  author    = {Guggenberger, Elisabeth and Hekker, Saskia and Basu, Sarbani and Bellinger, Earl P.},
  title     = {Significantly improving stellar mass and radius estimates: A new reference function for the {$\Delta\nu$} scaling relation},
  journal   = {Monthly Notices of the Royal Astronomical Society},
  year      = {2016},
  volume    = {460},
  number    = {4},
  pages     = {4277--4281},
  doi       = {10.1093/mnras/stw1326},
  eprint    = {1606.01917},
  archivePrefix = {arXiv},
  primaryClass = {astro-ph.SR}
}

@ARTICLE{2016ApJ...832..121G,
       author = {{Gaulme}, P. and {McKeever}, J. and {Jackiewicz}, J. and {Rawls}, M.~L. and {Corsaro}, E. and {Mosser}, B. and {Southworth}, J. and {Mahadevan}, S. and {Bender}, C. and {Deshpande}, R.},
        title = "{Testing the Asteroseismic Scaling Relations for Red Giants with Eclipsing Binaries Observed by Kepler}",
      journal = {\apj},
         year = 2016,
        month = dec,
       volume = {832},
       number = {2},
          eid = {121},
        pages = {121},
          doi = {10.3847/0004-637X/832/2/121},
archivePrefix = {arXiv},
       eprint = {1609.06645},
 primaryClass = {astro-ph.SR},
       adsurl = {https://ui.adsabs.harvard.edu/abs/2016ApJ...832..121G}
}

@ARTICLE{2022A&A...657A..31M,
       author = {{Mathur}, S. and {Garc{\'\i}a}, R.~A. and {Breton}, S. and {Santos}, A.~R.~G. and {Mosser}, B. and {Huber}, D. and {Sayeed}, M. and {Bugnet}, L. and {Chontos}, A.},
        title = "{Detections of solar-like oscillations in dwarfs and subgiants with Kepler DR25 short-cadence data}",
      journal = {\aap},
         year = 2022,
        month = jan,
       volume = {657},
          eid = {A31},
        pages = {A31},
          doi = {10.1051/0004-6361/202141168},
archivePrefix = {arXiv},
       eprint = {2109.14058},
 primaryClass = {astro-ph.SR},
       adsurl = {https://ui.adsabs.harvard.edu/abs/2022A&A...657A..31M}
}

@ARTICLE{2010ApJ...723.1583M,
       author = {{Metcalfe}, T.~S. and {Monteiro}, M.~J.~P.~F.~G. and {Thompson}, M.~J. and {Molenda-{\.Z}akowicz}, J. and {Appourchaux}, T. and {Chaplin}, W.~J. and {Do{\v{g}}an}, G. and {Eggenberger}, P. and {Bedding}, T.~R. and {Bruntt}, H. and {Creevey}, O.~L. and {Quirion}, P. -O. and {Stello}, D. and {Bonanno}, A. and {Silva Aguirre}, V. and {Basu}, S. and {Esch}, L. and {Gai}, N. and {Di Mauro}, M.~P. and {Kosovichev}, A.~G. and {Kitiashvili}, I.~N. and {Su{\'a}rez}, J.~C. and {Moya}, A. and {Piau}, L. and {Garc{\'\i}a}, R.~A. and {Marques}, J.~P. and {Frasca}, A. and {Biazzo}, K. and {Sousa}, S.~G. and {Dreizler}, S. and {Bazot}, M. and {Karoff}, C. and {Frandsen}, S. and {Wilson}, P.~A. and {Brown}, T.~M. and {Christensen-Dalsgaard}, J. and {Gilliland}, R.~L. and {Kjeldsen}, H. and {Campante}, T.~L. and {Fletcher}, S.~T. and {Handberg}, R. and {R{\'e}gulo}, C. and {Salabert}, D. and {Schou}, J. and {Verner}, G.~A. and {Ballot}, J. and {Broomhall}, A. -M. and {Elsworth}, Y. and {Hekker}, S. and {Huber}, D. and {Mathur}, S. and {New}, R. and {Roxburgh}, I.~W. and {Sato}, K.~H. and {White}, T.~R. and {Borucki}, W.~J. and {Koch}, D.~G. and {Jenkins}, J.~M.},
        title = "{A Precise Asteroseismic Age and Radius for the Evolved Sun-like Star KIC 11026764}",
      journal = {\apj},
         year = 2010,
        month = nov,
       volume = {723},
       number = {2},
        pages = {1583-1598},
          doi = {10.1088/0004-637X/723/2/1583},
archivePrefix = {arXiv},
       eprint = {1010.4329},
 primaryClass = {astro-ph.SR},
       adsurl = {https://ui.adsabs.harvard.edu/abs/2010ApJ...723.1583M}
}

@article{Mathur_2019,
   title={Revisiting the Impact of Stellar Magnetic Activity on the Detectability of Solar-Like Oscillations by Kepler},
   volume={6},
   ISSN={2296-987X},
   url={http://dx.doi.org/10.3389/fspas.2019.00046},
   DOI={10.3389/fspas.2019.00046},
   journal={Frontiers in Astronomy and Space Sciences},
   publisher={Frontiers Media SA},
   author={Mathur, Savita and García, Rafael A. and Bugnet, Lisa and Santos, Ângela R.G. and Santiago, Netsha and Beck, Paul G.},
   year={2019},
   month=jul }

@ARTICLE{2012A&A...543A..54A,
       author = {{Appourchaux}, T. and {Chaplin}, W.~J. and {Garc{\'\i}a}, R.~A. and {Gruberbauer}, M. and {Verner}, G.~A. and {Antia}, H.~M. and {Benomar}, O. and {Campante}, T.~L. and {Davies}, G.~R. and {Deheuvels}, S. and {Handberg}, R. and {Hekker}, S. and {Howe}, R. and {R{\'e}gulo}, C. and {Salabert}, D. and {Bedding}, T.~R. and {White}, T.~R. and {Ballot}, J. and {Mathur}, S. and {Silva Aguirre}, V. and {Elsworth}, Y.~P. and {Basu}, S. and {Gilliland}, R.~L. and {Christensen-Dalsgaard}, J. and {Kjeldsen}, H. and {Uddin}, K. and {Stumpe}, M.~C. and {Barclay}, T.},
        title = "{Oscillation mode frequencies of 61 main-sequence and subgiant stars observed by Kepler}",
      journal = {\aap},
         year = 2012,
        month = jul,
       volume = {543},
          eid = {A54},
        pages = {A54},
          doi = {10.1051/0004-6361/201218948},
archivePrefix = {arXiv},
       eprint = {1204.3147},
 primaryClass = {astro-ph.SR},
       adsurl = {https://ui.adsabs.harvard.edu/abs/2012A&A...543A..54A}
}

@ARTICLE{2010ApJ...713L.176B,
       author = {{Bedding}, T.~R. and {Huber}, D. and {Stello}, D. and {Elsworth}, Y.~P. and {Hekker}, S. and {Kallinger}, T. and {Mathur}, S. and {Mosser}, B. and {Preston}, H.~L. and {Ballot}, J. and {Barban}, C. and {Broomhall}, A.~M. and {Buzasi}, D.~L. and {Chaplin}, W.~J. and {Garc{\'\i}a}, R.~A. and {Gruberbauer}, M. and {Hale}, S.~J. and {De Ridder}, J. and {Frandsen}, S. and {Borucki}, W.~J. and {Brown}, T. and {Christensen-Dalsgaard}, J. and {Gilliland}, R.~L. and {Jenkins}, J.~M. and {Kjeldsen}, H. and {Koch}, D. and {Belkacem}, K. and {Bildsten}, L. and {Bruntt}, H. and {Campante}, T.~L. and {Deheuvels}, S. and {Derekas}, A. and {Dupret}, M. -A. and {Goupil}, M. -J. and {Hatzes}, A. and {Houdek}, G. and {Ireland}, M.~J. and {Jiang}, C. and {Karoff}, C. and {Kiss}, L.~L. and {Lebreton}, Y. and {Miglio}, A. and {Montalb{\'a}n}, J. and {Noels}, A. and {Roxburgh}, I.~W. and {Sangaralingam}, V. and {Stevens}, I.~R. and {Suran}, M.~D. and {Tarrant}, N.~J. and {Weiss}, A.},
        title = "{Solar-like Oscillations in Low-luminosity Red Giants: First Results from Kepler}",
      journal = {\apjl},
         year = 2010,
        month = apr,
       volume = {713},
       number = {2},
        pages = {L176-L181},
          doi = {10.1088/2041-8205/713/2/L176},
archivePrefix = {arXiv},
       eprint = {1001.0229},
 primaryClass = {astro-ph.SR},
       adsurl = {https://ui.adsabs.harvard.edu/abs/2010ApJ...713L.176B}
}

@ARTICLE{2018ApJS..236...42Y,
       author = {{Yu}, Jie and {Huber}, Daniel and {Bedding}, Timothy R. and {Stello}, Dennis and {Hon}, Marc and {Murphy}, Simon J. and {Khanna}, Shourya},
        title = "{Asteroseismology of 16,000 Kepler Red Giants: Global Oscillation Parameters, Masses, and Radii}",
      journal = {\apjs},
         year = 2018,
        month = jun,
       volume = {236},
       number = {2},
          eid = {42},
        pages = {42},
          doi = {10.3847/1538-4365/aaaf74},
archivePrefix = {arXiv},
       eprint = {1802.04455},
 primaryClass = {astro-ph.SR},
       adsurl = {https://ui.adsabs.harvard.edu/abs/2018ApJS..236...42Y}
}

@ARTICLE{2009CoAst.160...74H,
       author = {{Huber}, D. and {Stello}, D. and {Bedding}, T.~R. and {Chaplin}, W.~J. and {Arentoft}, T. and {Quirion}, P. -O. and {Kjeldsen}, H.},
        title = "{Automated extraction of oscillation parameters for Kepler observations of solar-type stars}",
      journal = {Communications in Asteroseismology},
         year = 2009,
        month = oct,
       volume = {160},
        pages = {74},
          doi = {10.48550/arXiv.0910.2764},
archivePrefix = {arXiv},
       eprint = {0910.2764},
 primaryClass = {astro-ph.SR},
       adsurl = {https://ui.adsabs.harvard.edu/abs/2009CoAst.160...74H}
}

@ARTICLE{2022JOSS....7.3331C,
       author = {{Chontos}, Ashley and {Huber}, Daniel and {Sayeed}, Maryum and {Yamsiri}, Pavadol},
        title = "{pySYD: Automated measurements of global asteroseismic parameters}",
      journal = {The Journal of Open Source Software},
         year = 2022,
        month = nov,
       volume = {7},
       number = {79},
          eid = {3331},
        pages = {3331},
          doi = {10.21105/joss.03331},
archivePrefix = {arXiv},
       eprint = {2108.00582},
 primaryClass = {astro-ph.SR},
       adsurl = {https://ui.adsabs.harvard.edu/abs/2022JOSS....7.3331C}
}

@ARTICLE{2010A&A...511A..46M,
       author = {{Mathur}, S. and {Garc{\'\i}a}, R.~A. and {R{\'e}gulo}, C. and {Creevey}, O.~L. and {Ballot}, J. and {Salabert}, D. and {Arentoft}, T. and {Quirion}, P. -O. and {Chaplin}, W.~J. and {Kjeldsen}, H.},
        title = "{Determining global parameters of the oscillations of solar-like stars}",
      journal = {\aap},
         year = 2010,
        month = feb,
       volume = {511},
          eid = {A46},
        pages = {A46},
          doi = {10.1051/0004-6361/200913266},
archivePrefix = {arXiv},
       eprint = {0912.3367},
 primaryClass = {astro-ph.SR},
       adsurl = {https://ui.adsabs.harvard.edu/abs/2010A&A...511A..46M}
}

@inproceedings{Mahabal_2017,
   title={Deep-learnt classification of light curves},
   url={http://dx.doi.org/10.1109/SSCI.2017.8280984},
   DOI={10.1109/ssci.2017.8280984},
   booktitle={2017 IEEE Symposium Series on Computational Intelligence (SSCI)},
   publisher={IEEE},
   author={Mahabal, A and Sheth, K and Gieseke, F and Pai, A and Djorgovski, S G and Drake, A J and Graham, M J},
   year={2017},
   month=nov, pages={1–8} }

@article{Aguirre_2018,
   title={Deep multi-survey classification of variable stars},
   volume={482},
   ISSN={1365-2966},
   url={http://dx.doi.org/10.1093/mnras/sty2836},
   DOI={10.1093/mnras/sty2836},
   number={4},
   journal={Monthly Notices of the Royal Astronomical Society},
   publisher={Oxford University Press (OUP)},
   author={Aguirre, C and Pichara, K and Becker, I},
   year={2018},
   month=nov, pages={5078–5092} }

@article{Becker_2020,
   title={Scalable end-to-end recurrent neural network for variable star classification},
   volume={493},
   ISSN={1365-2966},
   url={http://dx.doi.org/10.1093/mnras/staa350},
   DOI={10.1093/mnras/staa350},
   number={2},
   journal={Monthly Notices of the Royal Astronomical Society},
   publisher={Oxford University Press (OUP)},
   author={Becker, I and Pichara, K and Catelan, M and Protopapas, P and Aguirre, C and Nikzat, F},
   year={2020},
   month=feb, pages={2981–2995} }

@misc{hey2024precisetimedomainasteroseismologyrevised,
      title={Precise Time-Domain Asteroseismology and a Revised Target List for TESS Solar-Like Oscillators}, 
      author={Daniel Hey and Daniel Huber and Joel Ong and Dennis Stello and Daniel Foreman-Mackey},
      year={2024},
      eprint={2403.02489},
      archivePrefix={arXiv},
      primaryClass={astro-ph.SR},
      url={https://arxiv.org/abs/2403.02489}, 
}

@ARTICLE{2014A&A...570A..41K,
       author = {{Kallinger}, T. and {De Ridder}, J. and {Hekker}, S. and {Mathur}, S. and {Mosser}, B. and {Gruberbauer}, M. and {Garc{\'\i}a}, R.~A. and {Karoff}, C. and {Ballot}, J.},
        title = "{The connection between stellar granulation and oscillation as seen by the Kepler mission}",
      journal = {\aap},
         year = 2014,
        month = oct,
       volume = {570},
          eid = {A41},
        pages = {A41},
          doi = {10.1051/0004-6361/201424313},
archivePrefix = {arXiv},
       eprint = {1408.0817},
 primaryClass = {astro-ph.SR},
       adsurl = {https://ui.adsabs.harvard.edu/abs/2014A&A...570A..41K}
}

@ARTICLE{Mosser_2011,
       author = {{Mosser}, B. and {Elsworth}, Y. and {Hekker}, S. and {Huber}, D. and {Kallinger}, T. and {Mathur}, S. and {Belkacem}, K. and {Goupil}, M.~J. and {Samadi}, R. and {Barban}, C. and {Bedding}, T.~R. and {Chaplin}, W.~J. and {Garc{\'\i}a}, R.~A. and {Stello}, D. and {De Ridder}, J. and {Middour}, C.~K. and {Morris}, R.~L. and {Quintana}, E.~V.},
        title = "{Characterization of the power excess of solar-like oscillations in red giants with Kepler}",
      journal = {\aap},
         year = 2012,
        month = jan,
       volume = {537},
          eid = {A30},
        pages = {A30},
          doi = {10.1051/0004-6361/20111735210.1086/141952},
archivePrefix = {arXiv},
       eprint = {1110.0980},
 primaryClass = {astro-ph.SR},
       adsurl = {https://ui.adsabs.harvard.edu/abs/2012A&A...537A..30M}
}

@misc{hekker2019scalingrelationssolarlikeoscillations,
      title={Scaling relations for solar-like oscillations: a review}, 
      author={S. Hekker},
      year={2019},
      eprint={1907.10457},
      archivePrefix={arXiv},
      primaryClass={astro-ph.SR},
      url={https://arxiv.org/abs/1907.10457}, 
}

@misc{pérezgalarce2025selfregulatedconvolutionalneuralnetwork,
      title={A self-regulated convolutional neural network for classifying variable stars}, 
      author={Francisco Pérez-Galarce and Jorge Martínez-Palomera and Karim Pichara and Pablo Huijse and Márcio Catelan},
      year={2025},
      eprint={2505.14877},
      archivePrefix={arXiv},
      primaryClass={cs.LG},
      url={https://arxiv.org/abs/2505.14877}, 
}

@ARTICLE{2021MNRAS.505..515Z,
       author = {{Zhang}, Keming and {Bloom}, Joshua S.},
        title = "{Classification of periodic variable stars with novel cyclic-permutation invariant neural networks}",
      journal = {\mnras},
         year = 2021,
        month = jul,
       volume = {505},
       number = {1},
        pages = {515-522},
          doi = {10.1093/mnras/stab1248},
archivePrefix = {arXiv},
       eprint = {2011.01243},
 primaryClass = {astro-ph.IM},
       adsurl = {https://ui.adsabs.harvard.edu/abs/2021MNRAS.505..515Z}
}

@article{abdollahi_torabi_raeisi_rahvar_2022, address={IR},
   title={Hierarchical Classification of Variable Stars Using Deep Convolutional Neural Networks},
   volume={9},
   url={https://doi.org/10.22128/ijaa.2022.603.1131},
   DOI={10.22128/ijaa.2022.603.1131},
   number={1},
   journal={Iranian Journal of Astronomy and Astrophysics},
   publisher={Damghan University Astronomical Society of Iran},
   author={Abdollahi, Mahdi and Torabi, Nooshin and Raeisi, Sadegh and Rahvar, Sohrab},
   year={2022},
   month=Dec }

@ARTICLE{Benk__2016,
       author = {{Benk{\H{o}}}, J.~M. and {Szab{\'o}}, R. and {Derekas}, A. and {S{\'o}dor}, {\'A}.},
        title = "{Finest light curve details, physical parameters, and period fluctuations of CoRoT RR Lyrae stars}",
      journal = {\mnras},
         year = 2016,
        month = dec,
       volume = {463},
       number = {2},
        pages = {1769-1779},
          doi = {10.1093/mnras/stw2136},
archivePrefix = {arXiv},
       eprint = {1608.06418},
 primaryClass = {astro-ph.SR},
       adsurl = {https://ui.adsabs.harvard.edu/abs/2016MNRAS.463.1769B}
}

@ARTICLE{2013MNRAS.430.2986M,
       author = {{Murphy}, Simon J. and {Shibahashi}, Hiromoto and {Kurtz}, Donald W.},
        title = "{Super-Nyquist asteroseismology with the Kepler Space Telescope}",
      journal = {\mnras},
         year = 2013,
        month = apr,
       volume = {430},
       number = {4},
        pages = {2986-2998},
          doi = {10.1093/mnras/stt105},
archivePrefix = {arXiv},
       eprint = {1212.5603},
 primaryClass = {astro-ph.SR},
       adsurl = {https://ui.adsabs.harvard.edu/abs/2013MNRAS.430.2986M}
}

@ARTICLE{Campante_2024,
       author = {{Campante}, T.~L. and {Kjeldsen}, H. and {Li}, Y. and {Lund}, M.~N. and {Silva}, A.~M. and {Corsaro}, E. and {Gomes da Silva}, J. and {Martins}, J.~H.~C. and {Adibekyan}, V. and {Azevedo Silva}, T. and {Bedding}, T.~R. and {Bossini}, D. and {Buzasi}, D.~L. and {Chaplin}, W.~J. and {Costa}, R.~R. and {Cunha}, M.~S. and {Cristo}, E. and {Faria}, J.~P. and {Garc{\'\i}a}, R.~A. and {Huber}, D. and {Lundkvist}, M.~S. and {Metcalfe}, T.~S. and {Monteiro}, M.~J.~P.~F.~G. and {Neitzel}, A.~W. and {Nielsen}, M.~B. and {Poretti}, E. and {Santos}, N.~C. and {Sousa}, S.~G.},
        title = "{Expanding the frontiers of cool-dwarf asteroseismology with ESPRESSO. Detection of solar-like oscillations in the K5 dwarf $\epsilon$ Indi}",
      journal = {\aap},
         year = 2024,
        month = mar,
       volume = {683},
          eid = {L16},
        pages = {L16},
          doi = {10.1051/0004-6361/202449197},
archivePrefix = {arXiv},
       eprint = {2403.16333},
 primaryClass = {astro-ph.SR},
       adsurl = {https://ui.adsabs.harvard.edu/abs/2024A&A...683L..16C}
}

@article{Campante_2017,
   title={On the detectability of solar-like oscillations with the NASA TESS mission},
   volume={160},
   ISSN={2100-014X},
   url={http://dx.doi.org/10.1051/epjconf/201716001006},
   DOI={10.1051/epjconf/201716001006},
   journal={EPJ Web of Conferences},
   publisher={EDP Sciences},
   author={Campante, Tiago L.},
   editor={Monteiro, M.J.P.F.G. and Cunha, M.S. and Ferreira, J.M.T.S.},
   year={2017},
   pages={01006} }

@article{Chaplin_2011,
   title={Ensemble Asteroseismology of Solar-Type Stars with the NASA Kepler Mission},
   volume={332},
   ISSN={1095-9203},
   url={http://dx.doi.org/10.1126/science.1201827},
   DOI={10.1126/science.1201827},
   number={6026},
   journal={Science},
   publisher={American Association for the Advancement of Science (AAAS)},
   author={Chaplin, W. J. and Kjeldsen, H. and Christensen-Dalsgaard, J. and Basu, S. and Miglio, A. and Appourchaux, T. and Bedding, T. R. and Elsworth, Y. and García, R. A. and Gilliland, R. L. and Girardi, L. and Houdek, G. and Karoff, C. and Kawaler, S. D. and Metcalfe, T. S. and Molenda-Żakowicz, J. and Monteiro, M. J. P. F. G. and Thompson, M. J. and Verner, G. A. and Ballot, J. and Bonanno, A. and Brandão, I. M. and Broomhall, A.-M. and Bruntt, H. and Campante, T. L. and Corsaro, E. and Creevey, O. L. and Doğan, G. and Esch, L. and Gai, N. and Gaulme, P. and Hale, S. J. and Handberg, R. and Hekker, S. and Huber, D. and Jiménez, A. and Mathur, S. and Mazumdar, A. and Mosser, B. and New, R. and Pinsonneault, M. H. and Pricopi, D. and Quirion, P.-O. and Régulo, C. and Salabert, D. and Serenelli, A. M. and Aguirre, V. Silva and Sousa, S. G. and Stello, D. and Stevens, I. R. and Suran, M. D. and Uytterhoeven, K. and White, T. R. and Borucki, W. J. and Brown, T. M. and Jenkins, J. M. and Kinemuchi, K. and Van Cleve, J. and Klaus, T. C.},
   year={2011},
   month=apr, pages={213–216} }

@article{Huber_2011,
   title={TESTING SCALING RELATIONS FOR SOLAR-LIKE OSCILLATIONS FROM THE MAIN SEQUENCE TO RED GIANTS USINGKEPLERDATA},
   volume={743},
   ISSN={1538-4357},
   url={http://dx.doi.org/10.1088/0004-637X/743/2/143},
   DOI={10.1088/0004-637x/743/2/143},
   number={2},
   journal={The Astrophysical Journal},
   publisher={American Astronomical Society},
   author={Huber, D. and Bedding, T. R. and Stello, D. and Hekker, S. and Mathur, S. and Mosser, B. and Verner, G. A. and Bonanno, A. and Buzasi, D. L. and Campante, T. L. and Elsworth, Y. P. and Hale, S. J. and Kallinger, T. and Silva Aguirre, V. and Chaplin, W. J. and De Ridder, J. and García, R. A. and Appourchaux, T. and Frandsen, S. and Houdek, G. and Molenda-Żakowicz, J. and Monteiro, M. J. P. F. G. and Christensen-Dalsgaard, J. and Gilliland, R. L. and Kawaler, S. D. and Kjeldsen, H. and Broomhall, A. M. and Corsaro, E. and Salabert, D. and Sanderfer, D. T. and Seader, S. E. and Smith, J. C.},
   year={2011},
   month=dec, pages={143} }

@misc{sayeed2025homogeneouscatalogoscillatingsolartype,
      title={A Homogeneous Catalog of Oscillating Solar-Type Stars Observed by the Kepler Mission and a New Amplitude Scaling Relation Including Chromospheric Activity}, 
      author={Maryum Sayeed and Daniel Huber and Ashley Chontos and Yaguang Li},
      year={2025},
      eprint={2503.15599},
      archivePrefix={arXiv},
      primaryClass={astro-ph.SR},
      url={https://arxiv.org/abs/2503.15599}, 
}

@ARTICLE{2014ApJS..210....1C,
       author = {{Chaplin}, W.~J. and {Basu}, S. and {Huber}, D. and {Serenelli}, A. and {Casagrande}, L. and {Silva Aguirre}, V. and {Ball}, W.~H. and {Creevey}, O.~L. and {Gizon}, L. and {Handberg}, R. and {Karoff}, C. and {Lutz}, R. and {Marques}, J.~P. and {Miglio}, A. and {Stello}, D. and {Suran}, M.~D. and {Pricopi}, D. and {Metcalfe}, T.~S. and {Monteiro}, M.~J.~P.~F.~G. and {Molenda-{\.Z}akowicz}, J. and {Appourchaux}, T. and {Christensen-Dalsgaard}, J. and {Elsworth}, Y. and {Garc{\'\i}a}, R.~A. and {Houdek}, G. and {Kjeldsen}, H. and {Bonanno}, A. and {Campante}, T.~L. and {Corsaro}, E. and {Gaulme}, P. and {Hekker}, S. and {Mathur}, S. and {Mosser}, B. and {R{\'e}gulo}, C. and {Salabert}, D.},
        title = "{Asteroseismic Fundamental Properties of Solar-type Stars Observed by the NASA Kepler Mission}",
      journal = {\apjs},
         year = 2014,
        month = jan,
       volume = {210},
       number = {1},
          eid = {1},
        pages = {1},
          doi = {10.1088/0067-0049/210/1/1},
archivePrefix = {arXiv},
       eprint = {1310.4001},
 primaryClass = {astro-ph.SR},
       adsurl = {https://ui.adsabs.harvard.edu/abs/2014ApJS..210....1C}
}

@software{daniel_hey_2020_3629933,
author       = {Daniel Hey and
                Warrick Ball},
title        = {{Echelle: Dynamic echelle diagrams for
                asteroseismology}},
month        = jan,
year         = 2020,
publisher    = {Zenodo},
version      = {1.4},
doi          = {10.5281/zenodo.3629933},
url          = {https://doi.org/10.5281/zenodo.3629933}
}

@ARTICLE{2017ApJS..229...30M,
       author = {{Mathur}, Savita and {Huber}, Daniel and {Batalha}, Natalie M. and {Ciardi}, David R. and {Bastien}, Fabienne A. and {Bieryla}, Allyson and {Buchhave}, Lars A. and {Cochran}, William D. and {Endl}, Michael and {Esquerdo}, Gilbert A. and {Furlan}, Elise and {Howard}, Andrew and {Howell}, Steve B. and {Isaacson}, Howard and {Latham}, David W. and {MacQueen}, Phillip J. and {Silva}, David R.},
        title = "{Revised Stellar Properties of Kepler Targets for the Q1-17 (DR25) Transit Detection Run}",
      journal = {\apjs},
         year = 2017,
        month = apr,
       volume = {229},
       number = {2},
          eid = {30},
        pages = {30},
          doi = {10.3847/1538-4365/229/2/30},
archivePrefix = {arXiv},
       eprint = {1609.04128},
 primaryClass = {astro-ph.SR},
       adsurl = {https://ui.adsabs.harvard.edu/abs/2017ApJS..229...30M}
}

@ARTICLE{2008ApJ...687.1180A,
       author = {{Arentoft}, Torben and {Kjeldsen}, Hans and {Bedding}, Timothy R. and {Bazot}, Micha{\"e}l and {Christensen-Dalsgaard}, J{\o}rgen and {Dall}, Thomas H. and {Karoff}, Christoffer and {Carrier}, Fabien and {Eggenberger}, Patrick and {Sosnowska}, Danuta and {Wittenmyer}, Robert A. and {Endl}, Michael and {Metcalfe}, Travis S. and {Hekker}, Saskia and {Reffert}, Sabine and {Butler}, R. Paul and {Bruntt}, Hans and {Kiss}, L{\'a}szl{\'o} L. and {O'Toole}, Simon J. and {Kambe}, Eiji and {Ando}, Hiroyasu and {Izumiura}, Hideyuki and {Sato}, Bun'ei and {Hartmann}, Michael and {Hatzes}, Artie and {Bouchy}, Francois and {Mosser}, Benoit and {Appourchaux}, Thierry and {Barban}, Caroline and {Berthomieu}, Gabrielle and {Garcia}, Rafael A. and {Michel}, Eric and {Provost}, Janine and {Turck-Chi{\`e}ze}, Sylvaine and {Marti{\'c}}, Milena and {Lebrun}, Jean-Claude and {Schmitt}, Jerome and {Bertaux}, Jean-Loup and {Bonanno}, Alfio and {Benatti}, Serena and {Claudi}, Riccardo U. and {Cosentino}, Rosario and {Leccia}, Silvio and {Frandsen}, S{\o}ren and {Brogaard}, Karsten and {Glowienka}, Lars and {Grundahl}, Frank and {Stempels}, Eric},
        title = "{A Multisite Campaign to Measure Solar-like Oscillations in Procyon. I. Observations, Data Reduction, and Slow Variations}",
      journal = {\apj},
         year = 2008,
        month = nov,
       volume = {687},
       number = {2},
        pages = {1180-1190},
          doi = {10.1086/592040},
archivePrefix = {arXiv},
       eprint = {0807.3794},
 primaryClass = {astro-ph},
       adsurl = {https://ui.adsabs.harvard.edu/abs/2008ApJ...687.1180A}
}

@ARTICLE{2008ApJ...682.1370K,
       author = {{Kjeldsen}, Hans and {Bedding}, Timothy R. and {Arentoft}, Torben and {Butler}, R. Paul and {Dall}, Thomas H. and {Karoff}, Christoffer and {Kiss}, L{\'a}szl{\'o} L. and {Tinney}, C.~G. and {Chaplin}, William J.},
        title = "{The Amplitude of Solar Oscillations Using Stellar Techniques}",
      journal = {\apj},
         year = 2008,
        month = aug,
       volume = {682},
       number = {2},
        pages = {1370-1375},
          doi = {10.1086/589142},
archivePrefix = {arXiv},
       eprint = {0804.1182},
 primaryClass = {astro-ph},
       adsurl = {https://ui.adsabs.harvard.edu/abs/2008ApJ...682.1370K}
}

@article{Hon_2021,
   title={A “Quick Look” at All-sky Galactic Archeology with TESS: 158,000 Oscillating Red Giants from the MIT Quick-look Pipeline},
   volume={919},
   ISSN={1538-4357},
   url={http://dx.doi.org/10.3847/1538-4357/ac14b1},
   DOI={10.3847/1538-4357/ac14b1},
   number={2},
   journal={The Astrophysical Journal},
   publisher={American Astronomical Society},
   author={Hon, Marc and Huber, Daniel and Kuszlewicz, James S. and Stello, Dennis and Sharma, Sanjib and Tayar, Jamie and Zinn, Joel C. and Vrard, Mathieu and Pinsonneault, Marc H.},
   year={2021},
   month=oct, pages={131} }

@misc{marasco2025asteroseismologymetalpoorredgiants,
      title={Asteroseismology of Metal-Poor Red Giants Observed by TESS}, 
      author={Corin Marasco and Jamie Tayar and David Nidever},
      year={2025},
      eprint={2504.18642},
      archivePrefix={arXiv},
      primaryClass={astro-ph.SR},
      url={https://arxiv.org/abs/2504.18642}, 
}

@ARTICLE{2010ApJ...723.1607H,
       author = {{Huber}, D. and {Bedding}, T.~R. and {Stello}, D. and {Mosser}, B. and {Mathur}, S. and {Kallinger}, T. and {Hekker}, S. and {Elsworth}, Y.~P. and {Buzasi}, D.~L. and {De Ridder}, J. and {Gilliland}, R.~L. and {Kjeldsen}, H. and {Chaplin}, W.~J. and {Garc{\'\i}a}, R.~A. and {Hale}, S.~J. and {Preston}, H.~L. and {White}, T.~R. and {Borucki}, W.~J. and {Christensen-Dalsgaard}, J. and {Clarke}, B.~D. and {Jenkins}, J.~M. and {Koch}, D.},
        title = "{Asteroseismology of Red Giants from the First Four Months of Kepler Data: Global Oscillation Parameters for 800 Stars}",
      journal = {\apj},
         year = 2010,
        month = nov,
       volume = {723},
       number = {2},
        pages = {1607-1617},
          doi = {10.1088/0004-637X/723/2/1607},
archivePrefix = {arXiv},
       eprint = {1010.4566},
 primaryClass = {astro-ph.SR},
       adsurl = {https://ui.adsabs.harvard.edu/abs/2010ApJ...723.1607H}
}

@article{Metcalfe_2012,
   title={ASTEROSEISMOLOGY OF THE SOLAR ANALOGS 16 Cyg A AND B FROM
                    KEPLER
                    OBSERVATIONS},
   volume={748},
   ISSN={2041-8213},
   url={http://dx.doi.org/10.1088/2041-8205/748/1/L10},
   DOI={10.1088/2041-8205/748/1/l10},
   number={1},
   journal={The Astrophysical Journal},
   publisher={American Astronomical Society},
   author={Metcalfe, T. S. and Chaplin, W. J. and Appourchaux, T. and García, R. A. and Basu, S. and Brandão, I. and Creevey, O. L. and Deheuvels, S. and Doğan, G. and Eggenberger, P. and Karoff, C. and Miglio, A. and Stello, D. and Yıldız, M. and Çelik, Z. and Antia, H. M. and Benomar, O. and Howe, R. and Régulo, C. and Salabert, D. and Stahn, T. and Bedding, T. R. and Davies, G. R. and Elsworth, Y. and Gizon, L. and Hekker, S. and Mathur, S. and Mosser, B. and Bryson, S. T. and Still, M. D. and Christensen-Dalsgaard, J. and Gilliland, R. L. and Kawaler, S. D. and Kjeldsen, H. and Ibrahim, K. A. and Klaus, T. C. and Li, J.},
   year={2012},
   month=feb, pages={L10} }

@article{Anders_2016,
   title={Galactic archaeology with asteroseismology and spectroscopy: Red giants observed by CoRoT and APOGEE},
   volume={597},
   ISSN={1432-0746},
   url={http://dx.doi.org/10.1051/0004-6361/201527204},
   DOI={10.1051/0004-6361/201527204},
   journal={Astronomy \&; Astrophysics},
   publisher={EDP Sciences},
   author={Anders, F. and Chiappini, C. and Rodrigues, T. S. and Miglio, A. and Montalbán, J. and Mosser, B. and Girardi, L. and Valentini, M. and Noels, A. and Morel, T. and Johnson, J. A. and Schultheis, M. and Baudin, F. and de Assis Peralta, R. and Hekker, S. and Themeßl, N. and Kallinger, T. and García, R. A. and Mathur, S. and Baglin, A. and Santiago, B. X. and Martig, M. and Minchev, I. and Steinmetz, M. and da Costa, L. N. and Maia, M. A. G. and Allende Prieto, C. and Cunha, K. and Beers, T. C. and Epstein, C. and García Pérez, A. E. and García-Hernández, D. A. and Harding, P. and Holtzman, J. and Majewski, S. R. and Mészáros, Sz. and Nidever, D. and Pan, K. and Pinsonneault, M. and Schiavon, R. P. and Schneider, D. P. and Shetrone, M. D. and Stassun, K. and Zamora, O. and Zasowski, G.},
   year={2016},
   month=dec, pages={A30} }

@article{Valentini_2016,
   title={The CoRoT‐GES Collaboration: Improving red giants spectroscopic surface gravitity and abundances with asteroseismology},
   volume={337},
   ISSN={1521-3994},
   url={http://dx.doi.org/10.1002/asna.201612399},
   DOI={10.1002/asna.201612399},
   number={8–9},
   journal={Astronomische Nachrichten},
   publisher={Wiley},
   author={Valentini, M. and Chiappini, C. and Miglio, A. and Montalbán, J. and Rodrigues, T. and Mosser, B. and Anders, F. and CoRoT RG Group and GES Consortium},
   year={2016},
   month=sep, pages={970–975} }

@article{Rodr_guez_L_pez_2014,
   title={M dwarf search for pulsations within Kepler Guest Observer programme},
   volume={446},
   ISSN={0035-8711},
   url={http://dx.doi.org/10.1093/mnras/stu2211},
   DOI={10.1093/mnras/stu2211},
   number={3},
   journal={Monthly Notices of the Royal Astronomical Society},
   publisher={Oxford University Press (OUP)},
   author={Rodríguez-López, C. and Gizis, J. E. and MacDonald, J. and Amado, P. J. and Carosso, A.},
   year={2014},
   month=nov, pages={2613–2620} }

@ARTICLE{2019FrASS...6...76R,
       author = {{Rodr{\'\i}guez-L{\'o}pez}, Cristina},
        title = "{The quest for pulsating M dwarf stars}",
      journal = {Frontiers in Astronomy and Space Sciences},
         year = 2019,
        month = dec,
       volume = {6},
          eid = {76},
        pages = {76},
          doi = {10.3389/fspas.2019.00076},
       adsurl = {https://ui.adsabs.harvard.edu/abs/2019FrASS...6...76R}
}

@misc{simonyan2015deepconvolutionalnetworkslargescale,
      title={Very Deep Convolutional Networks for Large-Scale Image Recognition}, 
      author={Karen Simonyan and Andrew Zisserman},
      year={2015},
      eprint={1409.1556},
      archivePrefix={arXiv},
      primaryClass={cs.CV},
      url={https://arxiv.org/abs/1409.1556}, 
}

@misc{yu2016multiscalecontextaggregationdilated,
      title={Multi-Scale Context Aggregation by Dilated Convolutions}, 
      author={Fisher Yu and Vladlen Koltun},
      year={2016},
      eprint={1511.07122},
      archivePrefix={arXiv},
      primaryClass={cs.CV},
      url={https://arxiv.org/abs/1511.07122}, 
}

@misc{kinoshita2020fixedsmoothconvolutionallayer,
      title={Fixed smooth convolutional layer for avoiding checkerboard artifacts in CNNs}, 
      author={Yuma Kinoshita and Hitoshi Kiya},
      year={2020},
      eprint={2002.02117},
      archivePrefix={arXiv},
      primaryClass={eess.IV},
      url={https://arxiv.org/abs/2002.02117}, 
}

@ARTICLE{1986Natur.323..533R,
       author = {{Rumelhart}, David E. and {Hinton}, Geoffrey E. and {Williams}, Ronald J.},
        title = "{Learning representations by back-propagating errors}",
      journal = {\nat},
         year = 1986,
        month = oct,
       volume = {323},
       number = {6088},
        pages = {533-536},
          doi = {10.1038/323533a0},
       adsurl = {https://ui.adsabs.harvard.edu/abs/1986Natur.323..533R}
}

@ARTICLE{2020Natur.585..357H,
       author = {{Harris}, Charles R. and {Millman}, K. Jarrod and {van der Walt}, St{\'e}fan J. and {Gommers}, Ralf and {Virtanen}, Pauli and {Cournapeau}, David and {Wieser}, Eric and {Taylor}, Julian and {Berg}, Sebastian and {Smith}, Nathaniel J. and {Kern}, Robert and {Picus}, Matti and {Hoyer}, Stephan and {van Kerkwijk}, Marten H. and {Brett}, Matthew and {Haldane}, Allan and {del R{\'\i}o}, Jaime Fern{\'a}ndez and {Wiebe}, Mark and {Peterson}, Pearu and {G{\'e}rard-Marchant}, Pierre and {Sheppard}, Kevin and {Reddy}, Tyler and {Weckesser}, Warren and {Abbasi}, Hameer and {Gohlke}, Christoph and {Oliphant}, Travis E.},
        title = "{Array programming with NumPy}",
      journal = {\nat},
         year = 2020,
        month = sep,
       volume = {585},
       number = {7825},
        pages = {357-362},
          doi = {10.1038/s41586-020-2649-2},
archivePrefix = {arXiv},
       eprint = {2006.10256},
 primaryClass = {cs.MS},
       adsurl = {https://ui.adsabs.harvard.edu/abs/2020Natur.585..357H}
}

@software{reback2020pandas,
    author       = {The pandas development team},
    title        = {pandas-dev/pandas: Pandas},
    month        = feb,
    year         = 2020,
    publisher    = {Zenodo},
    version      = {latest},
    doi          = {10.5281/zenodo.3509134},
    url          = {https://doi.org/10.5281/zenodo.3509134}
}

@ARTICLE{2007CSE.....9...90H,
       author = {{Hunter}, John D.},
        title = "{Matplotlib: A 2D Graphics Environment}",
      journal = {Computing in Science and Engineering},
         year = 2007,
        month = jan,
       volume = {9},
       number = {3},
        pages = {90-95},
          doi = {10.1109/MCSE.2007.55},
       adsurl = {https://ui.adsabs.harvard.edu/abs/2007CSE.....9...90H}
}

@ARTICLE{2020NatMe..17..261V,
       author = {{Virtanen}, Pauli and {Gommers}, Ralf and {Oliphant}, Travis E. and {Haberland}, Matt and {Reddy}, Tyler and {Cournapeau}, David and {Burovski}, Evgeni and {Peterson}, Pearu and {Weckesser}, Warren and {Bright}, Jonathan and {van der Walt}, St{\'e}fan J. and {Brett}, Matthew and {Wilson}, Joshua and {Millman}, K. Jarrod and {Mayorov}, Nikolay and {Nelson}, Andrew R.~J. and {Jones}, Eric and {Kern}, Robert and {Larson}, Eric and {Carey}, C.~J. and {Polat}, {\.I}lhan and {Feng}, Yu and {Moore}, Eric W. and {VanderPlas}, Jake and {Laxalde}, Denis and {Perktold}, Josef and {Cimrman}, Robert and {Henriksen}, Ian and {Quintero}, E.~A. and {Harris}, Charles R. and {Archibald}, Anne M. and {Ribeiro}, Ant{\^o}nio H. and {Pedregosa}, Fabian and {van Mulbregt}, Paul and {SciPy 1. 0 Contributors}},
        title = "{SciPy 1.0: fundamental algorithms for scientific computing in Python}",
      journal = {Nature Medicine},
         year = 2020,
        month = feb,
       volume = {17},
        pages = {261-272},
          doi = {10.1038/s41592-019-0686-2},
archivePrefix = {arXiv},
       eprint = {1907.10121},
 primaryClass = {cs.MS},
       adsurl = {https://ui.adsabs.harvard.edu/abs/2020NatMe..17..261V}
}

@misc{paszke2019pytorchimperativestylehighperformance,
      title={PyTorch: An Imperative Style, High-Performance Deep Learning Library}, 
      author={Adam Paszke and Sam Gross and Francisco Massa and Adam Lerer and James Bradbury and Gregory Chanan and Trevor Killeen and Zeming Lin and Natalia Gimelshein and Luca Antiga and Alban Desmaison and Andreas Köpf and Edward Yang and Zach DeVito and Martin Raison and Alykhan Tejani and Sasank Chilamkurthy and Benoit Steiner and Lu Fang and Junjie Bai and Soumith Chintala},
      year={2019},
      eprint={1912.01703},
      archivePrefix={arXiv},
      primaryClass={cs.LG},
      url={https://arxiv.org/abs/1912.01703}, 
}

@ARTICLE{2014MNRAS.445..946C,
       author = {{Chaplin}, W.~J. and {Elsworth}, Y. and {Davies}, G.~R. and {Campante}, T.~L. and {Handberg}, R. and {Miglio}, A. and {Basu}, S.},
        title = "{Super-Nyquist asteroseismology of solar-like oscillators with Kepler and K2 - expanding the asteroseismic cohort at the base of the red giant branch}",
      journal = {\mnras},
         year = 2014,
        month = nov,
       volume = {445},
       number = {1},
        pages = {946-954},
          doi = {10.1093/mnras/stu1811},
archivePrefix = {arXiv},
       eprint = {1409.0696},
 primaryClass = {astro-ph.SR},
       adsurl = {https://ui.adsabs.harvard.edu/abs/2014MNRAS.445..946C}
}

@ARTICLE{2006A&A...450..695C,
       author = {{Carrier}, F. and {Eggenberger}, P.},
        title = "{Asteroseismology of the visual binary 70 Ophiuchi}",
      journal = {\aap},
         year = 2006,
        month = may,
       volume = {450},
       number = {2},
        pages = {695-699},
          doi = {10.1051/0004-6361:20054492},
archivePrefix = {arXiv},
       eprint = {astro-ph/0602341},
 primaryClass = {astro-ph},
       adsurl = {https://ui.adsabs.harvard.edu/abs/2006A&A...450..695C}
}

@ARTICLE{2022AJ....163...79H,
       author = {{Huber}, Daniel and {White}, Timothy R. and {Metcalfe}, Travis S. and {Chontos}, Ashley and {Fausnaugh}, Michael M. and {Ho}, Cynthia S.~K. and {Van Eylen}, Vincent and {Ball}, Warrick H. and {Basu}, Sarbani and {Bedding}, Timothy R. and {Benomar}, Othman and {Bossini}, Diego and {Breton}, Sylvain and {Buzasi}, Derek L. and {Campante}, Tiago L. and {Chaplin}, William J. and {Christensen-Dalsgaard}, J{\o}rgen and {Cunha}, Margarida S. and {Deal}, Morgan and {Garc{\'\i}a}, Rafael A. and {Garc{\'\i}a Mu{\~n}oz}, Antonio and {Gehan}, Charlotte and {Gonz{\'a}lez-Cuesta}, Luc{\'\i}a and {Jiang}, Chen and {Kayhan}, Cenk and {Kjeldsen}, Hans and {Lundkvist}, Mia S. and {Mathis}, St{\'e}phane and {Mathur}, Savita and {Monteiro}, M{\'a}rio J.~P.~F.~G. and {Nsamba}, Benard and {Ong}, Jia Mian Joel and {Pak{\v{s}}tien{\.{e}}}, Erika and {Serenelli}, Aldo M. and {Silva Aguirre}, Victor and {Stassun}, Keivan G. and {Stello}, Dennis and {Norgaard Stilling}, Sissel and {Lykke Winther}, Mark and {Wu}, Tao and {Barclay}, Thomas and {Daylan}, Tansu and {G{\"u}nther}, Maximilian N. and {Hermes}, J.~J. and {Jenkins}, Jon M. and {Latham}, David W. and {Levine}, Alan M. and {Ricker}, George R. and {Seager}, Sara and {Shporer}, Avi and {Twicken}, Joseph D. and {Vanderspek}, Roland K. and {Winn}, Joshua N.},
        title = "{A 20 Second Cadence View of Solar-type Stars and Their Planets with TESS: Asteroseismology of Solar Analogs and a Recharacterization of {\ensuremath{\pi}} Men c}",
      journal = {\aj},
         year = 2022,
        month = feb,
       volume = {163},
       number = {2},
          eid = {79},
        pages = {79},
          doi = {10.3847/1538-3881/ac3000},
archivePrefix = {arXiv},
       eprint = {2108.09109},
 primaryClass = {astro-ph.SR},
       adsurl = {https://ui.adsabs.harvard.edu/abs/2022AJ....163...79H}
}

@ARTICLE{2013ApJS..208....9P,
       author = {{Pecaut}, Mark J. and {Mamajek}, Eric E.},
        title = "{Intrinsic Colors, Temperatures, and Bolometric Corrections of Pre-main-sequence Stars}",
      journal = {\apjs},
         year = 2013,
        month = sep,
       volume = {208},
       number = {1},
          eid = {9},
        pages = {9},
          doi = {10.1088/0067-0049/208/1/9},
archivePrefix = {arXiv},
       eprint = {1307.2657},
 primaryClass = {astro-ph.SR},
       adsurl = {https://ui.adsabs.harvard.edu/abs/2013ApJS..208....9P}
}

@ARTICLE{2018AJ....156..102S,
       author = {{Stassun}, Keivan G. and {Oelkers}, Ryan J. and {Pepper}, Joshua and {Paegert}, Martin and {De Lee}, Nathan and {Torres}, Guillermo and {Latham}, David W. and {Charpinet}, St{\'e}phane and {Dressing}, Courtney D. and {Huber}, Daniel and {Kane}, Stephen R. and {L{\'e}pine}, S{\'e}bastien and {Mann}, Andrew and {Muirhead}, Philip S. and {Rojas-Ayala}, B{\'a}rbara and {Silvotti}, Roberto and {Fleming}, Scott W. and {Levine}, Al and {Plavchan}, Peter},
        title = "{The TESS Input Catalog and Candidate Target List}",
      journal = {\aj},
         year = 2018,
        month = sep,
       volume = {156},
       number = {3},
          eid = {102},
        pages = {102},
          doi = {10.3847/1538-3881/aad050},
archivePrefix = {arXiv},
       eprint = {1706.00495},
 primaryClass = {astro-ph.EP},
       adsurl = {https://ui.adsabs.harvard.edu/abs/2018AJ....156..102S}
}

@article{Hatt_2024,
   title={Asteroseismic signatures of core magnetism and rotation in hundreds of low-luminosity red giants},
   volume={534},
   ISSN={1365-2966},
   url={http://dx.doi.org/10.1093/mnras/stae2053},
   DOI={10.1093/mnras/stae2053},
   number={2},
   journal={Monthly Notices of the Royal Astronomical Society},
   publisher={Oxford University Press (OUP)},
   author={Hatt, Emily J and Ong, J M Joel and Nielsen, Martin B and Chaplin, William J and Davies, Guy R and Deheuvels, Sébastien and Ballot, Jérôme and Li, Gang and Bugnet, Lisa},
   year={2024},
   month=aug, pages={1060–1076} }

@INPROCEEDINGS{2025AAS...24511104K,
       author = {{Karim}, Waly M.~Z. and {Kiman}, Rocio and {Hallinan}, Gregg and {Buzasi}, Derek and {Garaffo}, Cecilia and {Wing}, Joshua and {Pedersen}, May and {Basu}, Sarbani and {Chaplin}, William and {Khalack}, Viktor},
        title = "{Searching for Pulsation in Low Mass Stars using Unsupervised Learning Techniques}",
    booktitle = {American Astronomical Society Meeting Abstracts \#245},
         year = 2025,
       series = {American Astronomical Society Meeting Abstracts},
       volume = {245},
        month = jan,
          eid = {111.04},
        pages = {111.04},
       adsurl = {https://ui.adsabs.harvard.edu/abs/2025AAS...24511104K}
}

@ARTICLE{2025A&A...701A.285L,
       author = {{Lund}, Mikkel N. and {Chontos}, Ashley and {Grundahl}, Frank and {Mathur}, Savita and {Garc{\'\i}a}, Rafael A. and {Huber}, Daniel and {Buzasi}, Derek and {Bedding}, Timothy R. and {Hon}, Marc and {Li}, Yaguang},
        title = "{Luminaries in the sky: The TESS legacy sample of bright stars: I. Asteroseismic detections in naked-eye main-sequence and subgiant solar-like oscillators}",
      journal = {\aap},
         year = 2025,
        month = sep,
       volume = {701},
          eid = {A285},
        pages = {A285},
          doi = {10.1051/0004-6361/202555485},
archivePrefix = {arXiv},
       eprint = {2508.08699},
 primaryClass = {astro-ph.SR},
       adsurl = {https://ui.adsabs.harvard.edu/abs/2025A&A...701A.285L}
}

@ARTICLE{2021AJ....162..209A,
       author = {{Audenaert}, J. and {Kuszlewicz}, J.~S. and {Handberg}, R. and {Tkachenko}, A. and {Armstrong}, D.~J. and {Hon}, M. and {Kgoadi}, R. and {Lund}, M.~N. and {Bell}, K.~J. and {Bugnet}, L. and {Bowman}, D.~M. and {Johnston}, C. and {Garc{\'\i}a}, R.~A. and {Stello}, D. and {Moln{\'a}r}, L. and {Plachy}, E. and {Buzasi}, D. and {Aerts}, C. and {T'DA collaboration}},
        title = "{TESS Data for Asteroseismology (T'DA) Stellar Variability Classification Pipeline: Setup and Application to the Kepler Q9 Data}",
      journal = {\aj},
         year = 2021,
        month = nov,
       volume = {162},
       number = {5},
          eid = {209},
        pages = {209},
          doi = {10.3847/1538-3881/ac166a},
archivePrefix = {arXiv},
       eprint = {2107.06301},
 primaryClass = {astro-ph.SR},
       adsurl = {https://ui.adsabs.harvard.edu/abs/2021AJ....162..209A}
}

@ARTICLE{2009A&A...494..237T,
       author = {{Teixeira}, T.~C. and {Kjeldsen}, H. and {Bedding}, T.~R. and {Bouchy}, F. and {Christensen-Dalsgaard}, J. and {Cunha}, M.~S. and {Dall}, T. and {Frandsen}, S. and {Karoff}, C. and {Monteiro}, M.~J.~P.~F.~G. and {Pijpers}, F.~P.},
        title = "{Solar-like oscillations in the G8 V star {\ensuremath{\tau}} Ceti}",
      journal = {\aap},
         year = 2009,
        month = jan,
       volume = {494},
       number = {1},
        pages = {237-242},
          doi = {10.1051/0004-6361:200810746},
archivePrefix = {arXiv},
       eprint = {0811.3989},
 primaryClass = {astro-ph},
       adsurl = {https://ui.adsabs.harvard.edu/abs/2009A&A...494..237T}
}

@ARTICLE{2015ApJ...798...73G,
       author = {{Gagn{\'e}}, Jonathan and {Lafreni{\`e}re}, David and {Doyon}, Ren{\'e} and {Malo}, Lison and {Artigau}, {\'E}tienne},
        title = "{BANYAN. V. A Systematic All-sky Survey for New Very Late-type Low-mass Stars and Brown Dwarfs in Nearby Young Moving Groups}",
      journal = {\apj},
         year = 2015,
        month = jan,
       volume = {798},
       number = {2},
          eid = {73},
        pages = {73},
          doi = {10.1088/0004-637X/798/2/73},
archivePrefix = {arXiv},
       eprint = {1410.4864},
 primaryClass = {astro-ph.SR},
       adsurl = {https://ui.adsabs.harvard.edu/abs/2015ApJ...798...73G}
}

@ARTICLE{2018ApJS..236...16V,
       author = {{VanderPlas}, Jacob T.},
        title = "{Understanding the Lomb-Scargle Periodogram}",
      journal = {\apjs},
         year = 2018,
        month = may,
       volume = {236},
       number = {1},
          eid = {16},
        pages = {16},
          doi = {10.3847/1538-4365/aab766},
archivePrefix = {arXiv},
       eprint = {1703.09824},
 primaryClass = {astro-ph.IM},
       adsurl = {https://ui.adsabs.harvard.edu/abs/2018ApJS..236...16V}
}

@INPROCEEDINGS{Audenaert2025,
       author = {{Audenaert}, Jeroen and {Hon}, Marc and {Jayaraman}, Rahul and {Kunimoto}, Michelle and {Tkachenko}, Andrew and {Buzasi}, Derek and {Lund}, Mikkel and {Ricker}, George and {The T'DA Collaboration}},
        title = "{Developing an Efficient and Modular Large-Scale Machine Learning Pipeline to Classify Millions of NASA TESS Light Curves in Search of Variable Stars}",
    booktitle = {Astronomical Society of the Pacific Conference Series},
         year = 2025,
       editor = {{Jacques}, Alice and {Seaman}, Robert and {Gandilo}, Natalie and {Linder}, Tyler},
       series = {Astronomical Society of the Pacific Conference Series},
       volume = {541},
        month = oct,
        pages = {181},
          doi = {10.26624/RXGX7087},
       adsurl = {https://ui.adsabs.harvard.edu/abs/2025ASPC..541..181A}
}

@ARTICLE{Boyle2025,
       author = {{Boyle}, Andrew W. and {Mann}, Andrew W. and {Bush}, Jonathan},
        title = "{Quantifying the Limits of TESS Stellar Rotation Measurements with the K2-TESS Overlap}",
      journal = {\apj},
         year = 2025,
        month = jun,
       volume = {985},
       number = {2},
          eid = {233},
        pages = {233},
          doi = {10.3847/1538-4357/adcecc},
archivePrefix = {arXiv},
       eprint = {2504.13262},
 primaryClass = {astro-ph.SR},
       adsurl = {https://ui.adsabs.harvard.edu/abs/2025ApJ...985..233B}
}

@ARTICLE{Brunt2006,
       author = {{Bruntt}, H. and {Buzasi}, D.~L.},
        title = "{High-precision photometry with the WIRE satellite .}",
      journal = {\memsai},
         year = 2006,
        month = jan,
       volume = {77},
        pages = {278},
          doi = {10.48550/arXiv.astro-ph/0509444},
archivePrefix = {arXiv},
       eprint = {astro-ph/0509444},
 primaryClass = {astro-ph},
       adsurl = {https://ui.adsabs.harvard.edu/abs/2006MmSAI..77..278B}
}

@ARTICLE{Matthews2003,
       author = {{Walker}, Gordon and {Matthews}, Jaymie and {Kuschnig}, Rainer and {Johnson}, Ron and {Rucinski}, Slavek and {Pazder}, John and {Burley}, Gregory and {Walker}, Andrew and {Skaret}, Kristina and {Zee}, Robert and {Grocott}, Simon and {Carroll}, Kieran and {Sinclair}, Peter and {Sturgeon}, Don and {Harron}, John},
        title = "{The MOST Asteroseismology Mission: Ultraprecise Photometry from Space}",
      journal = {\pasp},
         year = 2003,
        month = sep,
       volume = {115},
       number = {811},
        pages = {1023-1035},
          doi = {10.1086/377358},
       adsurl = {https://ui.adsabs.harvard.edu/abs/2003PASP..115.1023W}
}

@software{2025ascl.soft04002N,
       author = {{Nielsen}, M.~B. and {Davies}, G.~R. and {Ball}, W.~H. and {Lyttle}, A.~J. and {Li}, T. and {Hall}, O.~J. and {Chaplin}, W.~J. and {Gaulme}, P. and {Carboneau}, L. and {Ong}, J.~M.~J. and {Garc{\'\i}a}, R.~A. and {Mosser}, B. and {Roxburgh}, I.~W. and {Corsaro}, E. and {Benomar}, O. and {Moya}, A. and {Lund}, M.~N. and {Hatt}, E. and {Jones}, B.~P. and {Logue}, M.},
        title = "{PBjam: Automating asteroseismology of solar-like oscillators}",
 howpublished = {Astrophysics Source Code Library, record ascl:2504.002},
         year = 2025,
        month = apr,
          eid = {ascl:2504.002},
archivePrefix = {ascl},
       eprint = {2504.002},
       adsurl = {https://ui.adsabs.harvard.edu/abs/2025ascl.soft04002N}
}

@misc{shaw2025impactmastdataarchive,
      title={The Impact of the MAST Data Archive}, 
      author={Richard A. Shaw and Jenny L. Novacescu and Sarah Weissman and Travis A. Berger and Clara E. Brasseur and Jeff Chamblee and Brian Cherinka and Zachary R. Claytor and Theresa Dower and Chinwe Edeani and Scott W. Fleming and Jonathan R. Hargis and Julie Imig and Tim Januario and Karen Levay and Tim Kimball and Jenn Kotler and Hannah M. Lewis and Steve Lubow and Adrian Lucy and Brian McLean and Sunita G. Malla and Jacob Matuskey and Sophie J. Miller and Susan E. Mullally and Claire E. Murray and J. E. G. Peek and Carlita Phillip and Marc Rafelski and David R. Rodriguez and Gregory F. Snyder and Achu J. Usha and Richard L. White and Jinmi Yoon},
      year={2025},
      eprint={2512.18101},
      archivePrefix={arXiv},
      primaryClass={astro-ph.IM},
      url={https://arxiv.org/abs/2512.18101}, 
}


\section{Appendix}
\label{sec:Appendix}

\subsection{Diagnostic Plots for Candidates}
\label{sec:diagnostic-plots}
In this section, we present the diagnostic plot containing power spectrum, background-corrected power spectrum, autocorrelation function, and \'{e}chelle diagram of all the candidates in Figure~\ref{fig:teff-rad-plot} that were not shown in Figure~\ref{fig:pysyd-output}. 

\begin{figure*}
    \centering
    \includegraphics[width=0.9\linewidth]{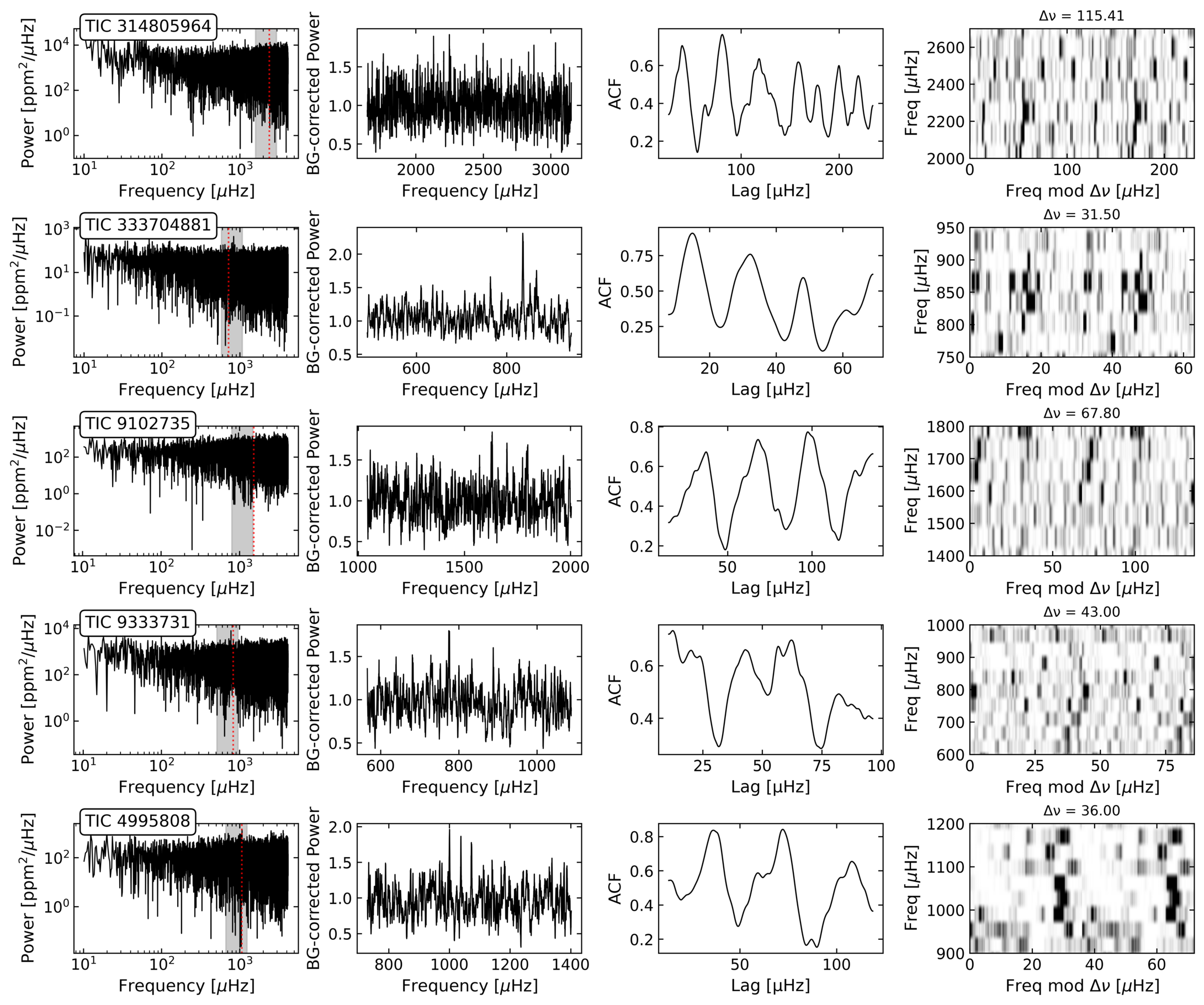}
    \caption{The same at at Figure~10, but for TIC 314805964, TIC 333704881, TIC 9102735, TIC 9333731, and TIC 4995808.}
    \label{fig:443}
\end{figure*}

\begin{figure*}
    \centering
    \includegraphics[width=0.9\linewidth]{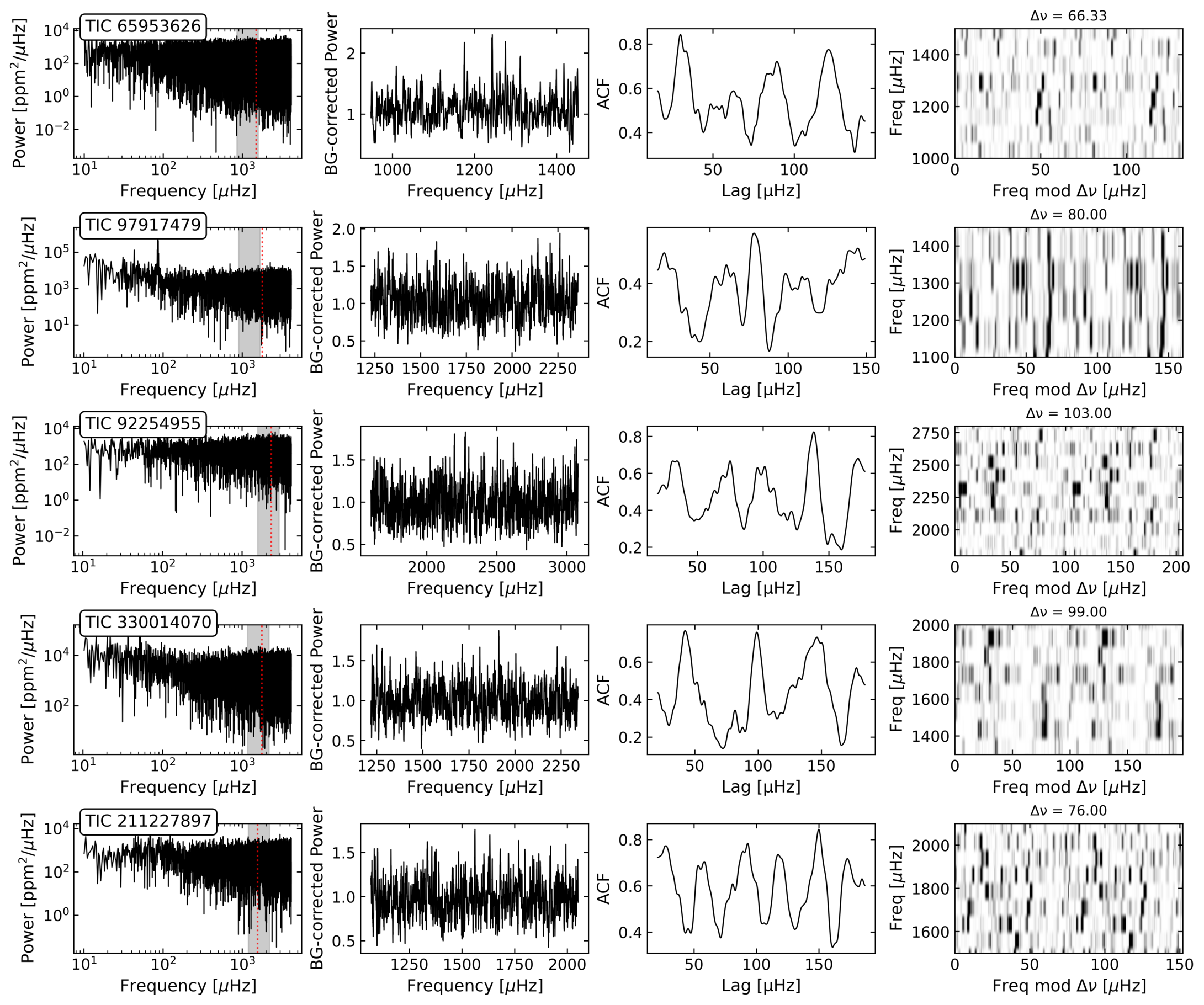}
    \caption{The same at at Figure~10, but for TIC 65953626, TIC 97917479, TIC 92254955, TIC 330014070, TIC 211227897.}
    \label{fig:placeholder}
\end{figure*}

\subsection{Neural Network}
\label{sec:encoder}
The neural network used in this study is composed of two main component: autoencoder and the classifier. The architecture of both these networks were discussed in great detail in Section~\ref{sec:network-and-training}. Here, we present a diagram of the encoder architecture, and the architecture of the classifier is the same as the classification network described in \cite{Jamal_2020}. 

\subsubsection{Autoencoder Architecture}
\label{sec:architecture}

\begin{figure}
    \centering
    \includegraphics[width=0.99\linewidth]{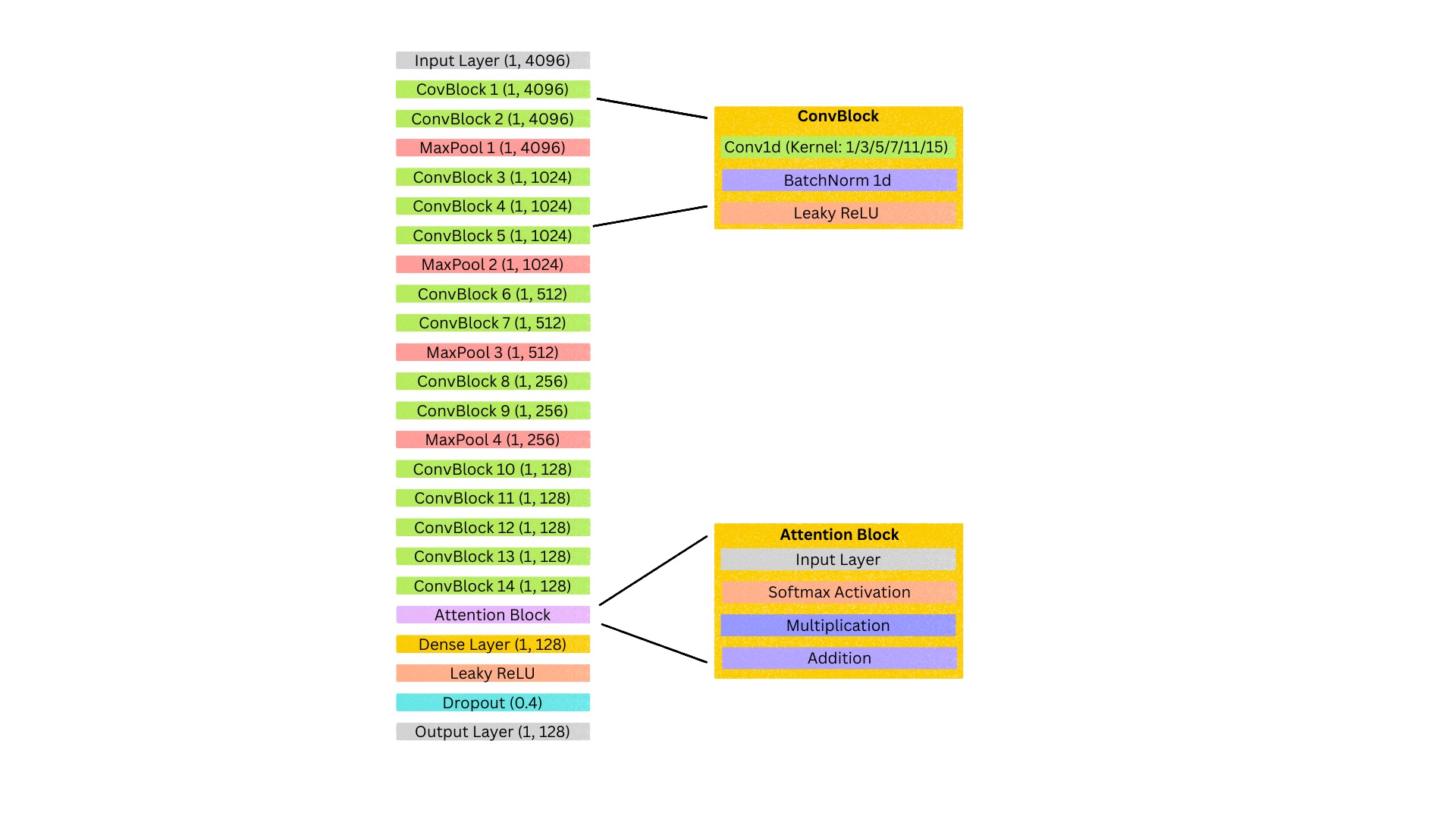}
    \caption{Detailed architecture of the encoder network}
    \label{fig:encoder}
\end{figure}

The autoencoder is comprised of an encoder and a decoder. As described in Section~\ref{sec:autoencoder}, the encoder network compresses the input signal of size 4096 down to 128 in the latent space. It is comprised of convolutional and dense layers that learn the features of input signal and pooling layers that compresses the signal to a lower dimension. The detailed architecture of the encoder network is presented in Figure~\ref{fig:encoder}.

The decoder is a mirror copy of encoder, except all the pooling layers are substituted by interpolation layers to reconstruct the original input length from low-dimensional latent space. To see the network hyperparameters used for training and description of each layers, please refer to Section~\ref{sec:network-and-training}.

\subsection{Latent Space Comparison}
\label{sec:latent-space}
Here, we present a comparison between the latent space representations of solar-like oscillators and other classes of stellar variability. The autoencoder described in Section~\ref{sec:network-and-training} takes a power spectrum binned to 4096 points as input, compresses this information into a 128-dimensional latent space, and passes this low-dimensional representation to the classifier. This latent representation effectively encodes salient frequency-domain features that enable discrimination between solar-like oscillators and other types of variables. 

\begin{figure}
    \centering
    \includegraphics[width=0.98\linewidth]{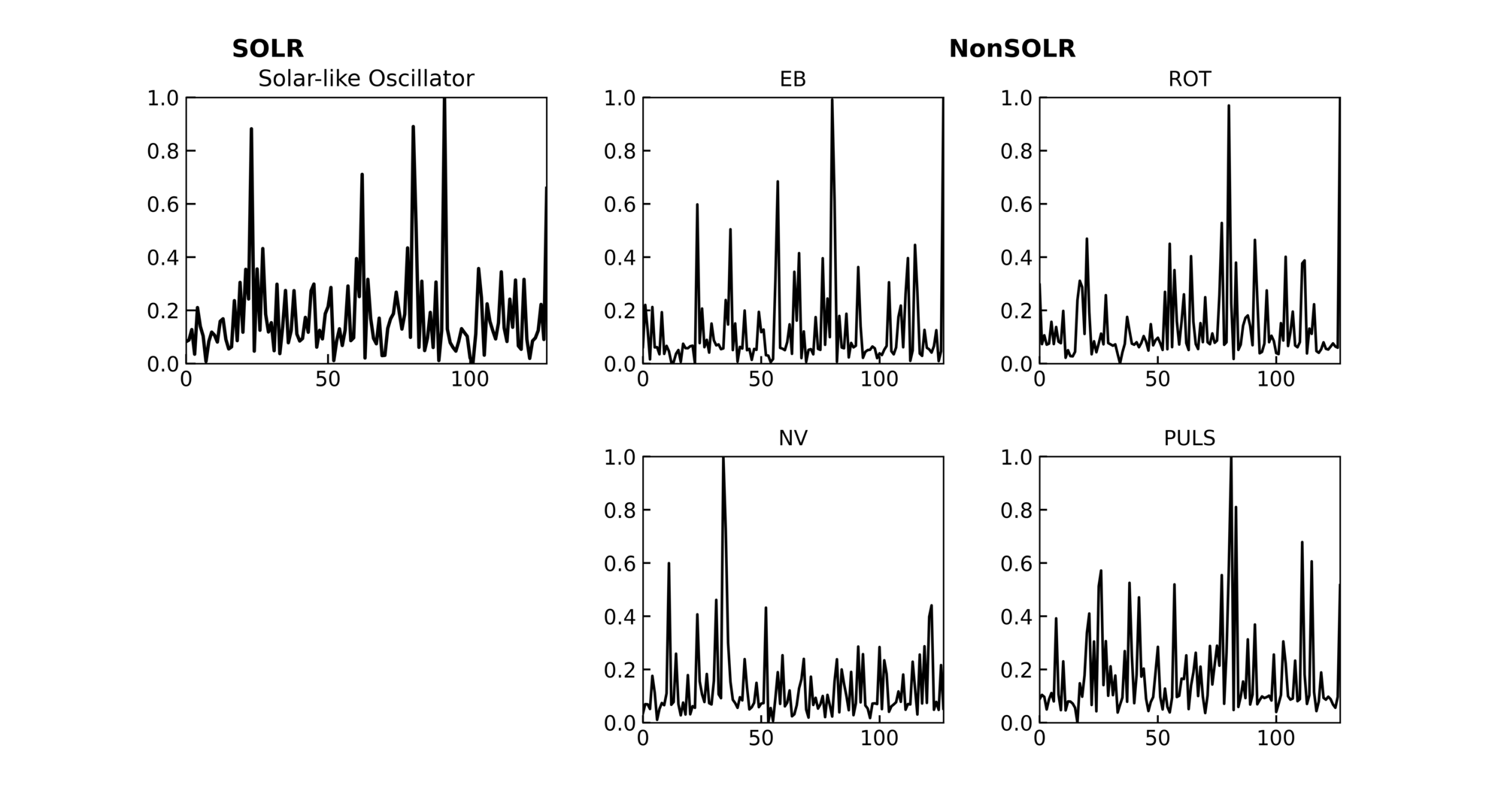}
    \caption{Comparison of latent space representation between a solar-like oscillator and other types of variables. The SOLR column represents solar-like oscillation and NonSOLR column represents other types of variables. All signals are normalized between 0 and 1.}
    \label{fig:latent-space}
\end{figure}

To verify the claim that latent space encodes features that discriminate solar-like oscillators from other types of variables, we have looked at the latent space representation of several solar-like oscillators and other types of variables from our validation set. The key distinguishing feature of solar-like oscillators in the latent space was the existence of 4 or more high-SNR peaks, while other types of variables usually had one or two high-SNR peaks followed by a cluster of low SNR peaks. An illustrative example is shown in Figure~\ref{fig:latent-space}, where we can notice the solar-like oscillator has 4 high-SNR peaks, whereas the NonSOLR class has usually one high-SNR peak followed by a groups of small peaks. This observation was consistent throughout the sample of latent spaces we have manually checked for different stars of these SOLR and NonSOLR classes.

\end{document}